 \documentclass[authoryear,preprint,12pt]{elsarticle}

\pdfoutput=1

\usepackage{amssymb}
\usepackage{amsthm}
\newtheorem{theorem}{Theorem}

\usepackage[left]{lineno} 
\usepackage[ruled, vlined,linesnumbered]{algorithm2e} 
\usepackage{mathrsfs} 
\usepackage{amsmath} 
\usepackage{MnSymbol} 
\usepackage{mathtools} 
\usepackage{rotating} 
\usepackage{booktabs} 
\usepackage{subfigure} 
\usepackage{ifpdf}
\usepackage[usenames]{color} 

\newif\ifpdf
\ifx\pdfoutput\undefined
  \pdffalse 
\else 
  \ifnum \pdfoutput=0
    \pdffalse
  \else
    \ifnum\pdfoutput=1 
    \pdftrue
    \fi
  \fi
\fi

\newcommand\independent{\protect\mathpalette{\protect\independenT}{\perp}}
\def\independenT#1#2{\mathrel{\rlap{$#1#2$}\mkern2mu{#1#2}}}

\definecolor{White}{rgb}{1,1,1}

\journal{Computational Statistics and Data Analysis}

\begin{document}

\begin{frontmatter}



  \title{Stable Graphical Model Estimation with Random Forests for
    Discrete, Continuous, and Mixed Variables}


\author[SPF,ETH]{Bernd
  Fellinghauer\corref{cor1}}
\cortext[cor1]{Corresponding author. Seminar f\"ur Statistik, ETH Z\"urich,
  R\"amistrasse 101, 8092
  Zurich, Switzerland. Tel. +41 44 632 3504; fax: +41 44 632 1228.}
\ead{bernd.fellinghauer@stat.math.ethz.ch}
\author[ETH]{Peter B\"uhlmann}
\author[ETH]{Martin Ryffel}
\author[Child]{Michael von Rhein}
\author[SPF,LU]{Jan D. Reinhardt}

\address[SPF]{Swiss Paraplegic Research, Nottwil, Switzerland}
\address[ETH]{Seminar f\"ur Statistik, ETH Z\"urich, Switzerland}
\address[Child]{Child Development Center, University Children's
  Hospital, Zurich, Switzerland}
\address[LU]{Department of Health Sciences and Health Policy,
  University of Lucerne, Lucerne, Switzerland}

\begin{abstract}
  A conditional independence graph is a concise representation of
  pairwise conditional independence among many variables. Graphical
  Random Forests (GRaFo) are a novel method for estimating pairwise
  conditional independence relationships among mixed-type,
  i.e.~continuous and discrete, variables. The number of edges is a
  tuning parameter in any graphical model estimator and there is no
  obvious number that constitutes a good choice. Stability Selection
  helps choosing this parameter with respect to a bound on the
  expected number of false positives (error control).

\textbf{The performance of GRaFo is evaluated and compared with
  various other methods for $p = 50$, $100$, and $200$ possibly
  mixed-type variables while sample size is $n = 100$ ($n = 500$ for
  maximum likelihood). Furthermore,} GRaFo is applied to data from the
Swiss Health Survey in order to evaluate how well it can reproduce the
interconnection of functional health components, personal, and
environmental factors, as hypothesized by the World Health
Organization's International Classification of Functioning, Disability
and Health (ICF). Finally, GRaFo is used to identify risk factors
which may be associated with adverse neurodevelopment of children who
suffer from trisomy 21 and experienced open-heart surgery.

GRaFo performs well with mixed data and thanks to Stability Selection
it provides an error control mechanism for false positive selection.
\end{abstract}


\end{frontmatter}

\linenumbers


\section{Introduction}

In many problems one is not confined to one response and a set of
predefined predictors. In turn, the interest is often in the association
structure of a whole set of $p$ variables, i.e.~asking whether two
variables are independent conditional on the remaining $p-2$ variables. A
conditional independence graph (CIG) is a concise representation of such
pairwise conditional independence among many possibly mixed,
i.e.~continuous and discrete, variables. In CIGs, variables appear as
nodes, whereas the presence (absence) of an edge among two nodes represents
their dependence (independence) conditional on all other
variables. Applications include among many others also the study of
functional health \citep{Strobl2009,Kalisch2010,Reinhardt2010}.

We largely focus on the high-dimensional case where the number of variables
(nodes in the graph) $p$ may be larger than sample size $n$. A popular
approach to graphical modeling is based on the Least Absolute Shrinkage and
Selection Operator \citep[LASSO;][]{Tibshirani1996}: see
\citet{Meinshausen2006} or \citet{Friedman2008} for the Gaussian case and
\citet{Ravikumar2010} for the binary case. However, empirical data often
involve both discrete and continuous variables. Conditional Gaussian
distributions were suggested to model such mixed-type data with maximum
likelihood inference \citep{Lauritzen1989}, but no corresponding
high-dimensional method has been suggested yet. Dichotomization, though
always applicable, comes at the cost of lost information
\citep{MacCallum2002}.

Tree-based methods are easy to use and accurate for dealing with
mixed-type data \citep{Breiman1984}. Random Forests
\citep{Breiman2001} evaluate an ensemble of trees often resulting in
notably improved performance compared to a single tree \citep[see
  also][]{Amit1997}. Furthermore, permutation importance in Random
Forests allows to rank the relevance of predictors for one specific
response. \textbf{However, Random Forests have also been criticized to
  perform possibly biased variable selection. We thus also consider
  Conditional Forests \citep{StroblC2007} and conditional variable
  importance \citep{StroblC2008}, which have been suggested to
  overcome this behavior.}

In general, the definition of \textbf{both the conditional and
  marginal} permutation importance differ for discrete and continuous
responses. Thus, ranking permutation importances across responses of
mixed-type is less obvious. However, such ranking is essential to
derive a network of the most relevant dependencies. Stability
Selection proposed by \citet{Meinshausen2010} is one possible
framework to rank the edges in the CIG across different types of
variables.  In addition, it allows to specify an upper bound on the
expected number of false positives, i.e.~the falsely selected edges,
and thus provides a means of error control.

We combine Random Forests estimation with appropriate ranking among
mixed-type variables and error control from Stability Selection. We
refer to the new method as Graphical Random Forests (GRaFo). The
specific aims of the paper are a) to evaluate and compare the
performance of GRaFo \textbf{with Stable LASSO (StabLASSO) and Stable
  Conditional Forests (StabcForests), which are LASSO- and conditional
  forest-based alternatives, and regular maximum likelihood (ML)
  estimation} across \textbf{various} simulated settings comprising
different distributions\textbf{, interactions, and nonlinear
  associations} for $p=50$, $100$, and $200$ possibly mixed-type
variables while sample size is $n=100$ ($n=500$ for ML), b) to apply
GRaFo to data from the Swiss Health Survey (SHS) to evaluate the
interconnection of functional health components, personal, and
environmental factors, as hypothesized by the World Health
Organization's (WHO) International Classification of Functioning,
Disability and Health (ICF), and c) to use GRaFo to identify risk
factors associated with adverse neurodevelopment in children with
trisomy 21 after open-heart surgery \textbf{and more generally to
  assess the plausibility of the suggested associations}.

\section{Graphical Modeling Based on Regression-Type Methods}

\subsection{Conditional Independence Graphs}
Let ${\bf X}=\{X_1,\ldots,X_p\}$ be a set of (possibly) mixed-type random
variables. The associated conditional independence graph of $\bf
X$ is the undirected graph $G_{\mbox{\tiny
    CIG}}=(\mathcal{V},\mathcal{E}(G_{\mbox{\tiny CIG}}))$, where the nodes
in $\mathcal{V}$ correspond to the $p$ variables in ${\bf X}$. The edges
represent the pairwise Markov property, i.e.~$i-j\not\in\mathcal{E}(G_{\mbox{\tiny CIG}})$ if and only if
$X_j\independent X_i|{\bf X}\setminus \{X_j,X_i\}$. For a rigorous
introduction to graphical models, see, for example, the monographs by
\citet{Whittaker1990} or \citet{Lauritzen1996}.

We will now show that the pairwise Markov property can, under certain
conditions, be inferred from conditional mean estimation.

\begin{theorem}
\label{TheoremCigMeanEst}
Assume that, for all $j=1,\ldots,p$, the conditional distribution of
$X_j$ given $\{X_h; h\neq j\}$ is depending on any realization $\{x_h;
h\neq j\}$ only through the conditional mean function
  \begin{align*}m_j(\{x_h; h\neq j\})=\mathbb{E}[X_j|\{x_h; h\neq j\}],
\end{align*}
 that is:
 \begin{align*} \mathbb{P}[X_j\leq x_j|\{x_h;
h\neq j\}]=F_j(x_j|m_j(\{x_h; h\neq j\})),\tag{A}
\end{align*} where
$F_j(\cdot|m)$ is a cumulative distribution function for all
$m\in\mathbb{R}$ (or $m\in\mathbb{R}^d$ if $X_j$ is
$d$-dimensional). (Thereby, we assume that the conditional mean
exists). Then
\begin{align*}
X_j\independent X_i|\{X_h; h\neq j,i\}
\end{align*}
 if and only if
\begin{align*}
m_j(\{x_h; h\neq j\})=m_j(\{x_h; h\neq j,i\})
\end{align*} does not depend on
$x_i$, for all $\{x_h; h\neq j\}$.
\end{theorem}
A proof is given in Section~\ref{Sec:Proofs}. Assumption $(A)$
trivially holds for a Bernoulli random variable $X_j$:
\begin{align*}
  \mathbb{P}[X_j=1|\{x_h; h\neq j\}]=\mathbb{E}[X_j|\{x_h; h\neq
  j\}]=m_j(\{x_h; h\neq j\}).
\end{align*}
Analogously, for a multinomial random variable
$X_j$ with $C$ levels, the probability that $X_j$ takes the level
$r\in\{1,\ldots,C\}$ can be expressed via a Bernoulli variable
$X_j^{(r)}$ with
\begin{align*}\mathbb{P}[X_j^{(r)}=1|\{x_h; h\neq
  j\}]=\mathbb{E}[X_j^{(r)}|\{x_h; h\neq j\}]=m_j(\{x_h; h\neq j\}).
 \end{align*}
Hence, $(A)$ holds. Moreover, if
$(X_1,\ldots,X_p)\sim\mathcal{N}_p({\bf 0},{\bf \Sigma})$, then $(A)$ holds
as well \citep[see for example][]{Lauritzen1996}. However, for the
  Conditional Gaussian distribution \citep[or CG distribution,
  see e.g.][]{Lauritzen1996}, we need to require for $(A)$ that the variance is
fixed and is not depending on the variables we condition on. For example,
let $X_1\sim \mathcal{B}(1,\pi)$ be Bernoulli distributed and let
\begin{align*}
X_2|X_1\sim\left\{\begin{matrix}\mathcal{N}(\mu_1,\sigma_1^2), \mbox{ if }
  X_1=1\\
\mathcal{N}(\mu_2,\sigma_2^2), \mbox{ if } X_1=0
\end{matrix}\right., \mbox{ where } \sigma_1^ 2\neq\sigma_2^2.
\end{align*}
Then the distribution of $X_2|X_1$ is not a function of the
conditional mean alone.

Theorem~1 motivates our approach to infer conditional dependences, or
edges in the CIG, via variable selection for many nonlinear
regressions, i.e.~determining whether a variable $X_i$ is relevant in
$\mathbb{E}[X_j|{\bf X}\setminus\{X_j\}]$ (regression of $X_j$ versus
all other variables).

\subsection{Ranking Edges}
In order to determine which edges should be included in the graphical
model, the edges suggested by the individual regressions need to be ranked
such that a smaller rank indicates a better candidate for
inclusion\footnote{For instance, if all variables are continuous, the size of the
  standardized regression coefficients from ordinary least squares is an
  obvious global ranking criterion. Analogously, in a situation where all
  variables are binary (and identically coded), coefficients from linear
  logistic regression lead to a global ranking.}. Note that each
edge $i-j$ is associated with two coefficients ($X_j$ regressed on $X_i$
and all other variables and vice versa for $X_i$ on $X_j$). To be
conservative, we rank each edge $i-j$ relative to the smaller one of the
two (absolute-valued) ranking coefficients.

If variables are mixed-type, a global ranking criterion is difficult
to find. For example, \textbf{continuous and categorical} response variables
  are not directly comparable. Instead, local rankings for each regression
are performed separately (where ``local'' means that we can rank the importance
of predictors for every individual regression). Analogous to global
ranking, each edge $i-j$ is associated with two possible ranks and the
worse among them is used.

When using Random Forests for performing the individual nonlinear
regressions, the ranking scheme is obtained from Random Forests'
variable importance measure. \textbf{For Conditional Forests, both the
  conditional and marginal variable importances can be used.} When
using the LASSO for individual linear or logistic regressions,
\textbf{the ranking scheme is obtained from the value of the penalty
  parameter $\lambda$ for which an estimated regression coefficient
  first becomes non-zero (i.e.~the value of the penalty parameter when
  a variable enters in a coefficient path plot).}

We then have to decide on the number of edges to select, i.e.~the
  tuning parameter. Say it is given as $q=11$. Then, for both global and
  local rankings, we select the $11$ best-ranked edges across all $p$
  individual regressions. If this is impossible due to tied ranks
  (e.g.~because the 11th and 12th best edges have a tied rank of 11.5), we
  neglect these (here: two) tied edges and select only the remainder of
  (here: 10) edges not in violation of the tuning parameter.

We next outline how Stability Selection can be used to guide the choice of
$q$.

\subsection{Aggregating Edge Ranks with Stability Selection}
Stability Selection \citep{Meinshausen2010} allows the specification
of an upper bound on the expected number $\mathbb{E}[V]$ of false
positives. It is based on subsampling
\citep{Politis1999,Buhlmann2002} random subsets ${\bf
  X}^{(1)},\ldots,{\bf X}^{(n_{\mbox{\tiny sub}})}$ of the original
sample ${\bf X}_1,\ldots,{\bf X}_n$, where each ${\bf X}^{(k)}$
contains $\lfloor n/2 \rfloor$ sample points.

Let $\mathcal{E}(\hat{G}_{\mbox{\tiny CIG}}({\bf X}^{(k)}))$ denote the
edges from a thresholded ranking based on ${\bf X}^{(k)}$,
$k=1,\ldots,n_{\mbox{\tiny sub}}$. Stability Selection suggests to
construct $\mathcal{E}(\hat{G}_{\mbox{\tiny CIG}}({\bf X}))$, the set of
all edges in the estimated CIG of ${\bf X}$, from all edges that were
``sufficiently stable'' across the $n_{\mbox{\tiny sub}}$
subsets. More concretely, we choose only edges $i-j$ which fulfill
\begin{align}
\label{pithr}
\frac{1}{n_{\mbox{\tiny sub}}}\sum_{k=1}^{n_{\mbox{\tiny
      sub}}}I_{\{i-j\in\mathcal{E}(\hat{G}_{\mbox{\tiny CIG}}({\bf X}^{(k)}))\}}\geq \pi_{\mbox{\tiny thr}},
\end{align}
where $\pi_{\mbox{\tiny thr}}$ imposes a threshold on the minimum
relative frequency of edges across the $n_{\mbox{\tiny sub}}$ subsets
to be included in $\mathcal{E}(\hat{G}_{\mbox{\tiny CIG}}({\bf X}))$
\textbf{and $I$ is the indicator function}.

In their Theorem 1, \citet{Meinshausen2010} relate $\mathbb{E}[V]$
  to the maximum number of selected edges $q$ per subset, the number of
  possible edges $p\cdot (p-1) / 2$ in $\mathcal{E}(\hat{G}_{\mbox{\tiny
      CIG}}({\bf X}))$, and the threshold $\pi_{\mbox{\tiny thr}}$ from
  formula~(\ref{pithr}) (requiring~$\pi_{\mbox{\tiny
      thr}}\in\left(\frac{1}{2},1\right)$):
\begin{align}
\label{StabSelErrorControl}
      \mathbb{E}[V]\leq\frac{q^2}{(2\pi_{\tiny
          \mbox{thr}}-1)\cdot p\cdot(p-1)/2}.
\end{align}

The expected number of false positives $\mathbb{E}[V]$, which is a
  type I error measure, needs to be specified a priori. The parameters
  $\pi_{\mbox{\tiny thr}}$ and $q$ are tuning parameters that depend on
  each other. More precisely, to obtain a stable graph estimate for a given
  $\mathbb{E}[V]$, the threshold $\pi_{\mbox{\tiny thr}}$ has to be large
  if the number of selected edges $q$ is large and vice versa. Consequently
  \citep[and as also argued by][]{Meinshausen2010} the actual values of
  $\pi_{\mbox{\tiny thr}}$ and $q$ are of minor importance for a given
  $\mathbb{E}[V]$ as the graph estimates do not vary much for different
  choices of $\pi_{\mbox{\tiny thr}}$ (results not shown). We thus fix
  $\pi_{\mbox{\tiny thr}}=0.75$ throughout the paper. Also, we follow the
  suggestion of \citet{Meinshausen2010} in choosing $n_{\mbox{\tiny sub}} =
  100$.

We can then use formula (\ref{StabSelErrorControl}) to derive
\begin{align*}
q=\lfloor\sqrt{(2\pi_{\mbox{\tiny thr}}-1)\mathbb{E}[V]\cdot p\cdot
  (p-1)/2}\rfloor
\end{align*}
by specifying the value of $\mathbb{E}[V]$ as desired (according to the
willingness to accept false positives).

Note that formula~(\ref{StabSelErrorControl}) is based on two assumptions:
1) the estimation procedure is better than random guessing and 2) the
probability of a false edge to be selected is exchangeable; for details we
refer to \citet{Meinshausen2010}. Also note that $\pi_{\mbox{\tiny
      thr}}$ is not to be interpreted as an edge probability threshold but
  solely as a means to assess stability which allows control of
  $\mathbb{E}[V]$. Finally, be aware that our method does not
  consider the goodness-of-fit of the model but instead leads to an
  undirected graph whose edges are controlled for false positive selections.

\section{Random Forests, Conditional Forests, LASSO Regression, and
  Maximum Likelihood}

\subsection{Random Forests}

Random Forests have, to date, not been used to estimate CIGs. They
perform a series of recursive binary partitions of the data and
construct the predictions from terminal nodes. Based on classification
and regression trees \citep{Breiman1984} they allow convenient
inference for mixed-type variables, also in the presence of
interaction effects. Incorporating bootstrap
\citep{Efron1979,Breiman1996} and random feature selection
\citep{Amit1997}, random subsets of both the observations and the
predictors are considered. The relevance of each predictor can be
assessed with permutation importance \citep{Breiman2002}, a measure of
the error difference between a regular Random Forests fit and a Random
Forests fit within which one predictor has been permuted at random to
purge its relationship with the response. An implementation of Random
Forests in R \citep{R2010} is available in the randomForest package
\citep{Liaw2002}. We chose the number of trees and the number of
features randomly selected per tree according to the package
defaults. \textbf{Further extensions (which we did not incorporate)
  allow to explicitly use the ordinal information of a categorical
  response: see e.g.~the R packages party \citep{Hothorn2006} and
  rpartOrdinal \citep{Archer2010}.}

Since the goodness-of-fit of continuous and categorical
responses is based on mean squared errors and majority votes, respectively,
the goodness-of-fit and importance measures are not directly comparable
across mixed-type responses. Thus a local ranking is derived, where each
edge $i-j$ is assigned either the rank of the permutation importance of
predictor $X_i^{(k)}$ for response $X_j^{(k)}$ or of predictor $X_j^{(k)}$
for response $X_i^{(k)}$ (whichever is more conservative, i.e.~assigns a
worse rank) and finally aggregated with Stability Selection; the upper
index $^{(k)}$ denotes the $k^{th}$ subsample in Stability Selection. We
refer to this procedure as Graphical Random Forests (GRaFo) henceforth.

\subsection{Conditional Forests}
\label{subsec:CondFor}
\textbf{
 \citet{StroblC2007} criticized Random Forests to favor variables with
 many categories. Furthermore, Random Forests have been criticized to
 favor correlated predictors, even if not all of them are influential
 for the response\footnote{This aspect though may be
 considered as both a source of bias and a beneficial effect as
 correlated predictors may help to localize relevant structures
 \citep{Nicodemus2010}} \citep{StroblC2008}.
}

\textbf{ To overcome the first limitation, Conditional Forests
  \citep{StroblC2007} were suggested, which are a modification of the
  original Random Forests implementation. They are based on
  conditional inference trees \citep{Hothorn2006}, an unbiased tree
  learning procedure, to obtain an unbiased ensemble of trees.}

\textbf{While the regular marginal permutation importance discussed in
  the previous section is also applicable to Conditional Forests, a
  conditional permutation importance, which aims to preserve the
  correlation structure among predictors, has been suggested by
  \citet{StroblC2008} to overcome the latter critique of forest
  ensembles favoring correlated predictors. An implementation of
  Conditional Forests, including the conditional variable importance,
  is available in the party package \citep{Hothorn2006} in R. However,
  we found that the computational cost to obtain the conditional
  variable importance is a lot higher than for the marginal
  permutation importance. When drastically reducing the number of
  trees to 10, the computations become feasible but the ensemble does
  hardly produce any true positives (likely due to instability of the
  small forest ensemble). As such, all calculations reported further
  below have been performed using the marginal permutation
  importance. To allow a fair comparison, we set the ensemble size to
  500 trees (as with Random Forests).}

\textbf{
The same ranking rule as for Random Forests can then be used to
construct a Stable Conditional Forest (StabcForests) algorithm.
}

\subsection{Least Absolute Shrinkage and Selection Operator (LASSO)}

In the case of linear regression for continuous responses and
predictors, the LASSO \citep{Tibshirani1996} penalizes with the
$\ell_1$-norm and corresponding penalty parameter $\lambda$ the
coefficients of some less relevant predictors to zero. The larger
$\lambda$ is chosen, the more coefficients will be set to zero. This
concept has also been extended to logistic regression
\citep{Lokhorst1999} and implemented in R in the glmnet package
\citep{Friedman2010}. In the case of multinomial and mixed-type
data, no eligible off-the-shelf implementation of the LASSO is
available.  We hence dichotomize these data according to a median
split for continuous variables and aggregate categories such that the
resulting frequency of the -1 and 1 categories was as balanced as
possible for discrete variables. Consequently, a loss of information is to
be expected \citep[cf.,][]{MacCallum2002,Altman2006,Royston2006}.

CIG estimation via the LASSO with Stability Selection was suggested
for Gaussian data by \citet{Meinshausen2010} and can be represented as
a global ranking. For each response $X_j^{(k)}$, we estimate LASSO
regressions with all remaining ${\bf X}^{(k)}\setminus\{X_j^{(k)}\}$
as predictors and with a decreasing sequence of penalties
$\lambda_j^{(k),\mbox{\tiny max}},\ldots,\lambda_j^{(k),\mbox{\tiny
    min}}$. Let $\lambda_{ij}^{(k)}$ denote the largest penalty value
of the sequence for which the coefficient of predictor $X_i^{(k)}$ for
response $X_j^{(k)}$ is non-zero, and if no such penalty exists let
$\lambda_{ij}^{(k)}=0$. For each edge $i-j$ we select the more
conservative penalty
$\lambda_{i-j}^{(k)}=\min\left(\lambda_{ij}^{(k)},\lambda_{ji}^{(k)}\right)$
and rank $i-j$ relative to the global rank from the absolute-valued
estimated regression coefficient corresponding to
$\lambda_{i-j}^{(k)}$. As before, the upper index $^{(k)}$ denotes the
$k^{th}$ subsample from Stability Selection. We denote this procedure
in combination with Stability Selection as Stable LASSO (StabLASSO).

\subsection{Maximum Likelihood}
\textbf{Ordinary maximum likelihood (ML) estimation does neither
  impose a penalty (such as the LASSO) nor does it use subsampling to
  reduce the number of predictors to consider in each run (such as the
  Forest-type algorithms). Consequently, ordinary ML inference can
  only be applied in the case, where the number of parameters to be
  estimated is at most as large as the sample size $n$.}

\textbf{ If the dependent variable is continuous, we use the ordinary
  linear model, otherwise the multinomial log-linear model. Local
  rankings are obtained from the F-Test for each of the predictor
  variables. The calculations were performed with the regr0 package
  (available from R-Forge) in R.}

\textbf{We could wrap a Stability Selection scheme around ML
  estimation which is computationally demanding in the case of mixed
  continuous and categorical variables. Our main goal here, however,
  is to compare with plain ML estimation.}

\section{Simulation Study}

\subsection{Simulating Data from Directed Acyclic Graphs}
We use a directed acyclic graph \citep[DAG; cf.,][]{Whittaker1990} to embed
conditional dependence statements among nodes representing the $p$ random
variables. The associated CIG follows by moralization, i.e.~connecting any
two parents with a common child that are not already connected and removing
all arrowheads \citep{Lauritzen1988}.

Let $\mathcal{A}$ be a $(p\times p)$-dimensional weight matrix with
entries $a_{ij}\in\{[-1,-0.1]\cup\{0\}\cup[0.1,1]\}$ if $i<j$ and
$a_{ij}=0$ otherwise. In addition, we sample $\mathcal{A}$ to be
sparse, i.e.~we expect only one percent of its entries to deviate from
$0$. The non-zeros in $\mathcal{A}$ encode the directed edges in a DAG
we simulate from similarly as in \citet{Kalisch2007}; see also
Table~\ref{tab:SimModels}. \textbf{For the Gaussian setting with
  interaction effects, we furthermore sample
  $b_{ikj}\in\{[-1,-0.1]\cup\{0\}\cup[0.1,1]\}$ for all indices
  $i,k,j$ where main effects between $i,j$ and $k,j$ are present
  (cf.,~Table~\ref{tab:SimModels})}. Also, for all
$i,j\in\{1,\ldots,p\}$ \textbf{in the multinomial and mixed setting}
with $a_{ij}\neq 0$ let $u_{ij}$ and $v_{ij}$ be vectors that we use
to impose some additional structure on multinomial variables: 1) at
least one category of a multinomial predictor $X_i$ should have an
effect opposite to the remainder, 2) the (total) effect of the
categories of a multinomial predictor $X_i$ should be positive on some
categories of a multinomial response $X_j$ and negative on others. For
this purpose, we restrict
$u_{ij}=(u_{ij}^{(1)},\ldots,u_{ij}^{(C_i)})$ and
$v_{ij}=(v_{ij}^{(1)},\ldots,v_{ij}^{(C_j)})$:
\begin{align*}
u_{ij}^{(l)}&\in\{-1,1\}\ \forall l=1,\ldots,C_i \mbox{ s.t. }-C_i<\sum_{l=1}^{C_i} u_{ij}^{(l)}<C_i,\\
v_{ij}^{(s)}&\in\{-1,1\}\ \forall s=1,\ldots,C_j \mbox{ s.t. }-C_j<\sum_{s=1}^{C_j} v_{ij}^{(s)}<C_j.
\end{align*}

With these definitions, we sample data from different distributions
using the inverse link function to relate the conditional mean to all
previously sampled predictors. Table~\ref{tab:SimModels} describes the
settings in detail, covering models with purely Gaussian, purely
Bernoulli, purely multinomial, and an alternating sequence of Gaussian
and multinomial variables (``mixed'' setting). \textbf{The Gaussian
  setting can be further distinguished into a main effects only
  setting, a main plus interaction effects setting, and a nonlinear
  effects setting. For the nonlinear setting the signal was amplified
  by a factor of 5 to obtain comparable results to the other Gaussian
  settings. The exact specifications are given in
  Table~\ref{tab:SimModels}.}

\begin{sidewaystable}
\begin{center}
\begin{tabular}[H]{lll}
  \toprule
  Distribution & Model & Conditional Mean\\
  \midrule
  Gaussian   &  $X_j\sim \mathcal{N}(\mu_j,\sigma^2=1)$ &
  $\mu_j=\sum_{i<j}a_{ij}x_i$\\
  Gaussian   &  $X_j\sim \mathcal{N}(\mu_j,\sigma^2=1)$, with~$I_j\subseteq\{(i,k):a_{ij}\neq 0,a_{kj}\neq 0\}$ &
  $\mu_j=\sum_{i<j}a_{ij}x_i$\\
  \quad +Interactions & s.t.~$|I_j|\approx|\{(i,k):a_{ij}\neq
  0,a_{kj}\neq 0\}|/2$ & $\quad+ \sum_{(i,k)\in I_j}b_{ikj}x_ix_k$\\
  Gaussian   &  $X_j\sim \mathcal{N}(\mu_j,\sigma^2=1)$,
  with~$L_j\subseteq\{1,\ldots,j\}$ s.t.~$|L_j|\approx j/2$ &
  $\mu_j=\sum_{i\in L_j}5a_{ij}x_i$\\
  \quad +Nonlinear & and $\bar L_j=\{1,\ldots,j\}\setminus L_j$
  & $\quad+ \sum_{i\in\bar L_j}5a_{ij}\log(|x_i|)$\\
  Bernoulli & $X_j=2\widetilde{X}_j-1$,  &
  $\pi_j=\frac{\exp(\sum_{i<j}a_{ij}x_i)}{1+\exp(\sum_{i<j}a_{ij}x_i)}$\\
& $\widetilde{X}_j\sim\mathcal{B}(1,\pi_j)$ &\\
Multinomial &
$X_j\sim\mathcal{M}(\boldsymbol\pi_j=(\pi_j^{(1)},\ldots,\pi_j^{(C_j)})),$ &
$\pi_j^{(s)}=\frac{\exp(\eta_j^{(s)})}{\sum_{r=1}^{C_j}\exp(\eta_j^{(r)})}$\\
&
$\eta_j^{(s)}=\sum_{i<j}v_{ij}^{(s)}a_{ij}\sum_{l=1}^{C_i}u_{ij}^{(l)}(2I_{\{x_i=l\}}-1)$,
 &\\
& $C_j\sim\mathcal{U}\{3,4,5\},\ s=1,\ldots,C_j$
 &\\
Mixed & $ X_j\sim\left\{\begin{matrix*}[l]
\mathcal{N}(\mu_j,\sigma^2=1),&  \mbox{ if } \frac{j}{2}\not\in\mathbb{N}\\
\mathcal{M}(\boldsymbol\pi_j=(\pi_j^{(1)},\ldots,\pi_j^{(C_j)})),& \mbox{ else}
\end{matrix*}\right.$ & $\begin{matrix*}[l] \mu_j=\eta_j^{(1)}
\\ \pi_j^{(s)}=\frac{\exp(\eta_j^{(s)})}{\sum_{r=1}^{C_j}\exp(\eta_j^{(r)})}
\end{matrix*}$
\\
&
$\eta_j^{(s)}=\sum_{i:i<j\wedge\frac{i}{2}\not\in\mathbb{N}}v_{ij}^{(s)}a_{ij}x_{ij}+$
\\
& $\phantom{\eta_j^{(s)}}+\sum_{i:i<j\wedge\frac{i}{2}\in\mathbb{N}}v_{ij}^{(s)}a_{ij}\sum_{l=1}^{C_i}u_{ij}^{(l)}(2I_{\{x_i=l\}}-1)$ &\\
& $C_j\sim\mathcal{U}\{3,4,5\},\ s=1,\ldots,C_j$ & \\
  \bottomrule
\end{tabular}
\caption{The table shows the six simulation models based on
  DAGs. $\mathcal{N},\ \mathcal{B},\ \mathcal{M},\mbox{and }
  \mathcal{U}$ are the Gaussian, Bernoulli, multinomial, and discrete
  uniform distribution, respectively. Initial values for $X_1$ are
  sampled with $\mu_1=0$, $\pi_1=\frac{1}{2}$, and
  $\boldsymbol\pi_1=(\frac{1}{C_1},\ldots,\frac{1}{C_1})$,
  respectively, where $C_1\sim\mathcal{U}\{3,4,5\}$. The weights
  $a_{ij}$ and $b_{ikj}$ are chosen from
  $\{[-1,-0.1]\cup\{0\}\cup[0.1,1]\}$ to determine the dependence
  relationships among the random variables. The scalars $u_{ij}^{(l)}$
  and $v_{ij}^{(s)}$ are chosen from $\{-1,1\}$ to impose additional
  structures on multinomial random variables. \textbf{$I_j$ is a
    random set of index numbers, s.t.~the number of interactions is
    about half as big as the number of associations with a non-zero
    coefficient $a_{ij}$. $L_j$ is a random set of index numbers,
    s.t.~about half of the associations are linear and the other half are
    nonlinear.}}
\label{tab:SimModels}
\end{center}
\end{sidewaystable}

\subsection{Simulating Data from the Ising Model}

A common approach to model pairwise dependencies between a set of binary
variables is the Ising model with probability function
\begin{align}
\label{eq:IsingModel}
p({\bf
  x},\Theta)=\exp\left(\sum\theta_{ii}x_i+\sum\theta_{ij}x_ix_j-\Gamma(\Theta)\right)
\end{align}
for realizations ${\bf x}\in {\bf X}$, normalization constant
$\Gamma(\Theta)$, and $(p\times p)$-dimensional symmetric parameter matrix
$\Theta=\{\theta_{ij}\}_{i,j\in\{1,\ldots,p\}}$. From the conditional
densities of equation (\ref{eq:IsingModel}) if follows that $\theta_{ij}=0\
(\theta_{ij}\neq 0)$ implies the absence (presence) of edge $i-j$ in the
associated CIG. See also \citet{Ravikumar2010}.

We sample the diagonal and the upper-triangular matrix of $\Theta$
uniformly from $\{-1,0,1\}$ such that the average neighborhood size
for each node equals $4$. The lower-triangular matrix equals its upper
counterpart. We use the Gibbs sampler \citep[cf.,][]{Givens2005} to
sample realizations from equation
(\ref{eq:IsingModel}). \citet{Hoefling2009} provide an implementation
in the BMN package in R.

\subsection{Simulation Results: Gaussian, Binomial, Multinomial,
  Mixed, and Ising}
\label{Sec:SimRes}

For $p\in\{50,100,200\}$ variables and samples of size $n=100$, each
of the $5$ simulation models\footnote{In this section, the Gaussian
  setting refers to the first model in Table~\ref{tab:SimModels},
  i.e.~the Gaussian setting without interaction effects and without
  nonlinear effects.} was averaged over $50$ repetitions. More
precisely, for a given $q$, the number of observed true and false
positives across the 50 repetitions was averaged. The results are
shown in Figures \ref{FigGBI50}-\ref{FigMMi200}. Error control for
small bounds on the expected number of false positives $\mathbb{E}[V]$
could be achieved for both GRaFo and StabLASSO in all but the mixed
setting with $p=200$ in Figure \ref{FigMMi200}.

In the Gaussian, Bernoulli and Ising settings, StabLASSO seems to
perform slightly better than GRaFo for small error bounds and rather
similar across the figures for the true/false positive rates (third
column of Figures \ref{FigGBI50}-\ref{FigGBI200}). Note that StabLASSO
sets many coefficients to 0. As a consequence, a large proportion of
edges cannot be selected for false positive rates smaller than $1$
resulting in some StabLASSO curves not covering the entire range of
the rates.

In the multinomial and mixed setting
  (Figures~\ref{FigMMi50}-\ref{FigMMi200}), GRaFo returned satisfactory
  results while StabLASSO performed poorly, presumably caused by
  dichotomization. In general, both procedures seem to perform best in the
Gaussian setting, followed by the mixed, multinomial, Bernoulli, and Ising
setting, respectively. The latter seems especially hard for both procedures
if the upper error bound in formula (\ref{StabSelErrorControl}) for
$\mathbb{E}[V]$ is chosen small. Nevertheless, given one's willingness to
expect more errors, the rate figures indicate the potential to recover
(parts of) the true structure \citep[cf.,][]{Ravikumar2010,Hoefling2009}.

The ``raw'' counterparts, Random Forests and LASSO, correspond to
estimations and rankings performed on the full data set without Stability
Selection.  Consequently, these approaches lack any guidance on choosing
$q$. The rate figures were obtained by evaluation of the graphs arising
from various values of $q$. We provide them as a means to check if
introducing Stability Selection has any additional (positive or negative)
effect on the performance of the Random Forests and LASSO methods besides
enabling us to choose $q$. From the rate figures, we can deduce that the
raw methods perform quite similar to GRaFo and StabLASSO across all
settings. Hence, the use of Stability Selection did not introduce any
surprising new behavior of Random Forests or LASSO.

A violation of condition (A) of Theorem 1 in the mixed setting could
explain the failure of both GRaFo and StabLASSO to achieve error control
for $p=200$. However, both the mixed setting with $p=50$ and $p=100$
returned very few observed errors and remained well below the error bounds
indicating the problematic behavior may be linked to larger values of
$p$. Also, for any setting it is unlikely that the exchangeability
assumption holds. \citet{Meinshausen2010} argue that Stability Selection
appears to be robust to violations, but did not study mixed data which may
be particularly affected. \textbf{We study this aspect more closely
  further below.}

The computational cost is growing rather quickly with growing $p$. The
runtime of a single of the 50 repetitions per setting is in the order
of $15$ minutes for GRaFo and $20$ minutes for StabLASSO for $p=50$
and increases to several hours for GRaFo and $30$ minutes for
StabLASSO in the case of $p=200$. Each batch of 50 repetitions was run
in parallel on 50 cores of the BRUTUS high-performance cluster
comprising quad-core AMD Opteron 8380 2.5 Ghz CPUs with 1 GB of RAM
per core using the Rmpi package \citep{Yu2010} available in R.


\ifpdf
  \begin{figure}
    \centering
    \textbf{Gaussian, Bernoulli, and Ising models, $p=50$}\\\vspace{0.5cm}
     \subfigure[Gaussian: GRaFo]{\includegraphics[width=0.3\textwidth]{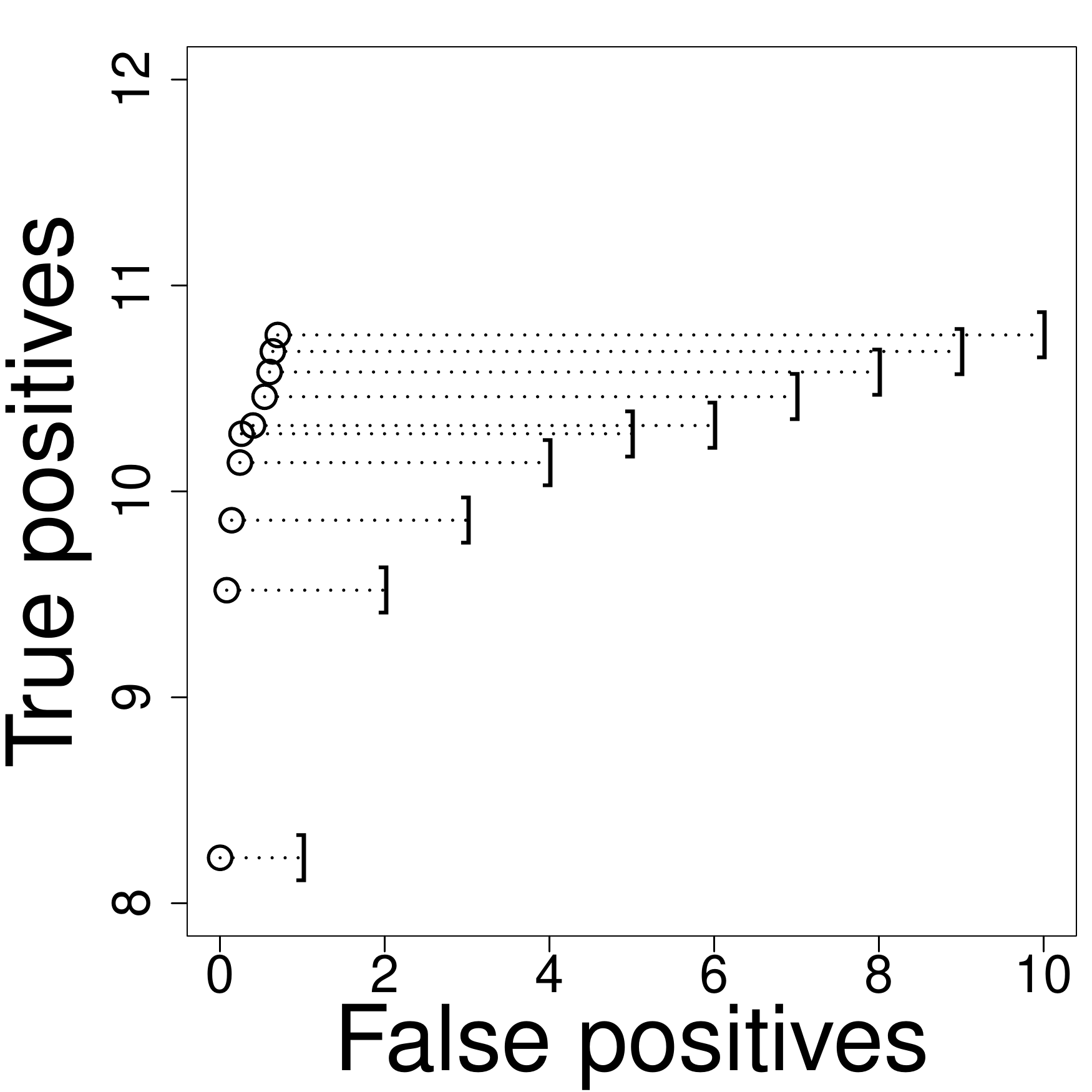}}
    \subfigure[Gaussian: StabLASSO]{\includegraphics[width=0.3\textwidth]{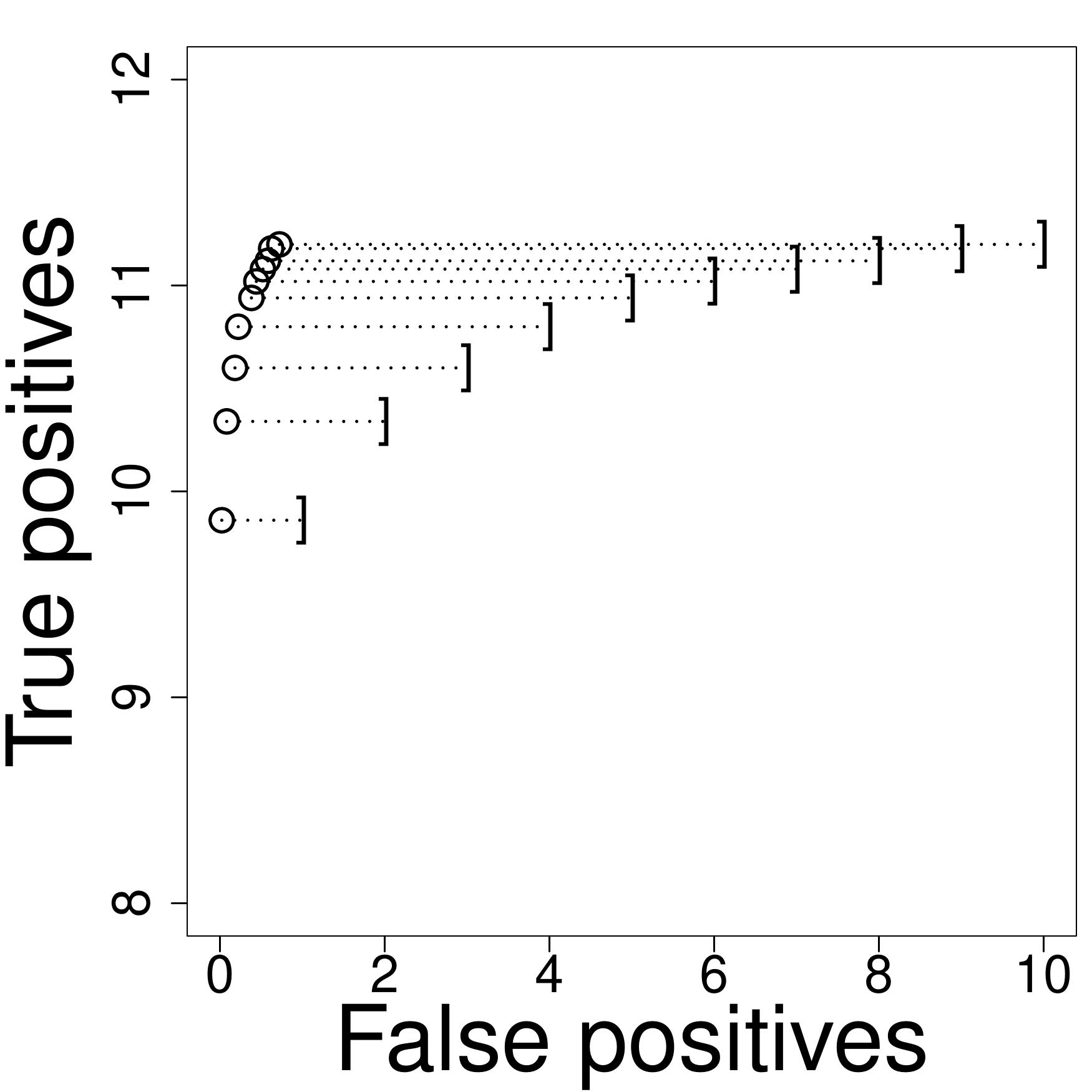}}
    \subfigure[Gaussian: Rates]{\includegraphics[width=0.3\textwidth]{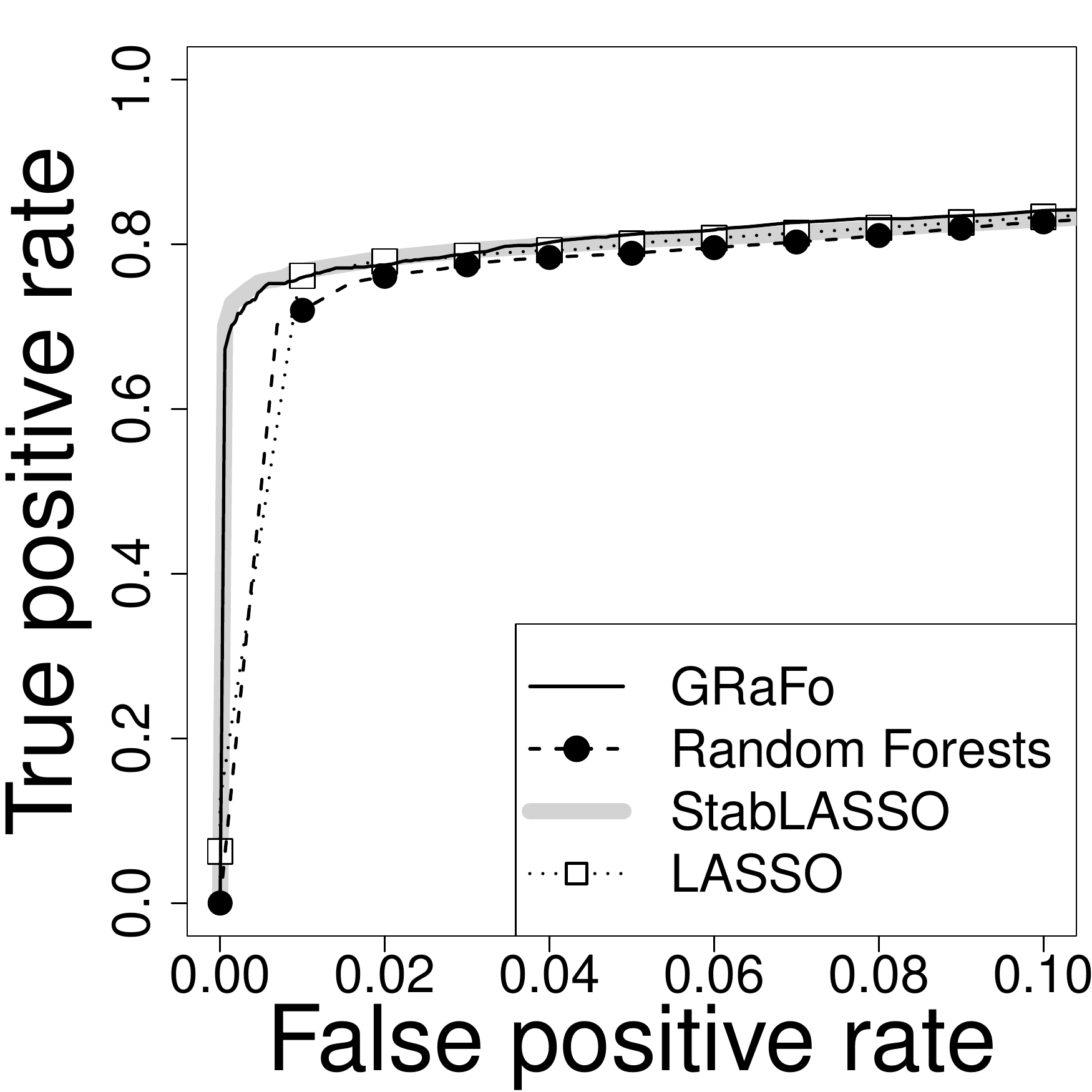}}\\
    \subfigure[Bernoulli: GRaFo]{\includegraphics[width=0.3\textwidth]{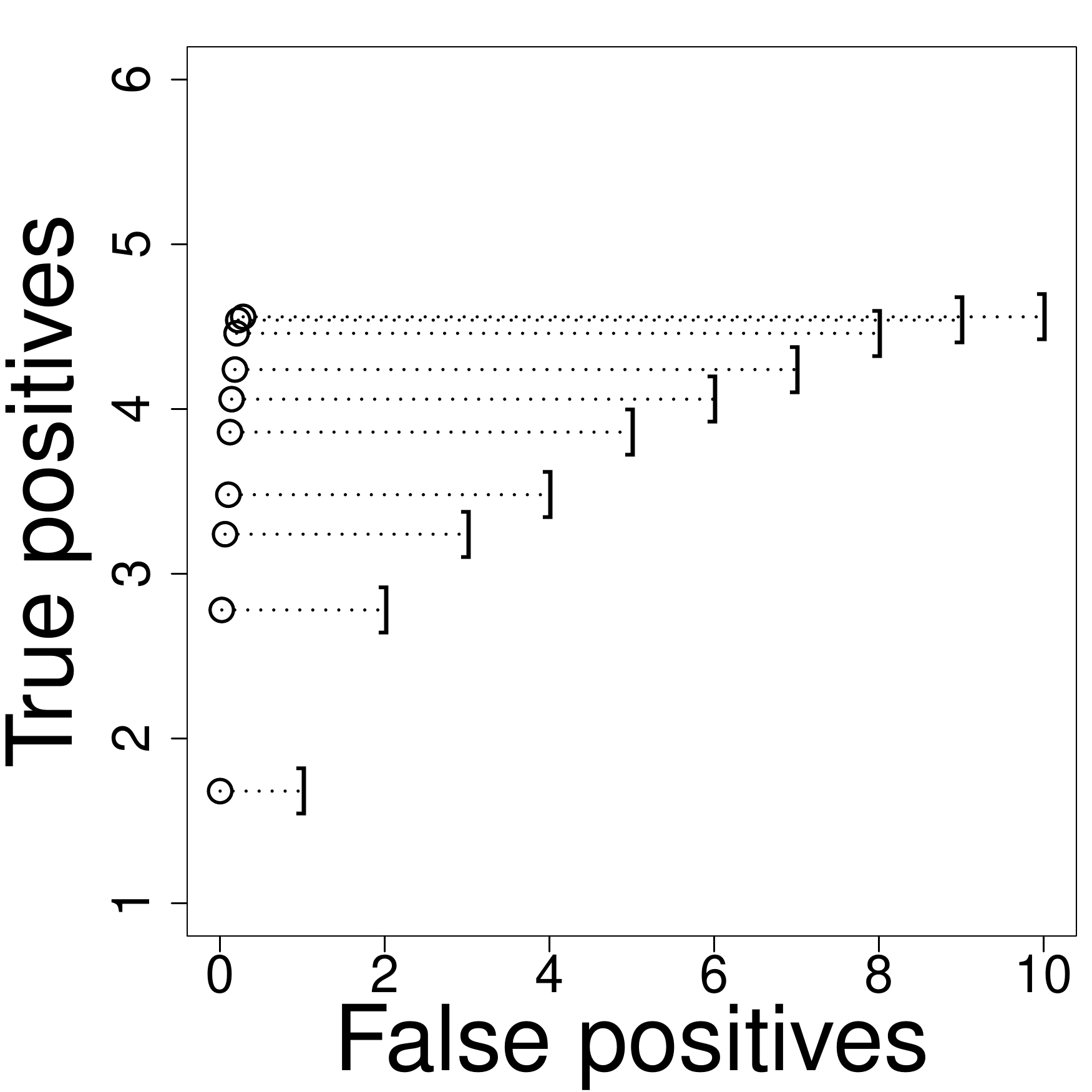}}
    \subfigure[Bernoulli:
    StabLASSO]{\includegraphics[width=0.3\textwidth]{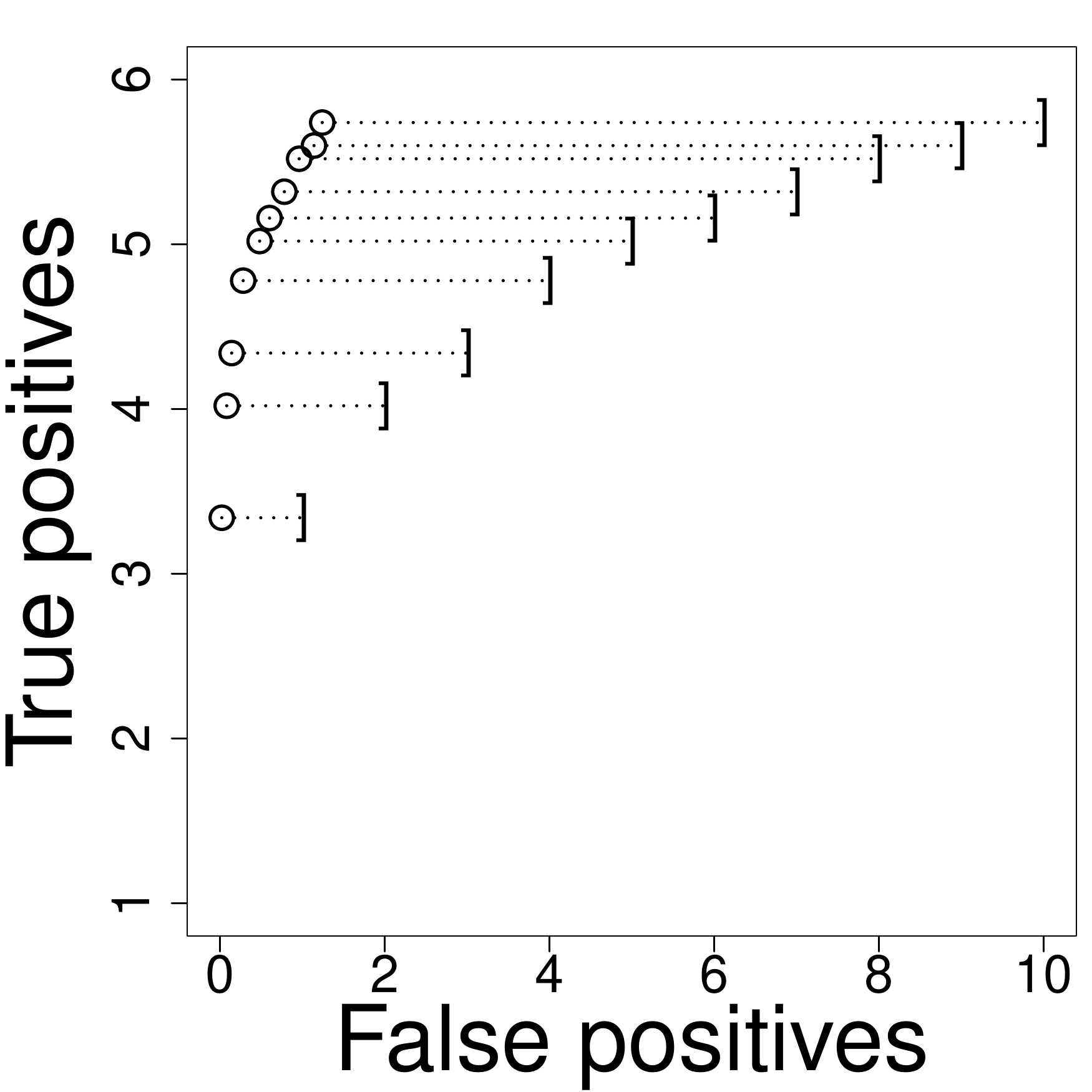}}
    \subfigure[Bernoulli: Rates]{\includegraphics[width=0.3\textwidth]{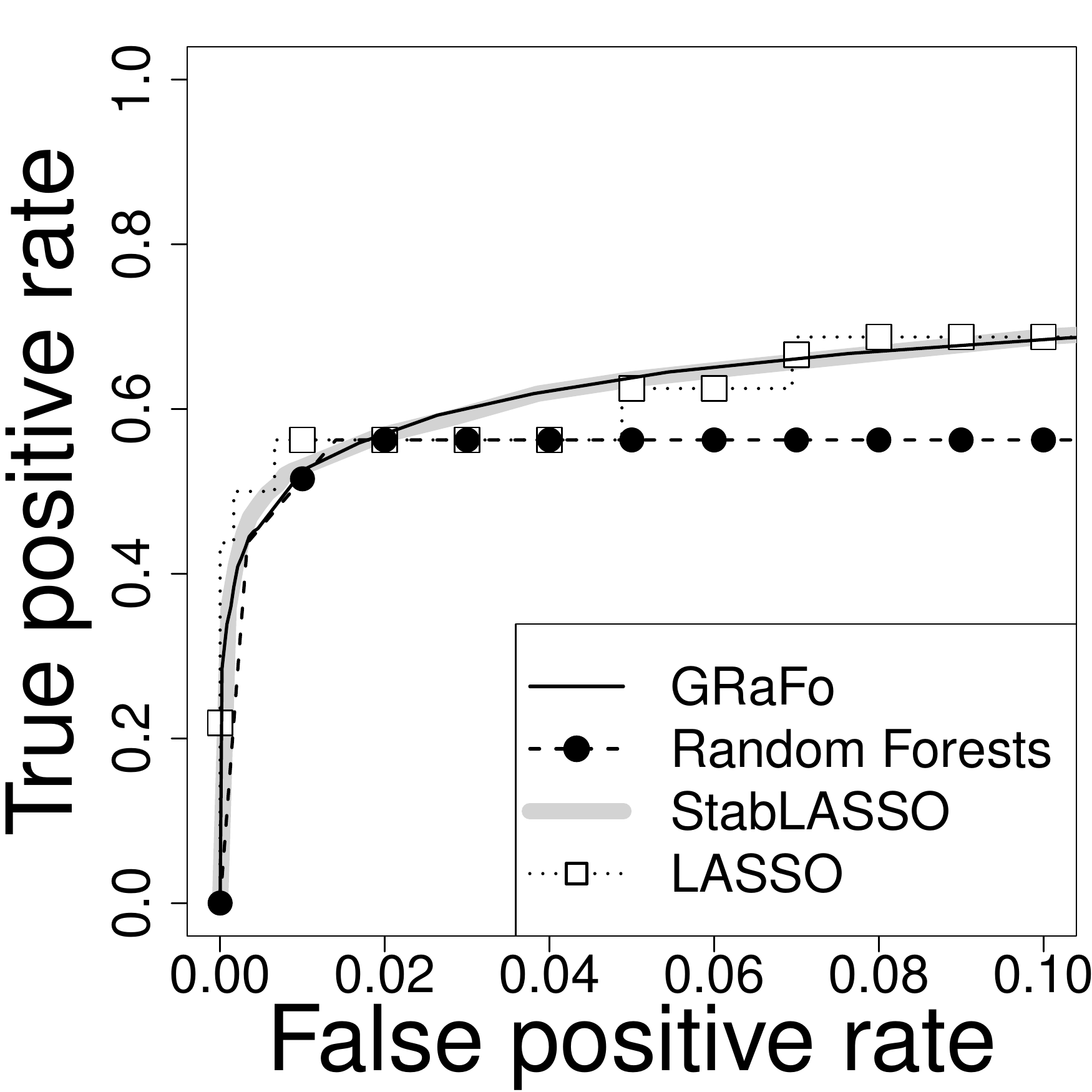}}\\
    \subfigure[Ising: GRaFo]{\includegraphics[width=0.3\textwidth]{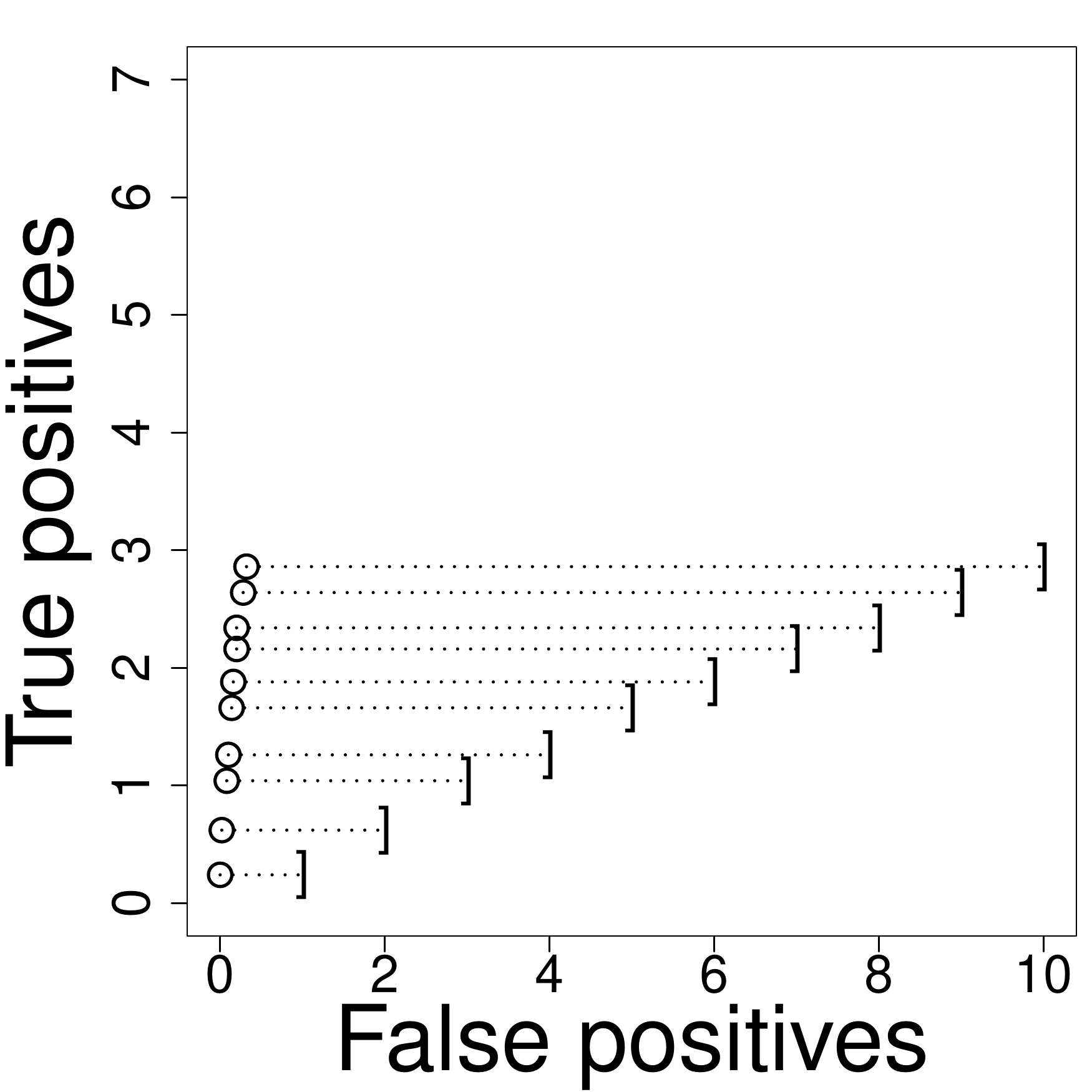}}
    \subfigure[Ising:
    StabLASSO]{\includegraphics[width=0.3\textwidth]{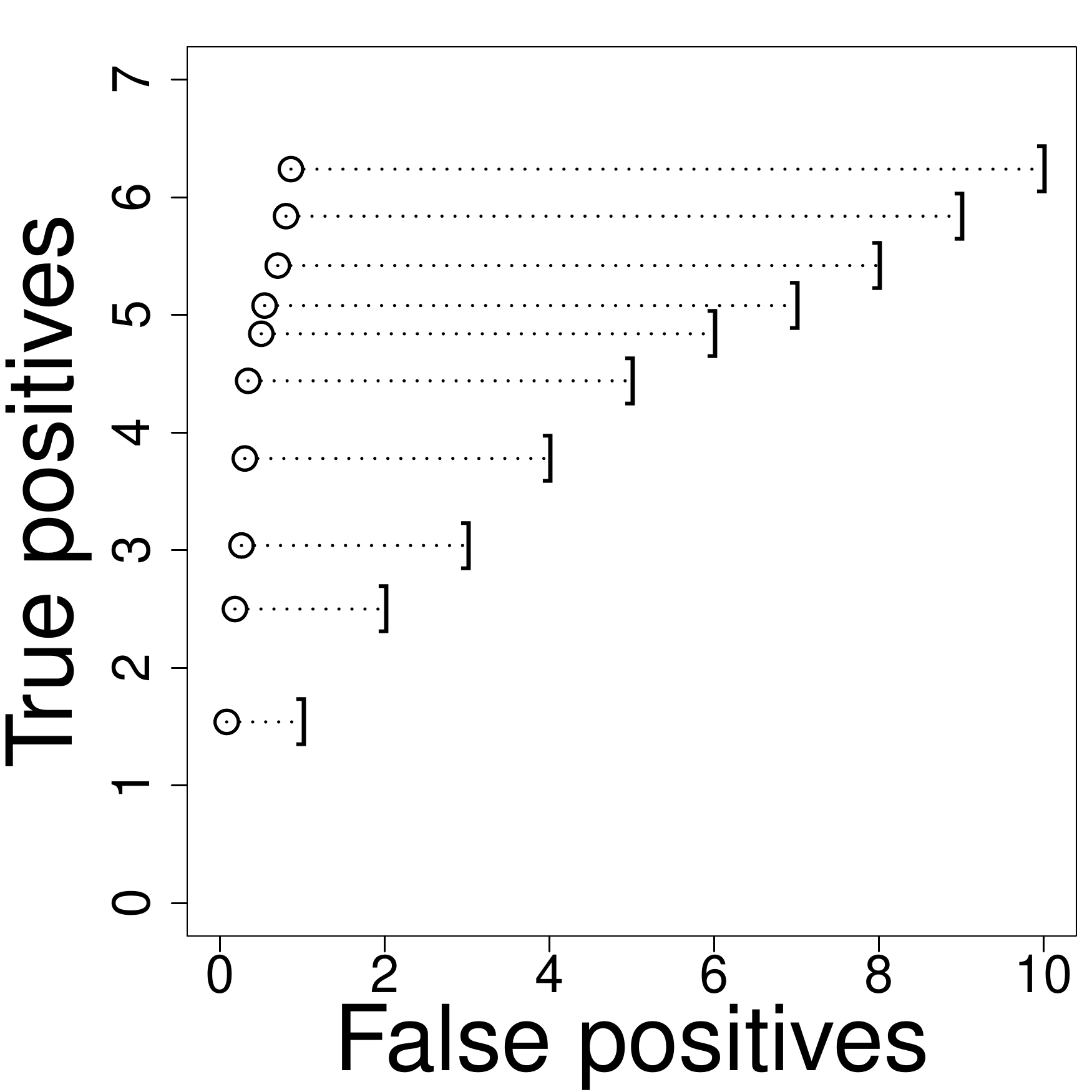}}
    \subfigure[Ising: Rates]{\includegraphics[width=0.3\textwidth]{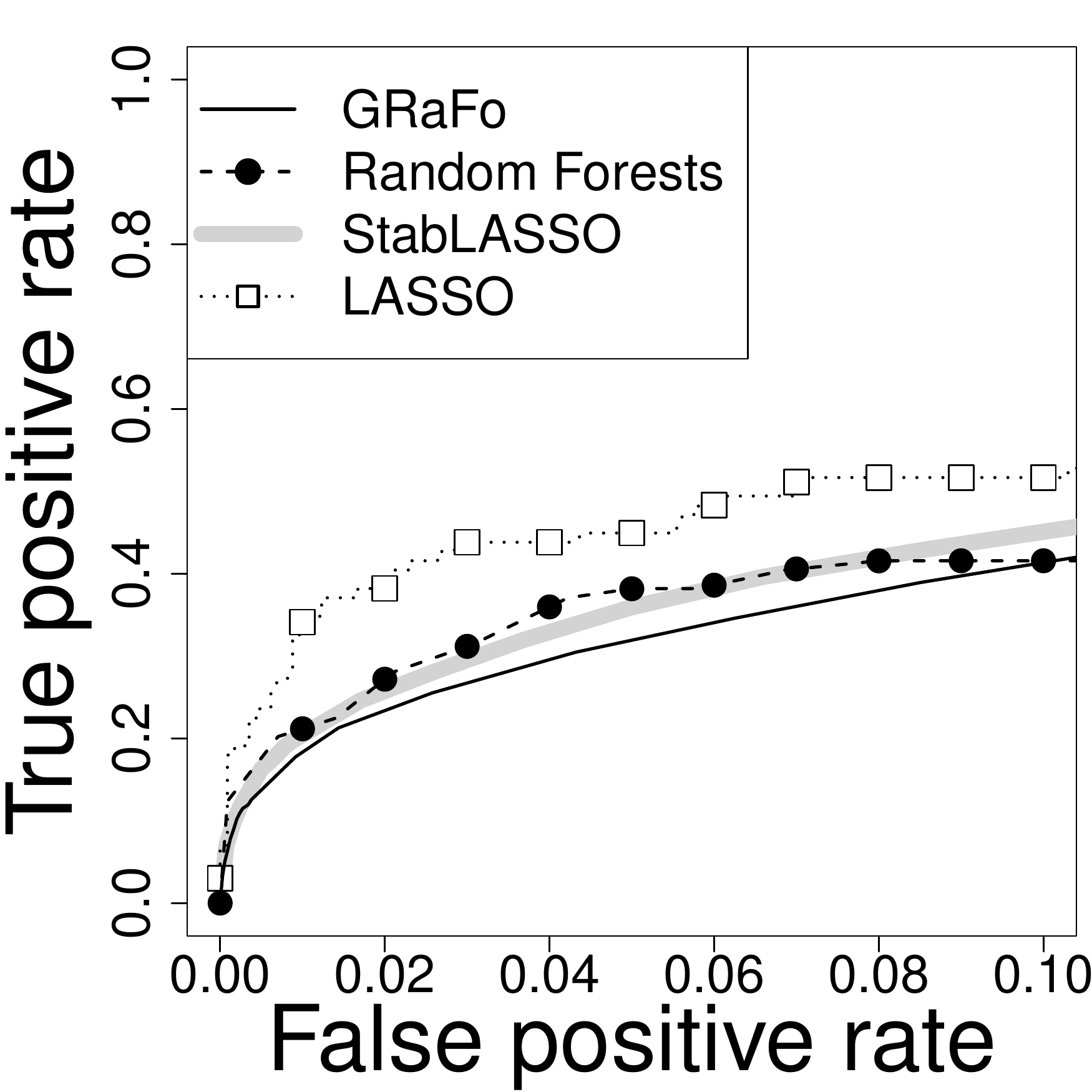}}
    \caption{The rows correspond to the Gaussian, Bernoulli, and Ising
      model with $p=50$. Their true CIGs have $16$, $16$ and $89$
      edges, respectively. The first two columns report the observed
      number of true and false positives (``o'') relative to the bound
      in (\ref{StabSelErrorControl}) for the expected number
      $\mathbb{E}[V]$ of false positives (``$]$'') for GRaFo and
      StabLASSO, respectively, averaged over 50 simulations.  The
      third column reports the averaged true and false positive rates
      of GRaFo and StabLASSO relative to the performance of their
      ``raw'' counterparts without Stability Selection.}
    \label{FigGBI50}
  \end{figure}

  \begin{figure}
    \centering
    \textbf{Gaussian, Bernoulli, and Ising models, $p=100$}\\\vspace{0.5cm}
    \subfigure[Gaussian: GRaFo]{\includegraphics[width=0.3\textwidth]{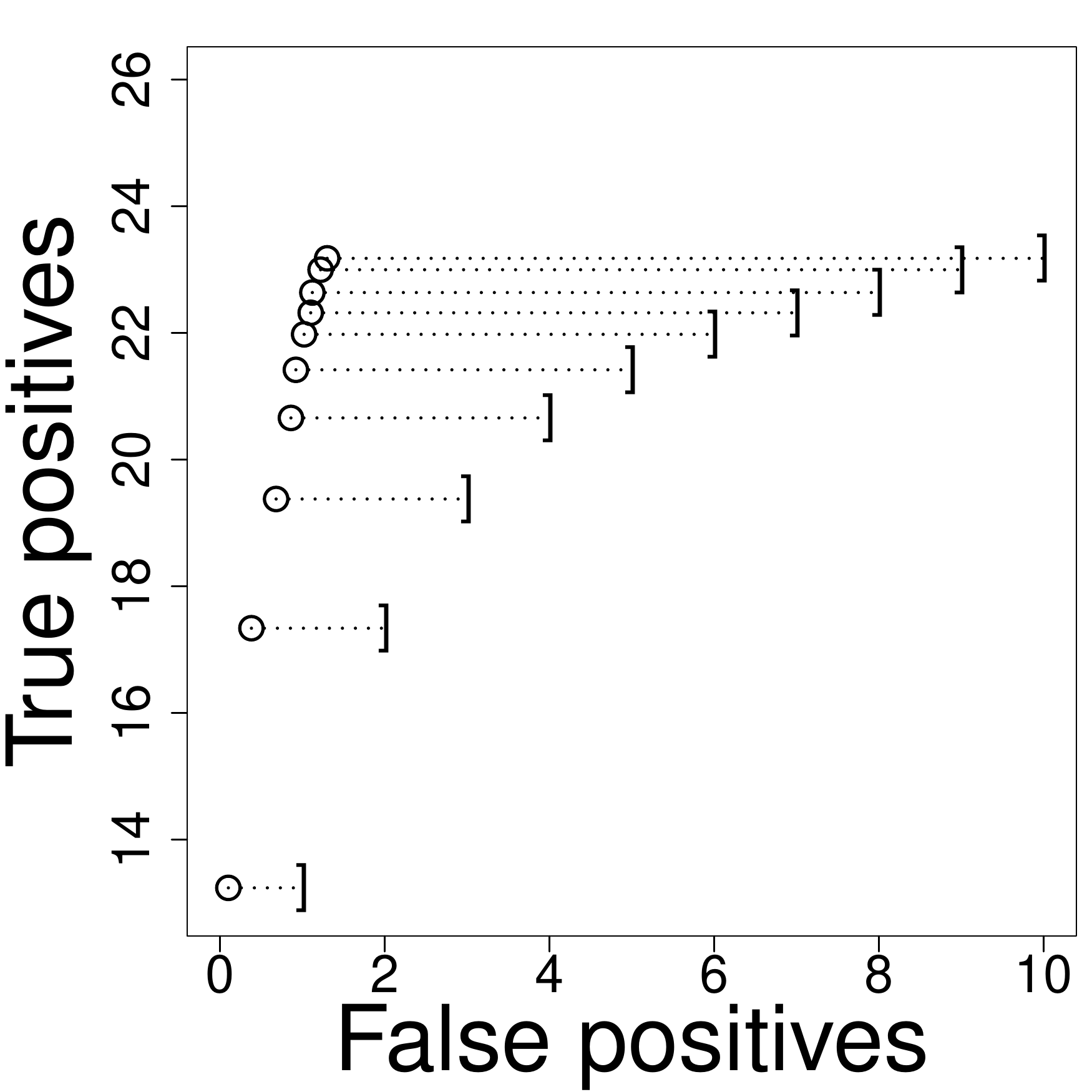}}
    \subfigure[Gaussian:
    StabLASSO]{\includegraphics[width=0.3\textwidth]{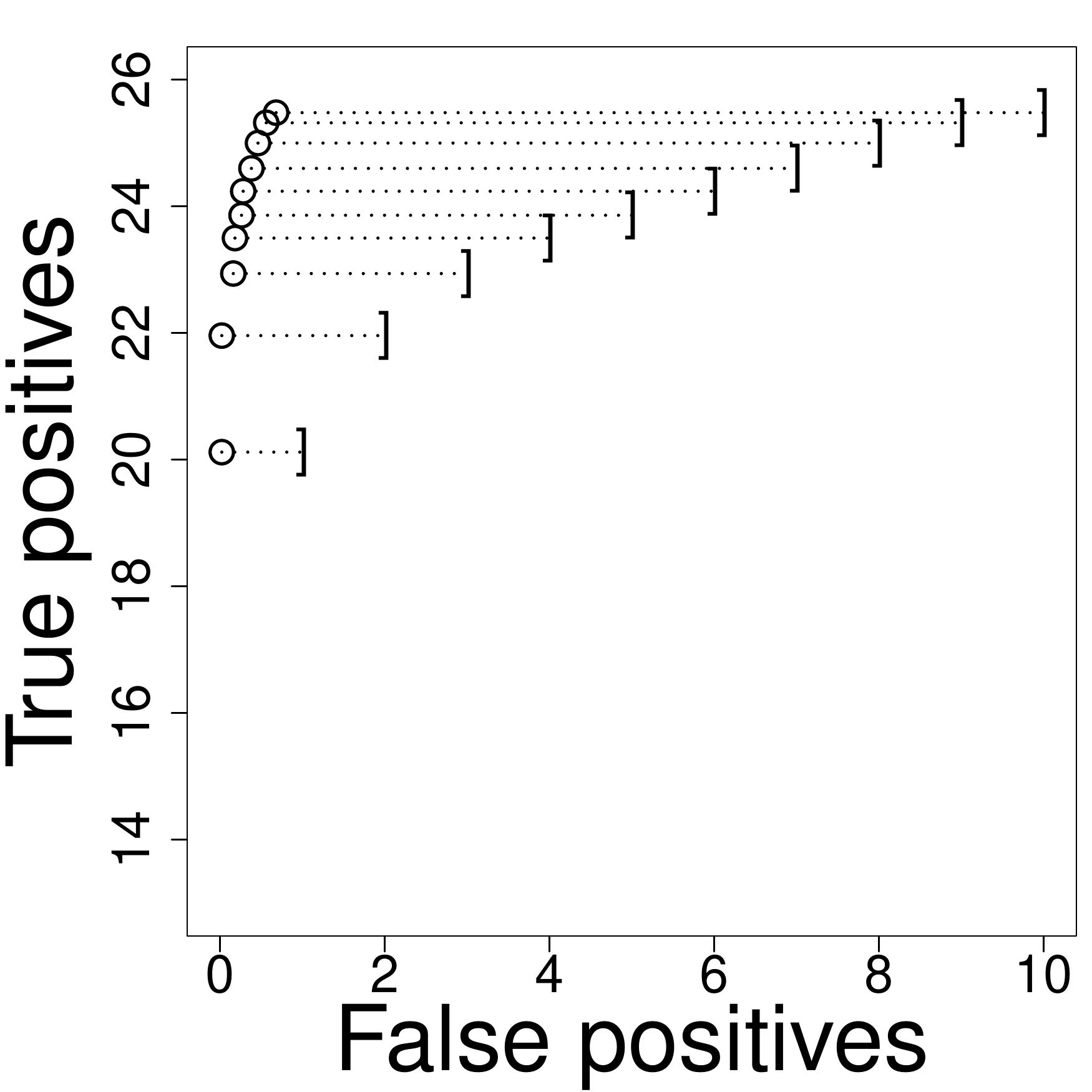}}
    \subfigure[Gaussian: Rates]{\includegraphics[width=0.3\textwidth]{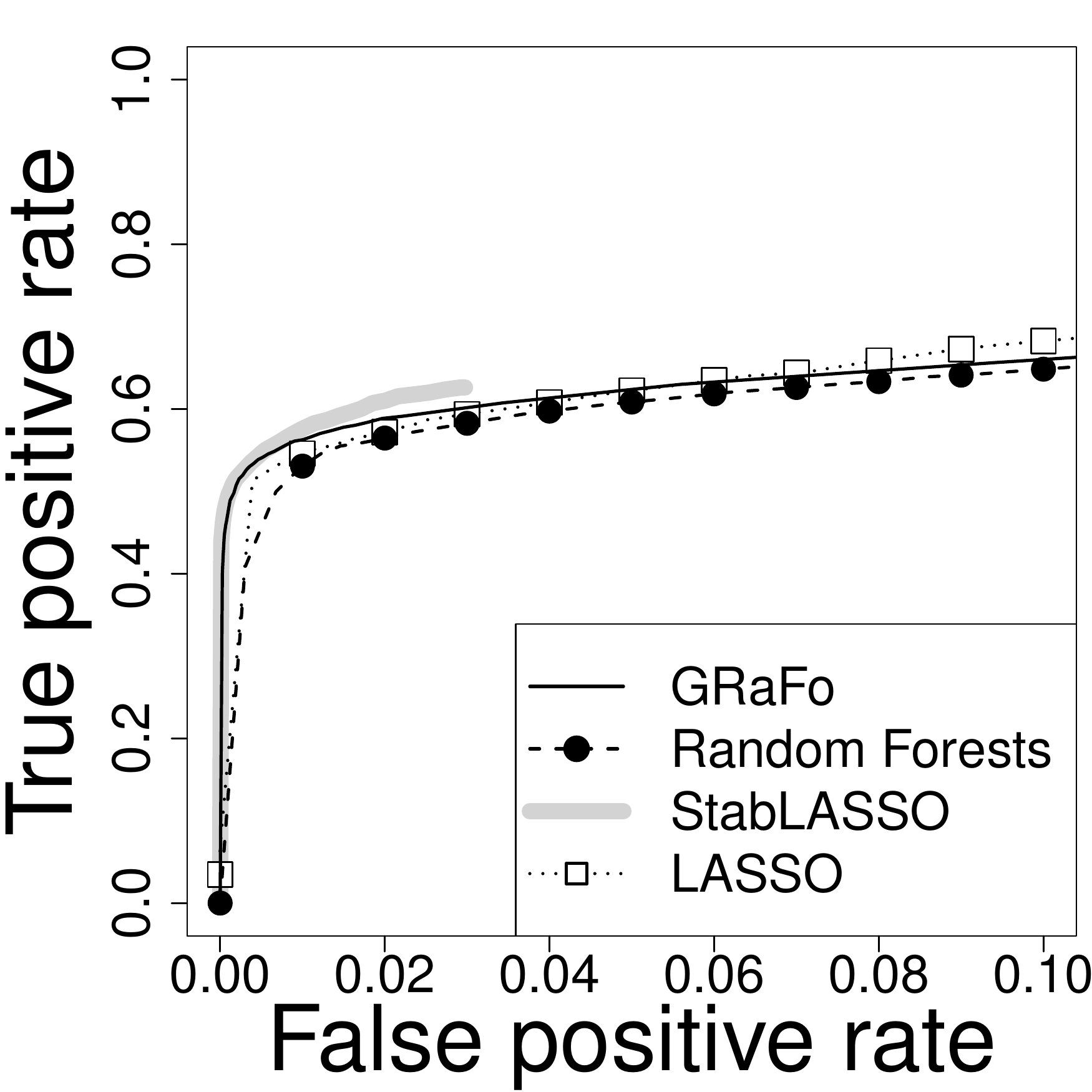}}\\
    \subfigure[Bernoulli: GRaFo]{\includegraphics[width=0.3\textwidth]{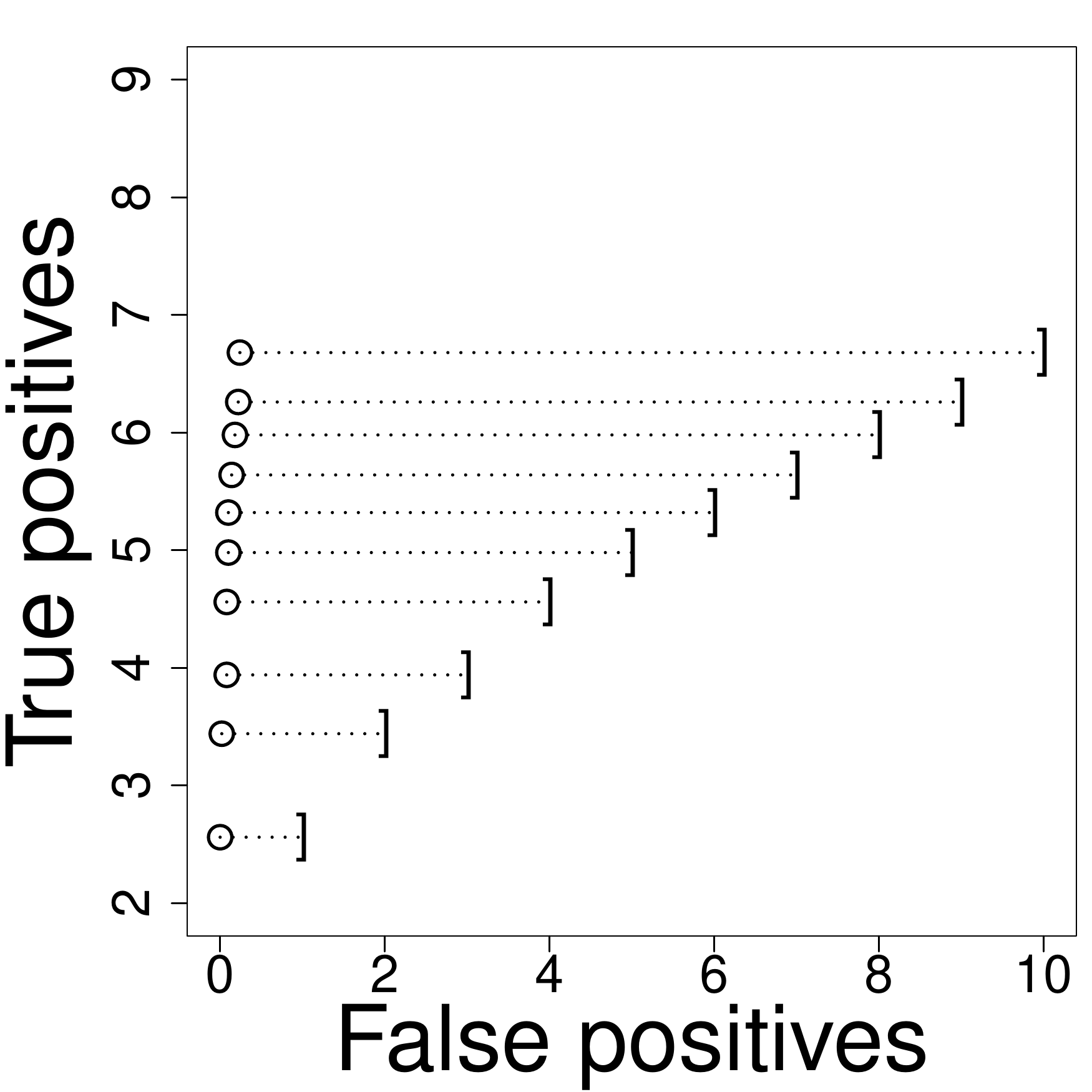}}
    \subfigure[Bernoulli:
    StabLASSO]{\includegraphics[width=0.3\textwidth]{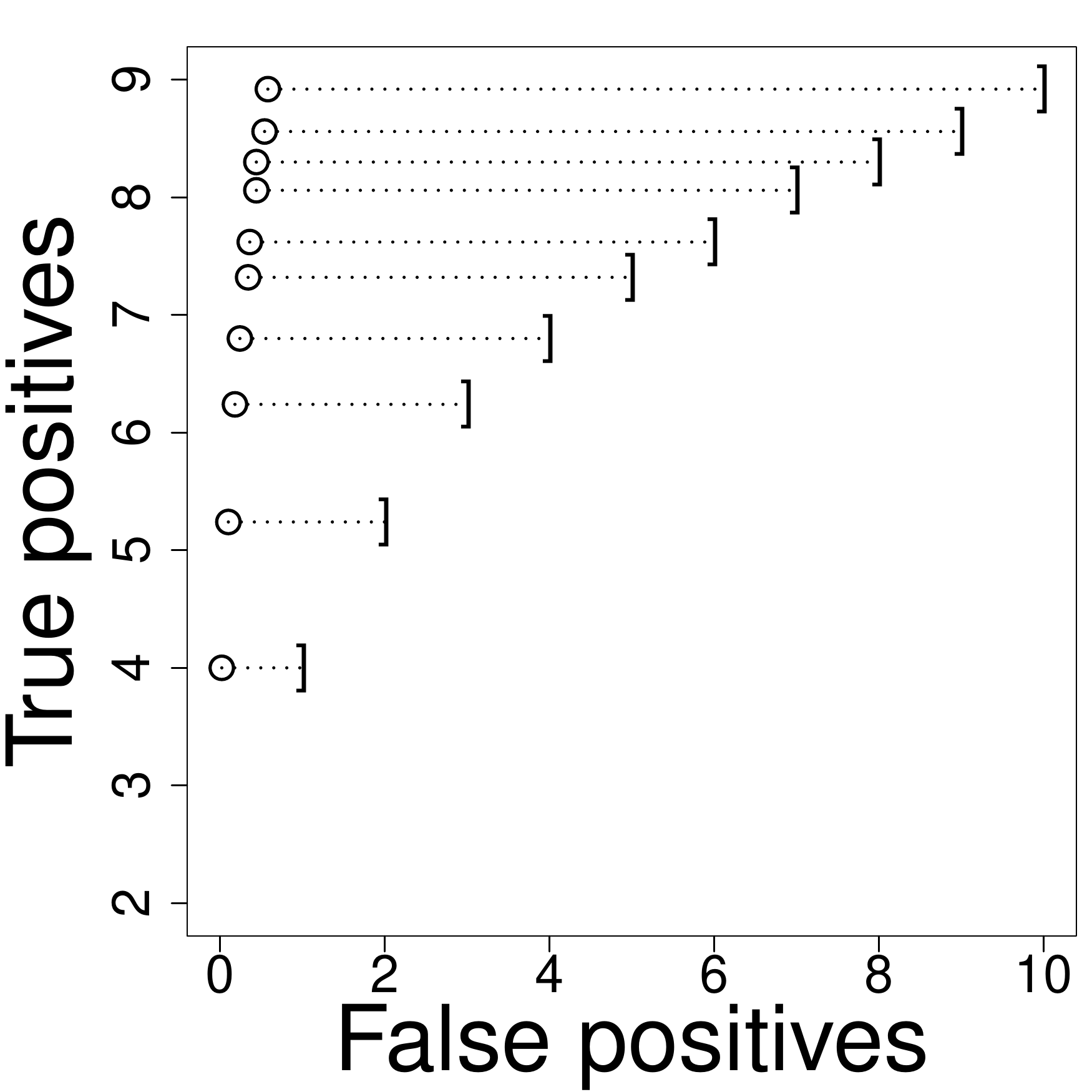}}
    \subfigure[Bernoulli: Rates]{\includegraphics[width=0.3\textwidth]{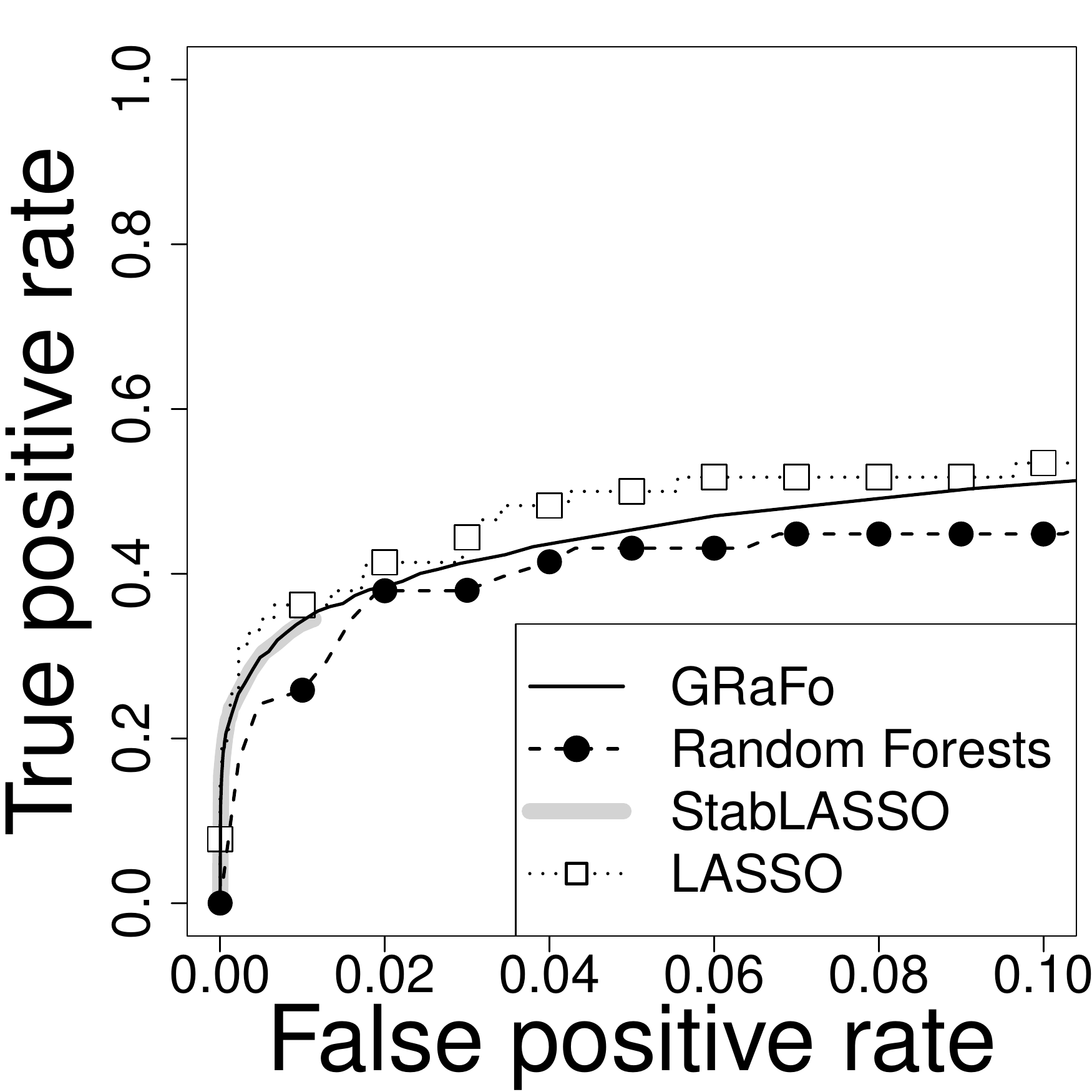}}\\
    \subfigure[Ising: GRaFo]{\includegraphics[width=0.3\textwidth]{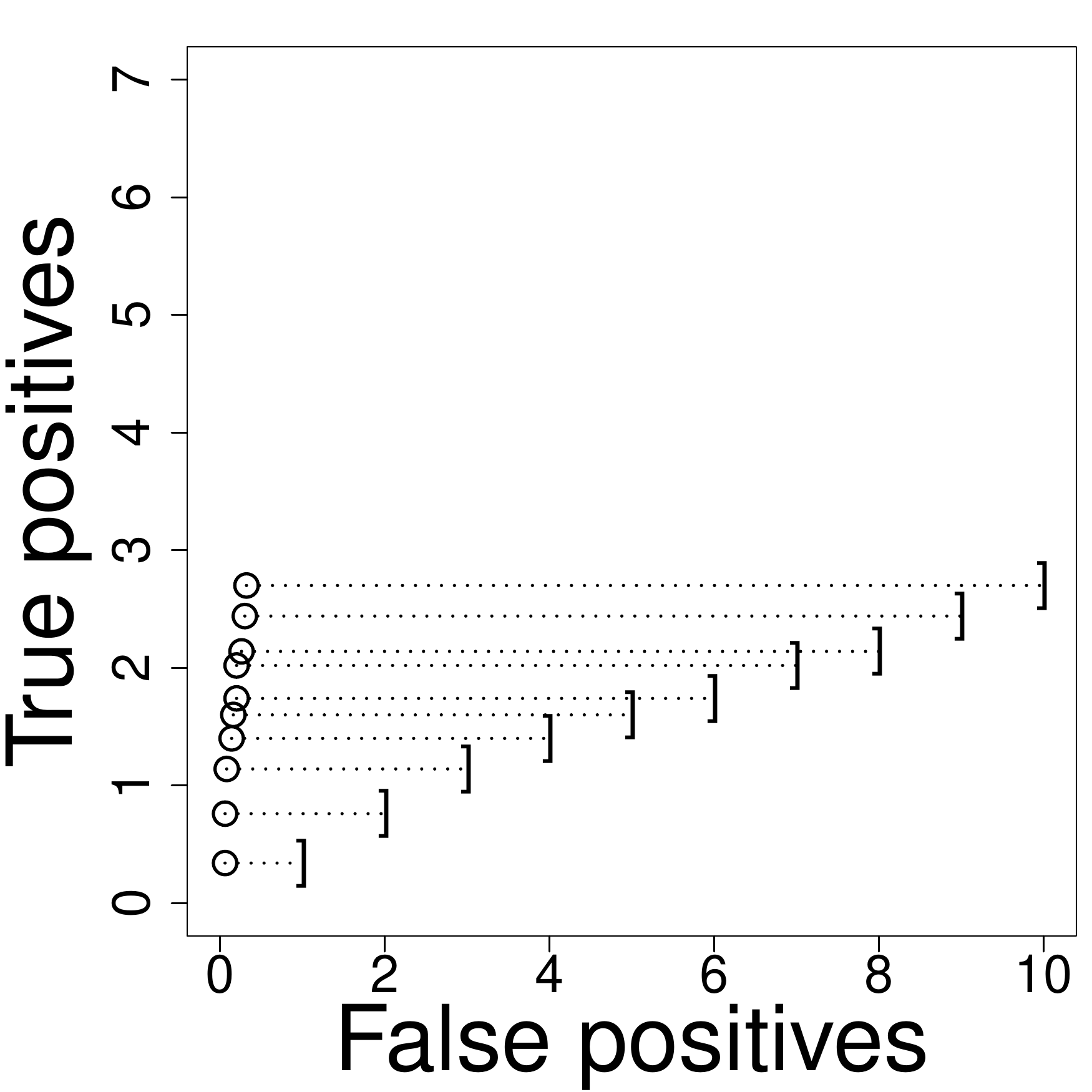}}
    \subfigure[Ising: StabLASSO]{\includegraphics[width=0.3\textwidth]{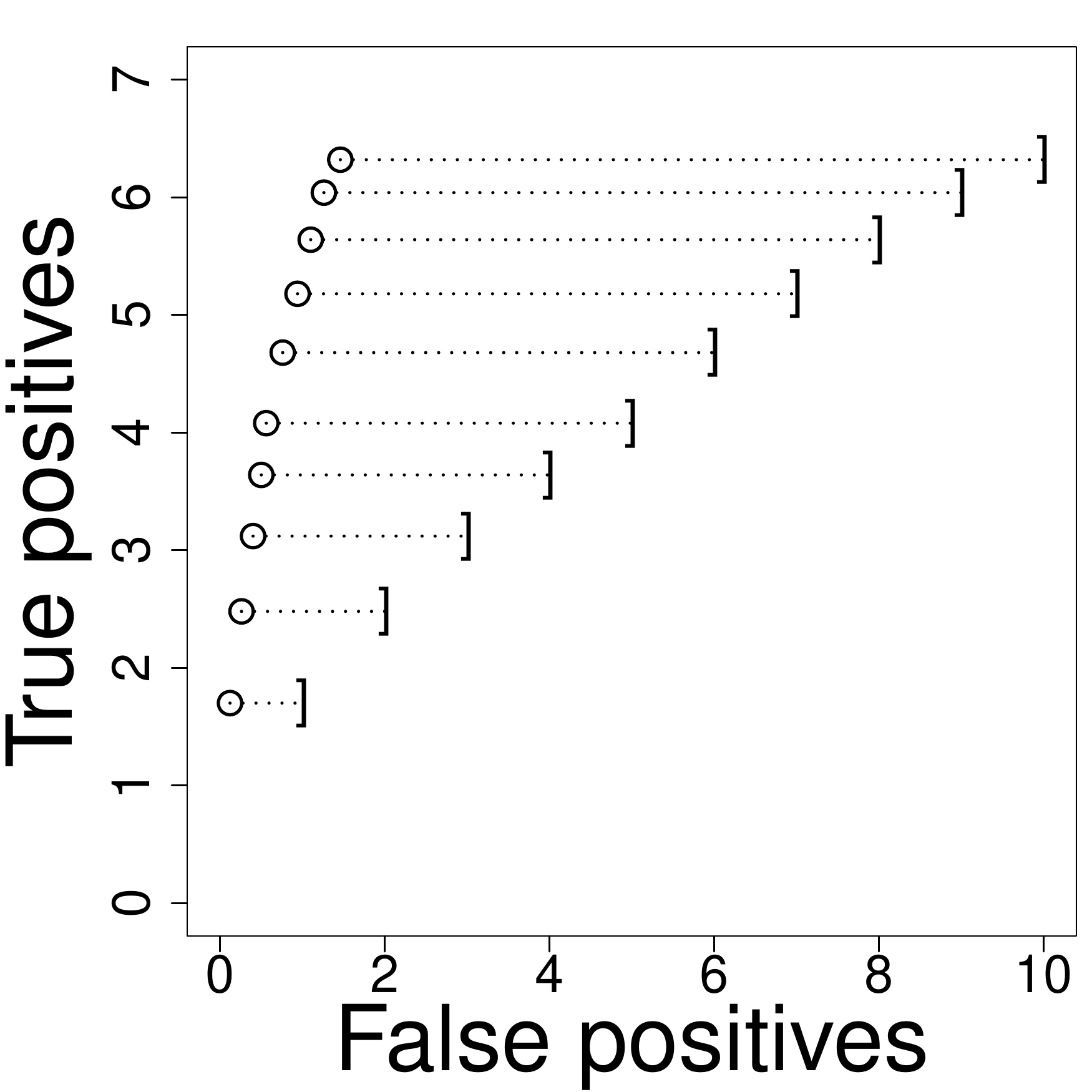}}
    \subfigure[Ising:
    Rates]{\includegraphics[width=0.3\textwidth]{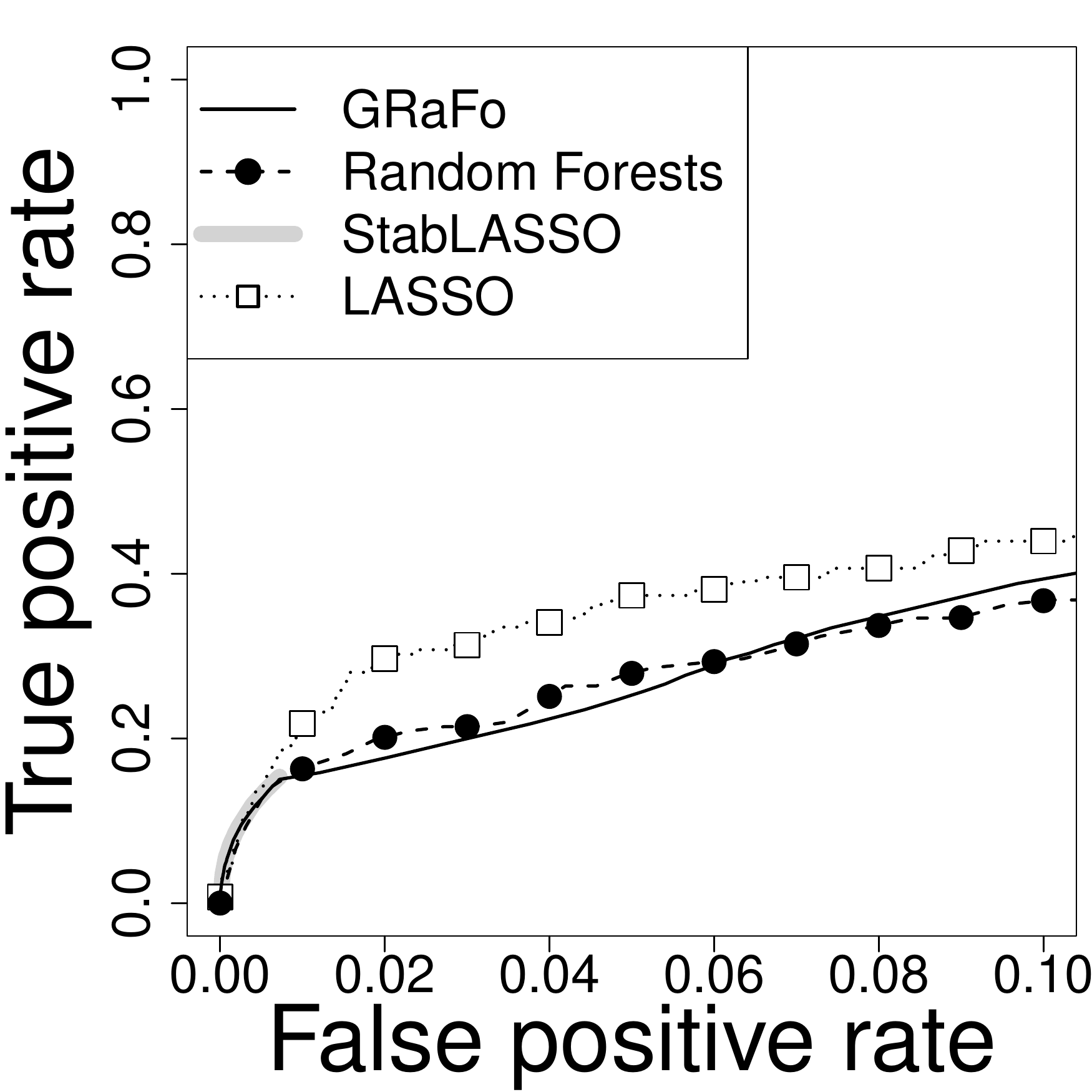}}
    \caption{The rows correspond to the Gaussian, Bernoulli, and Ising
      model with $p=100$. Their true CIGs have $58$, $58$ and $182$
      edges, respectively. The first two columns report the observed
      number of true and false positives (``o'') relative to the bound
      in (\ref{StabSelErrorControl}) for the expected number
      $\mathbb{E}[V]$ of false positives (``$]$'') for GRaFo and
      StabLASSO, respectively, averaged over 50 simulations.  The
      third column reports the averaged true and false positive rates
      of GRaFo and StabLASSO relative to the performance of their
      ``raw'' counterparts without Stability Selection.}
    \label{FigGBI100}
  \end{figure}

  \begin{figure}
    \centering
    \textbf{Gaussian, Bernoulli, and Ising models, $p=200$}\\\vspace{0.5cm}
    \subfigure[Gaussian: GRaFo]{\includegraphics[width=0.3\textwidth]{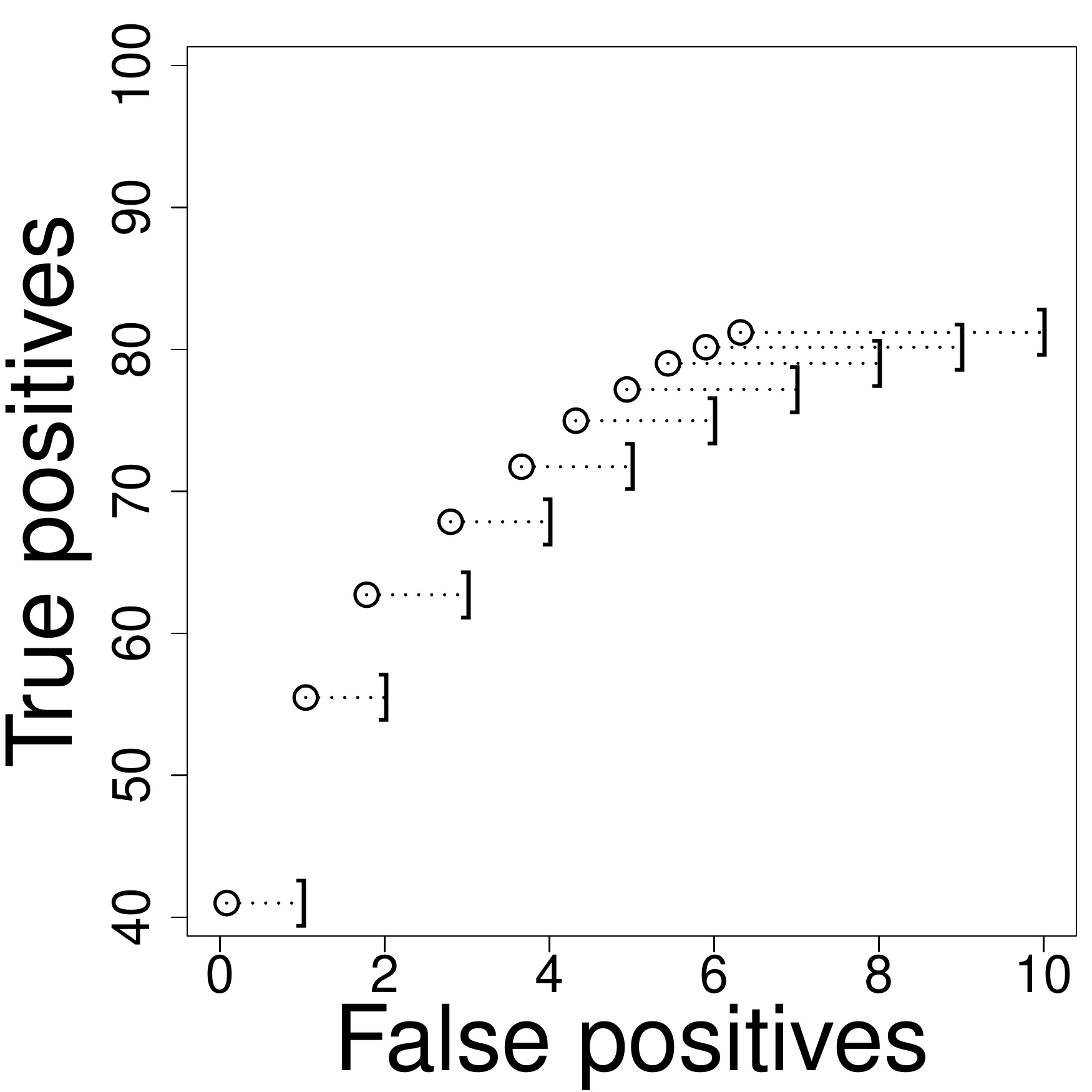}}
    \subfigure[Gaussian:
    StabLASSO]{\includegraphics[width=0.3\textwidth]{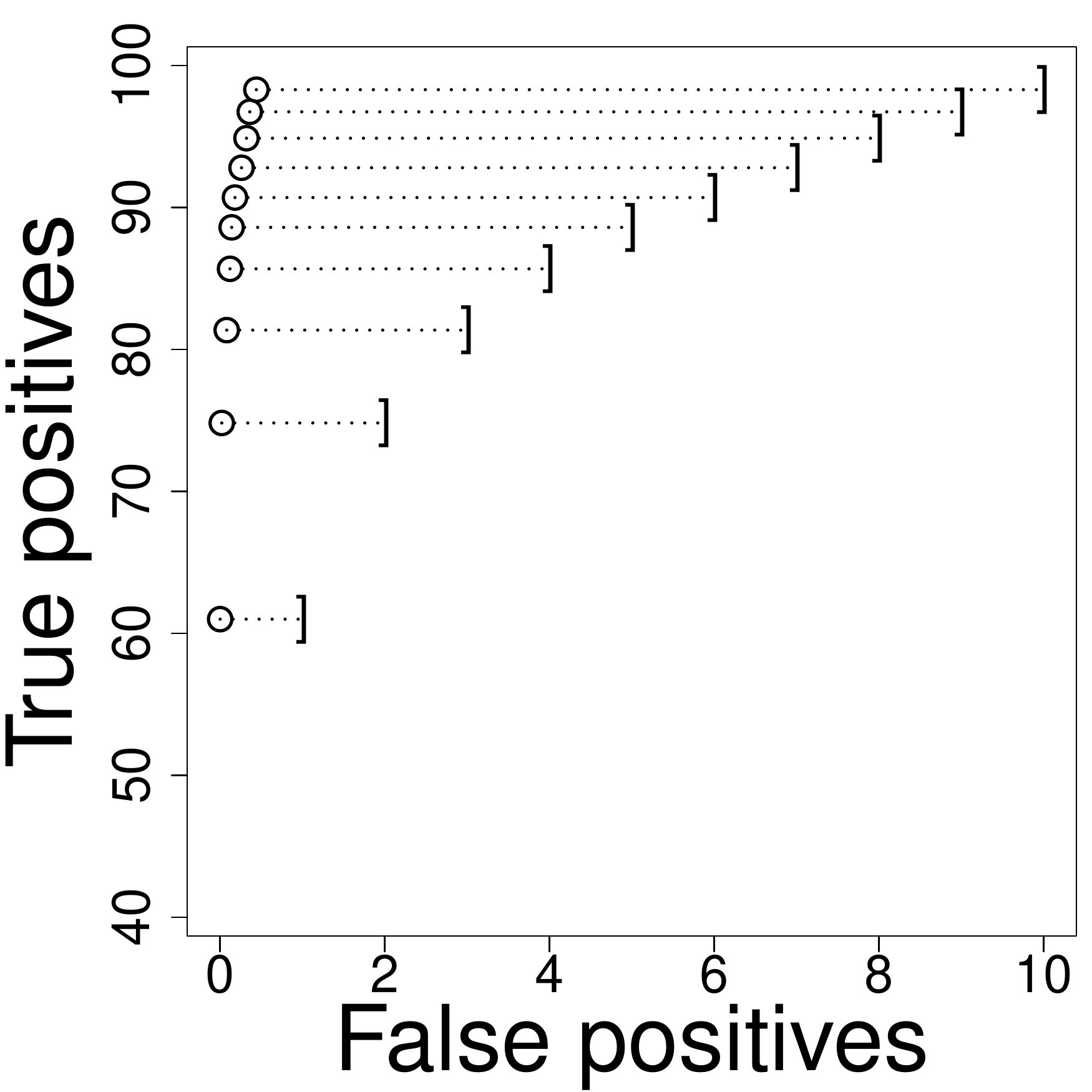}}
    \subfigure[Gaussian: Rates]{\includegraphics[width=0.3\textwidth]{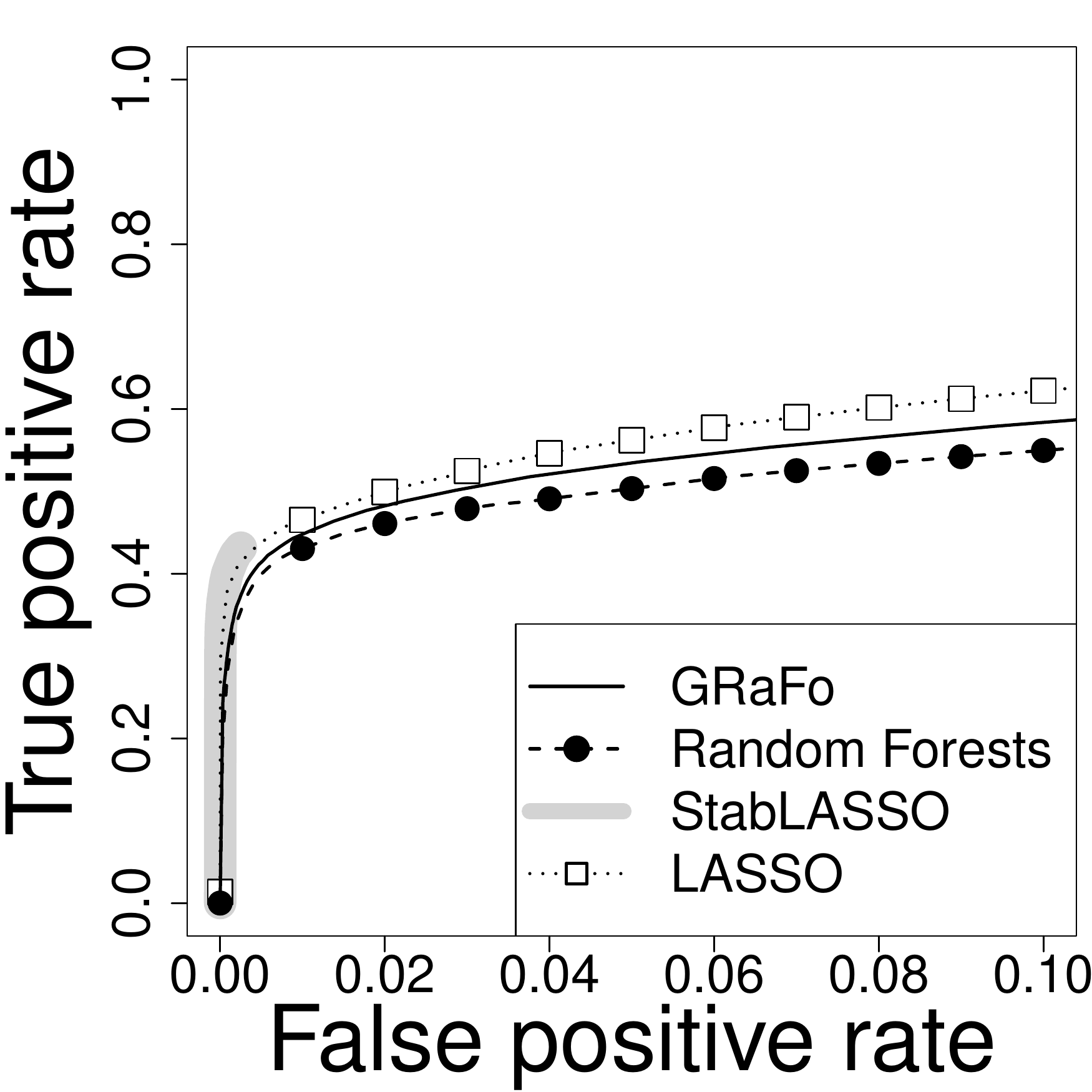}}\\
    \subfigure[Bernoulli: GRaFo]{\includegraphics[width=0.3\textwidth]{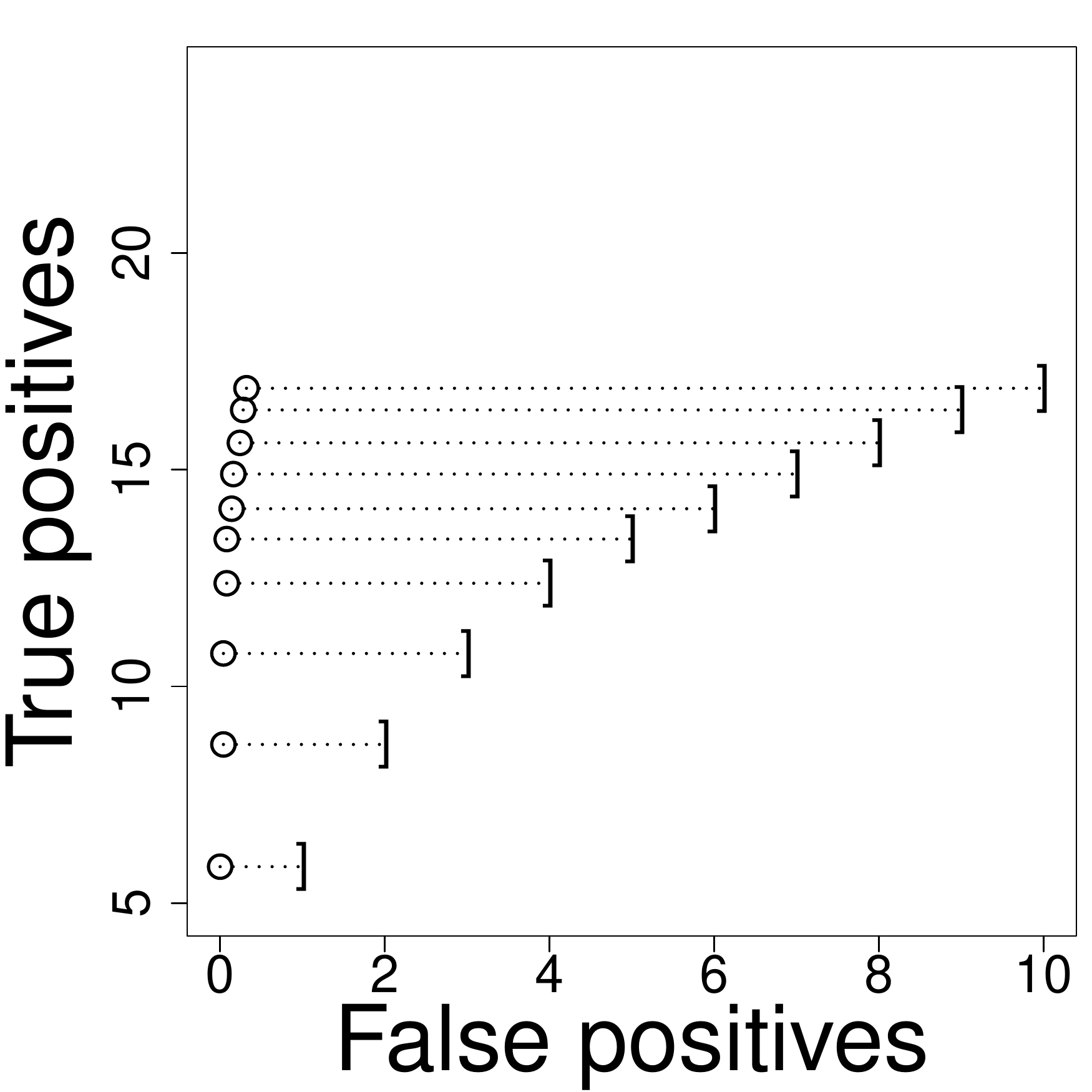}}
    \subfigure[Bernoulli:
    StabLASSO]{\includegraphics[width=0.3\textwidth]{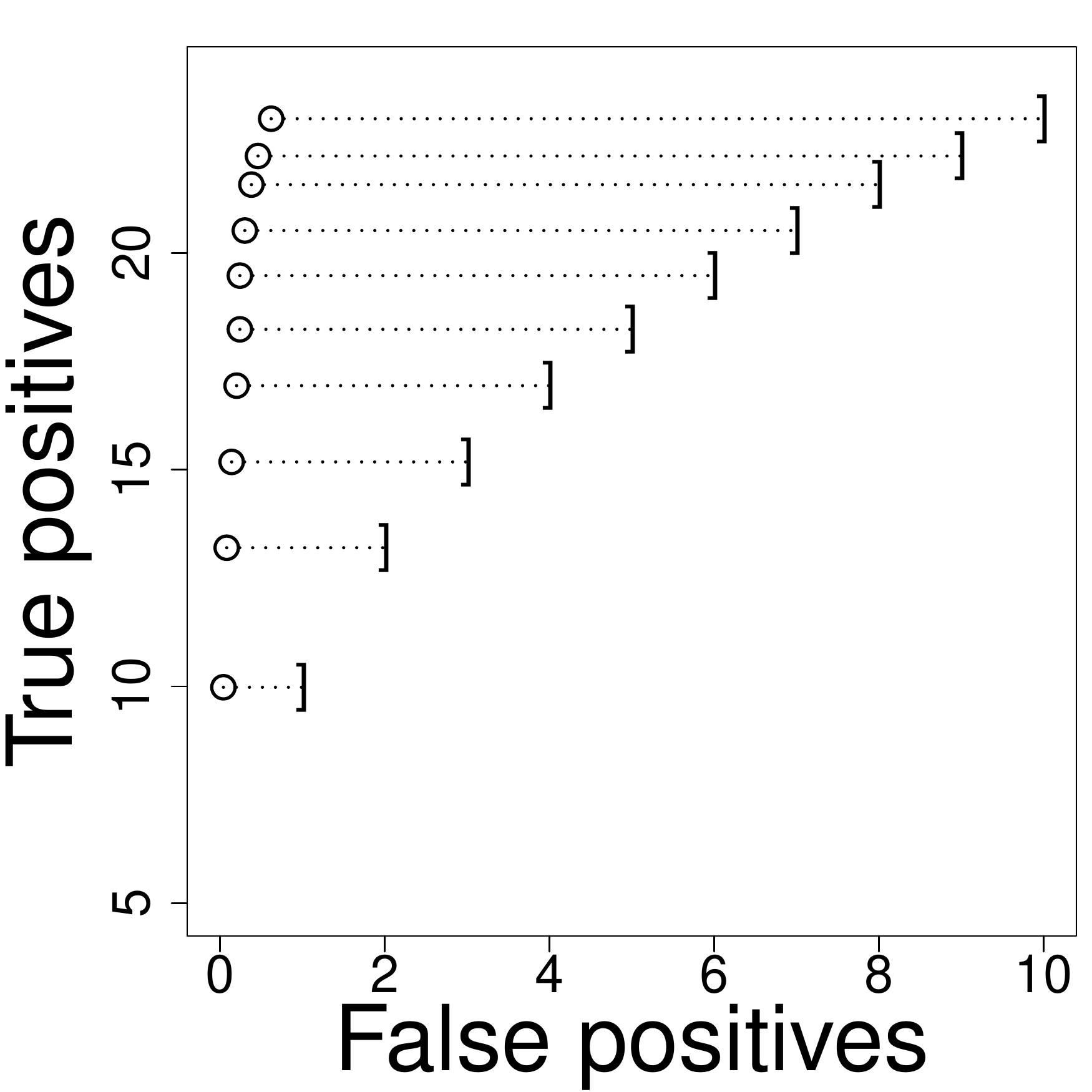}}
    \subfigure[Bernoulli: Rates]{\includegraphics[width=0.3\textwidth]{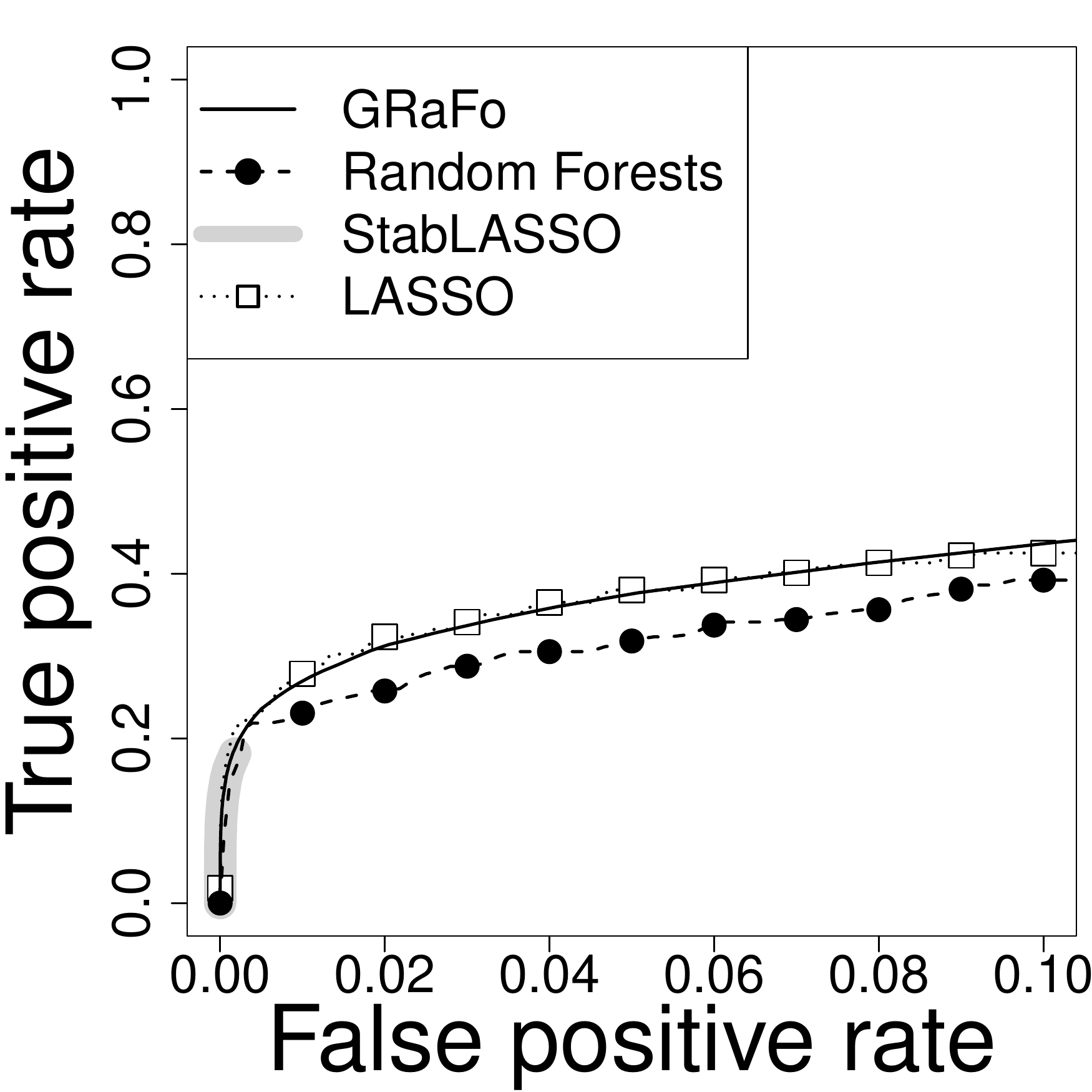}}\\
    \subfigure[Ising: GRaFo]{\includegraphics[width=0.3\textwidth]{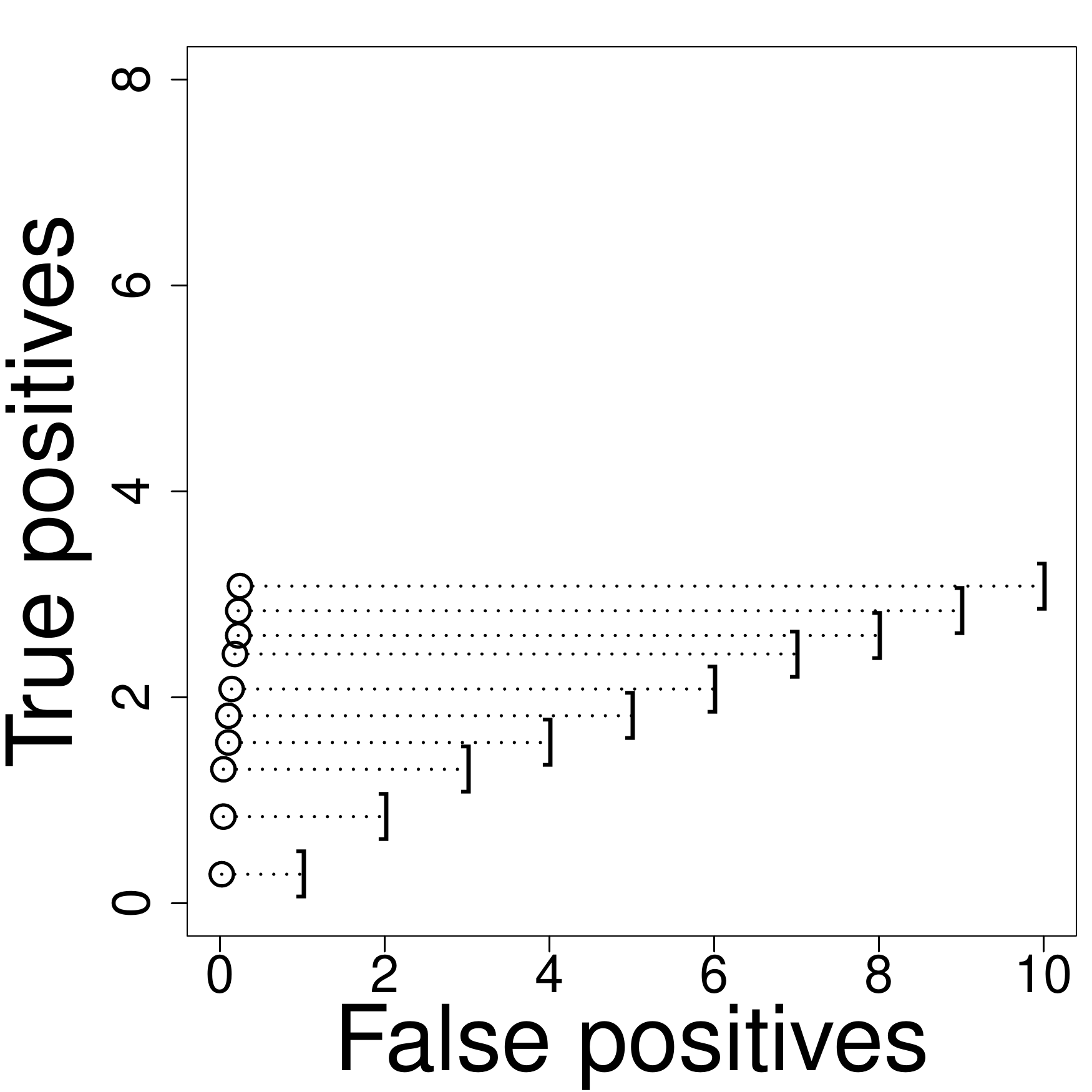}}
    \subfigure[Ising:
    StabLASSO]{\includegraphics[width=0.3\textwidth]{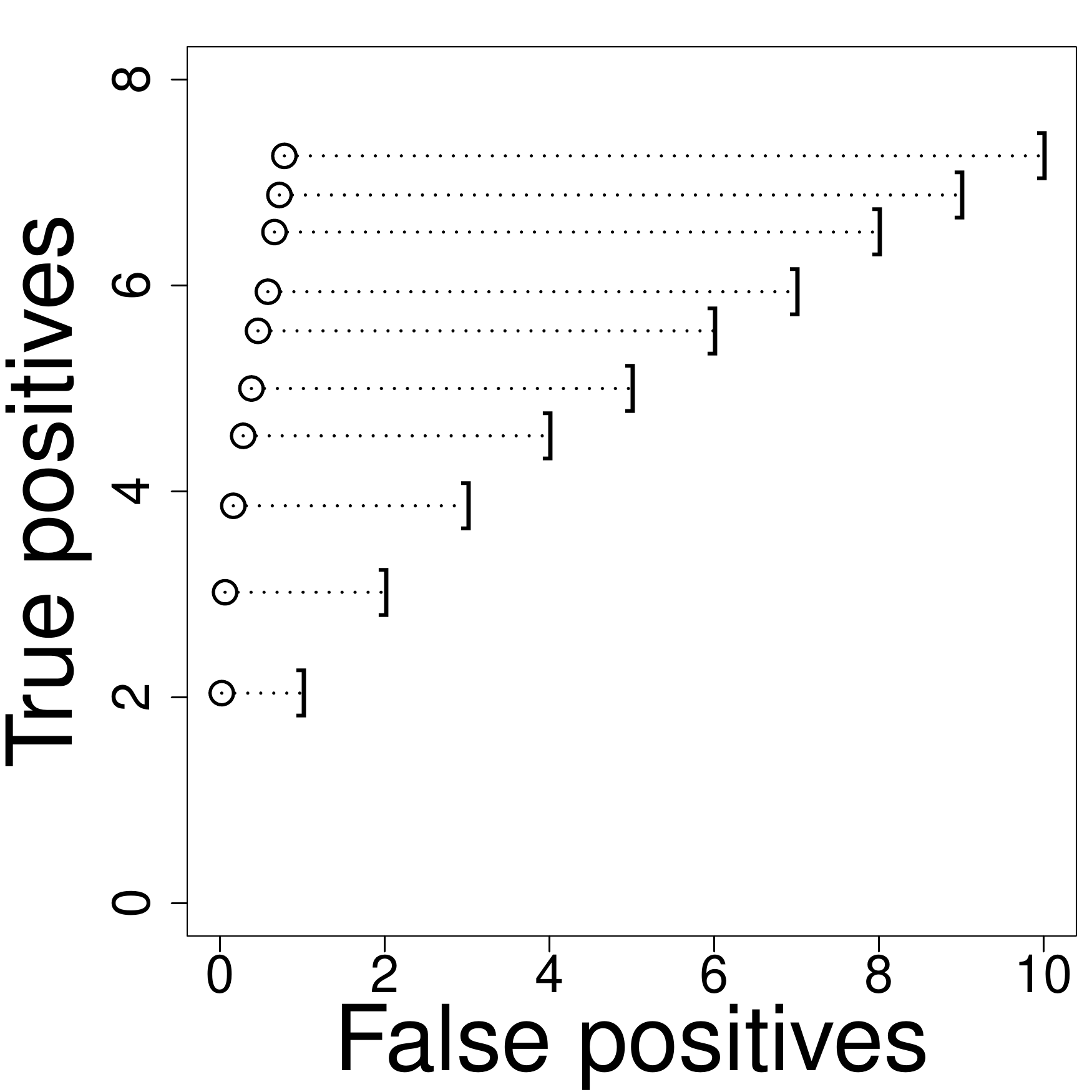}}
    \subfigure[Ising:
    Rates]{\includegraphics[width=0.3\textwidth]{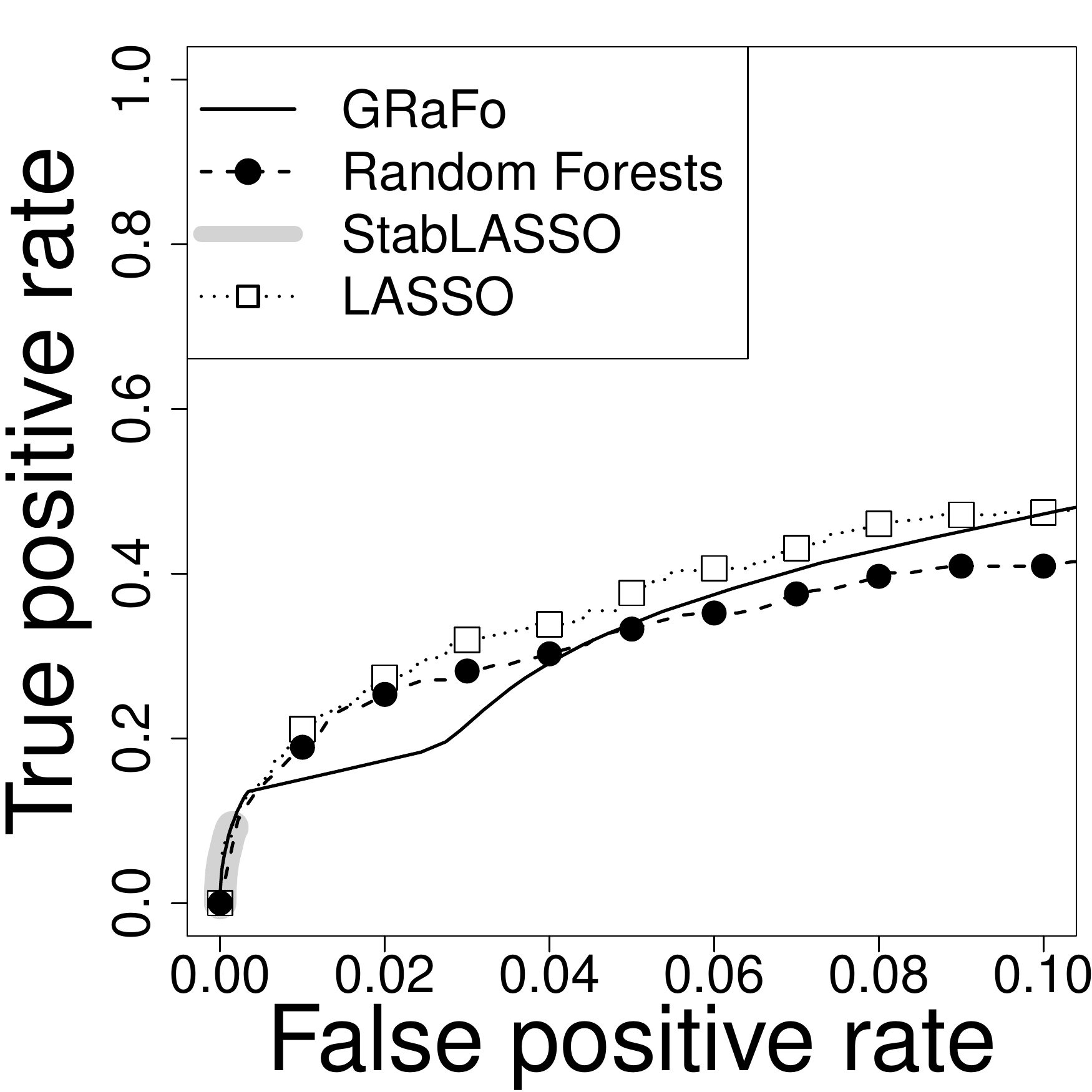}}
    \caption{The rows correspond to the Gaussian, Bernoulli, and Ising
      model with $p=200$. Their true CIGs have $334$, $334$ and $369$
      edges, respectively. The first two columns report the observed
      number of true and false positives (``o'') relative to the bound
      in (\ref{StabSelErrorControl}) for the expected number
      $\mathbb{E}[V]$ of false positives (``$]$'') for GRaFo and
      StabLASSO, respectively, averaged over 50 simulations.  The
      third column reports the averaged true and false positive rates
      of GRaFo and StabLASSO relative to the performance of their
      ``raw'' counterparts without Stability Selection.}
    \label{FigGBI200}
  \end{figure}

  \begin{figure}
    \centering
\textbf{Multinomial and mixed-type models, $p=50$}\\\vspace{0.5cm}
\subfigure[Multinomial:
    GRaFo]{\includegraphics[width=0.3\textwidth]{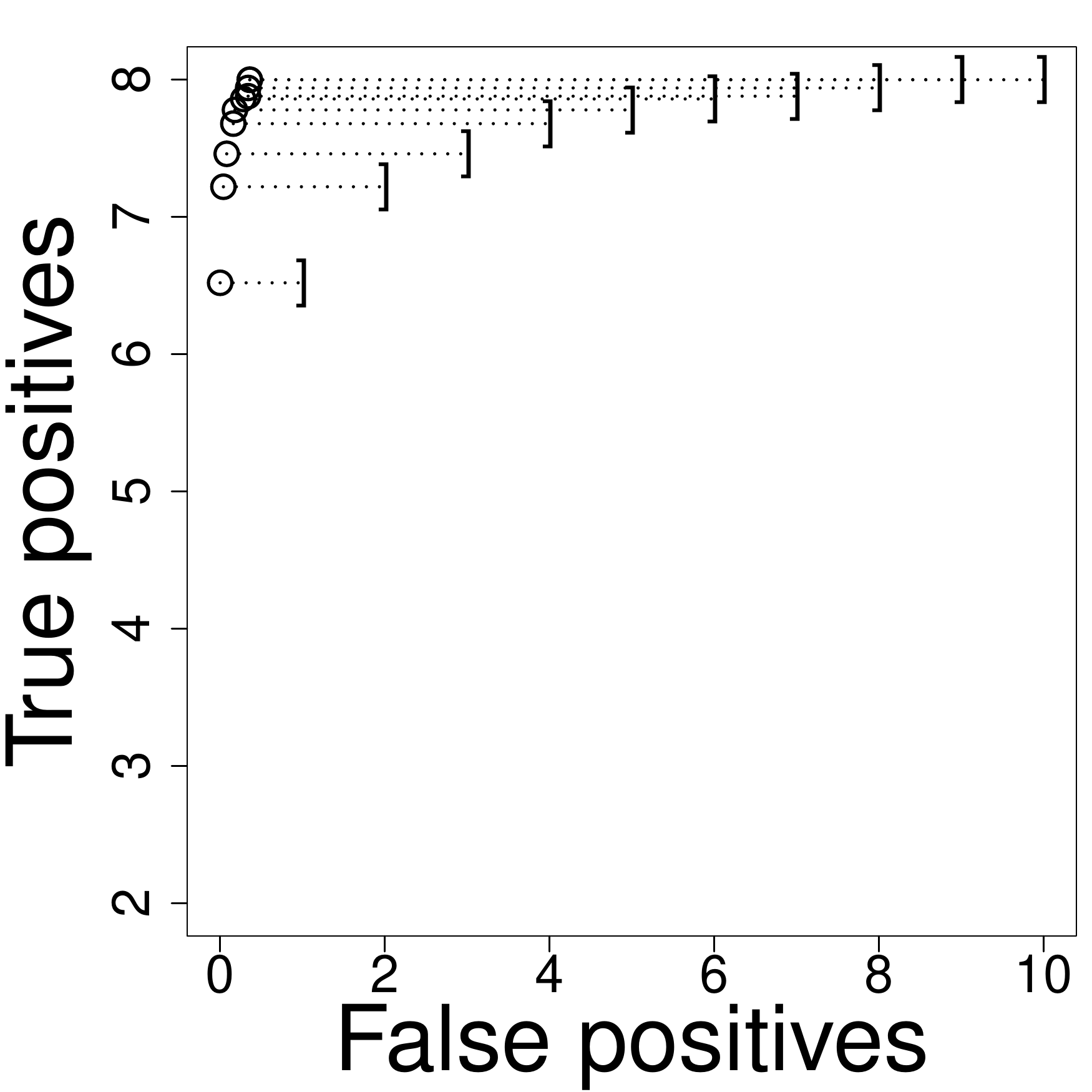}}
    \subfigure[Multinomial:\newline \textcolor{White}{(b) }StabLASSO (-1/1)]{\includegraphics[width=0.3\textwidth]{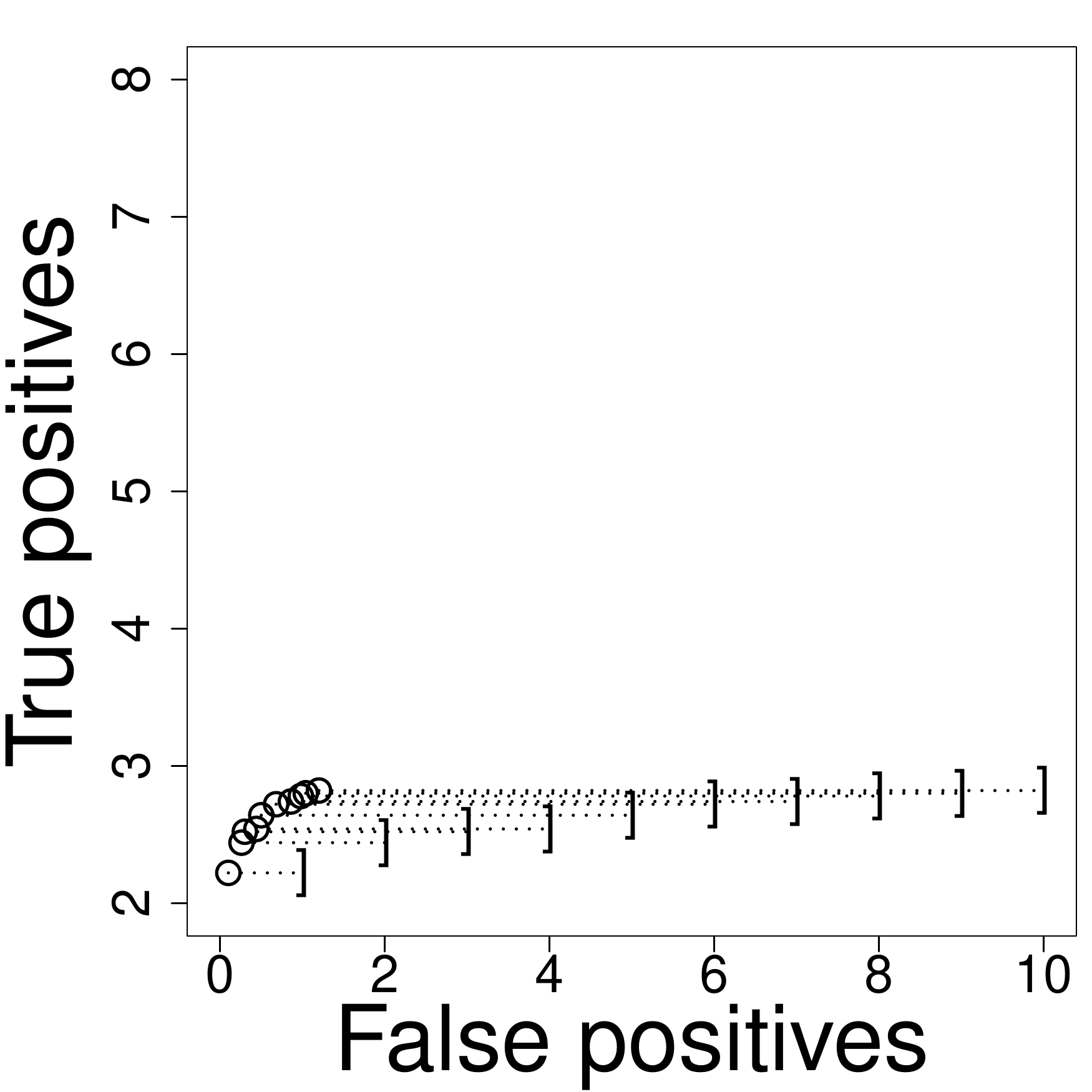}}
    \subfigure[Multinomial: Rates]{\includegraphics[width=0.3\textwidth]{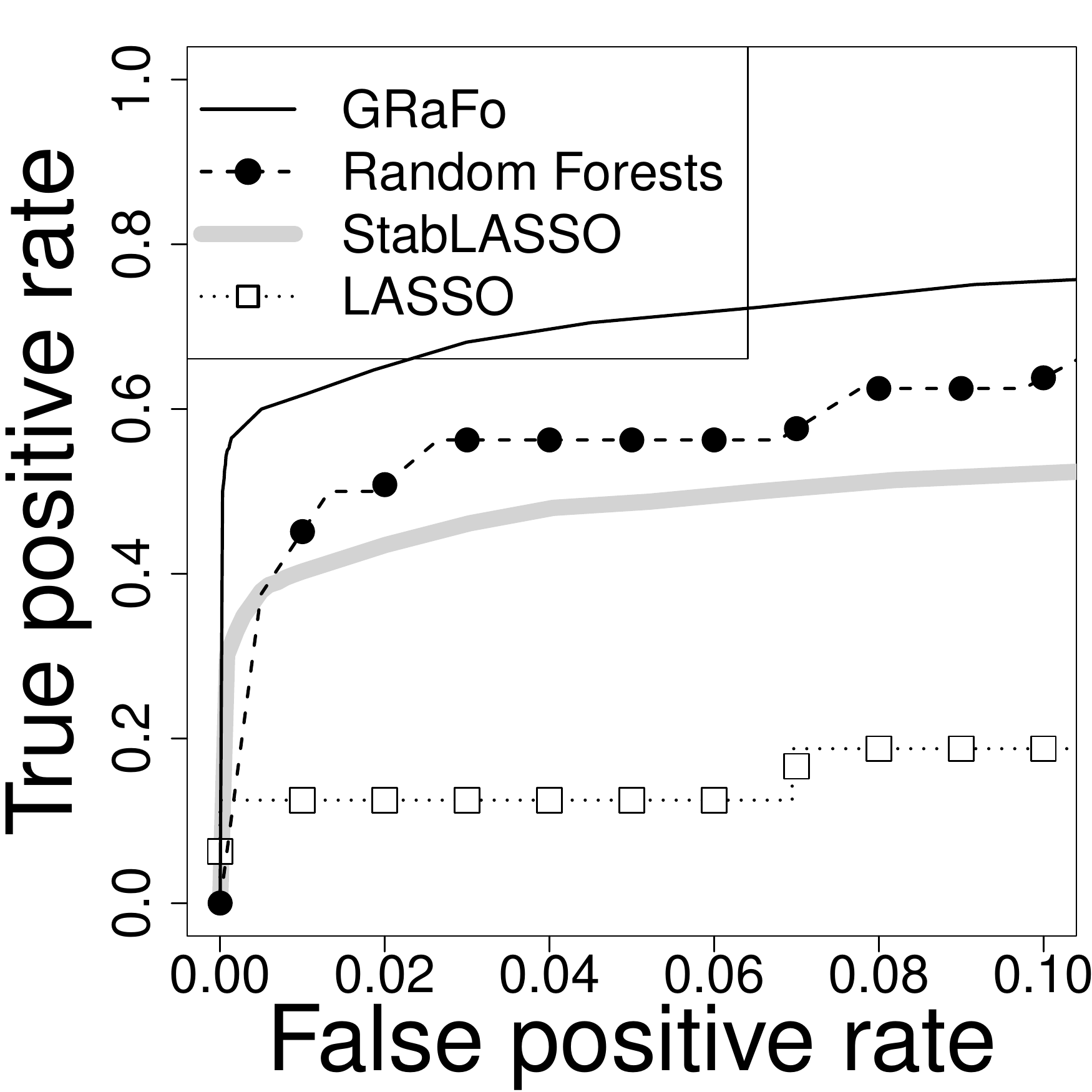}}\\
    \subfigure[Mixed:
    GRaFo]{\includegraphics[width=0.3\textwidth]{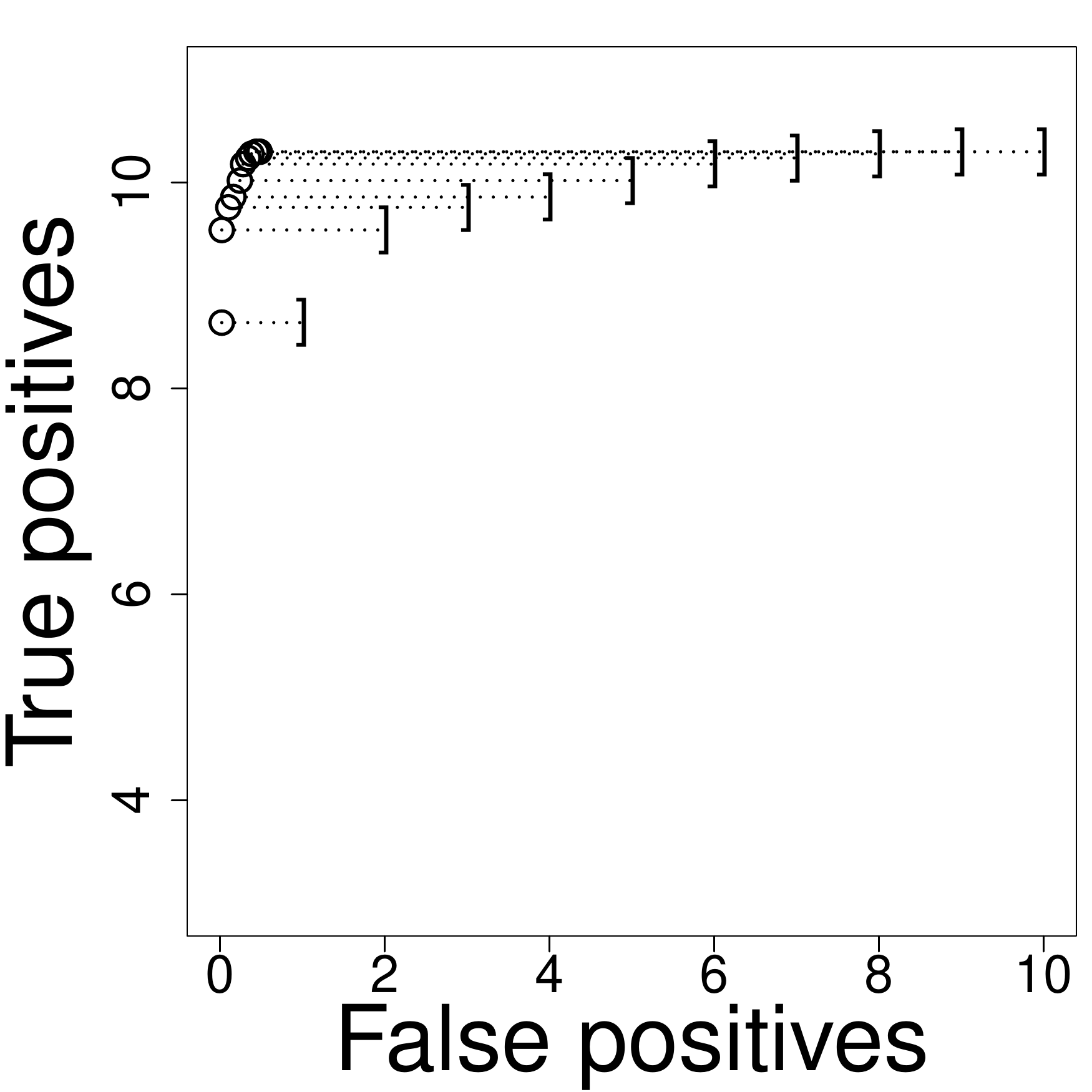}}
    \subfigure[Mixed: \newline \textcolor{White}{(b) }StabLASSO (-1/1)]{\includegraphics[width=0.3\textwidth]{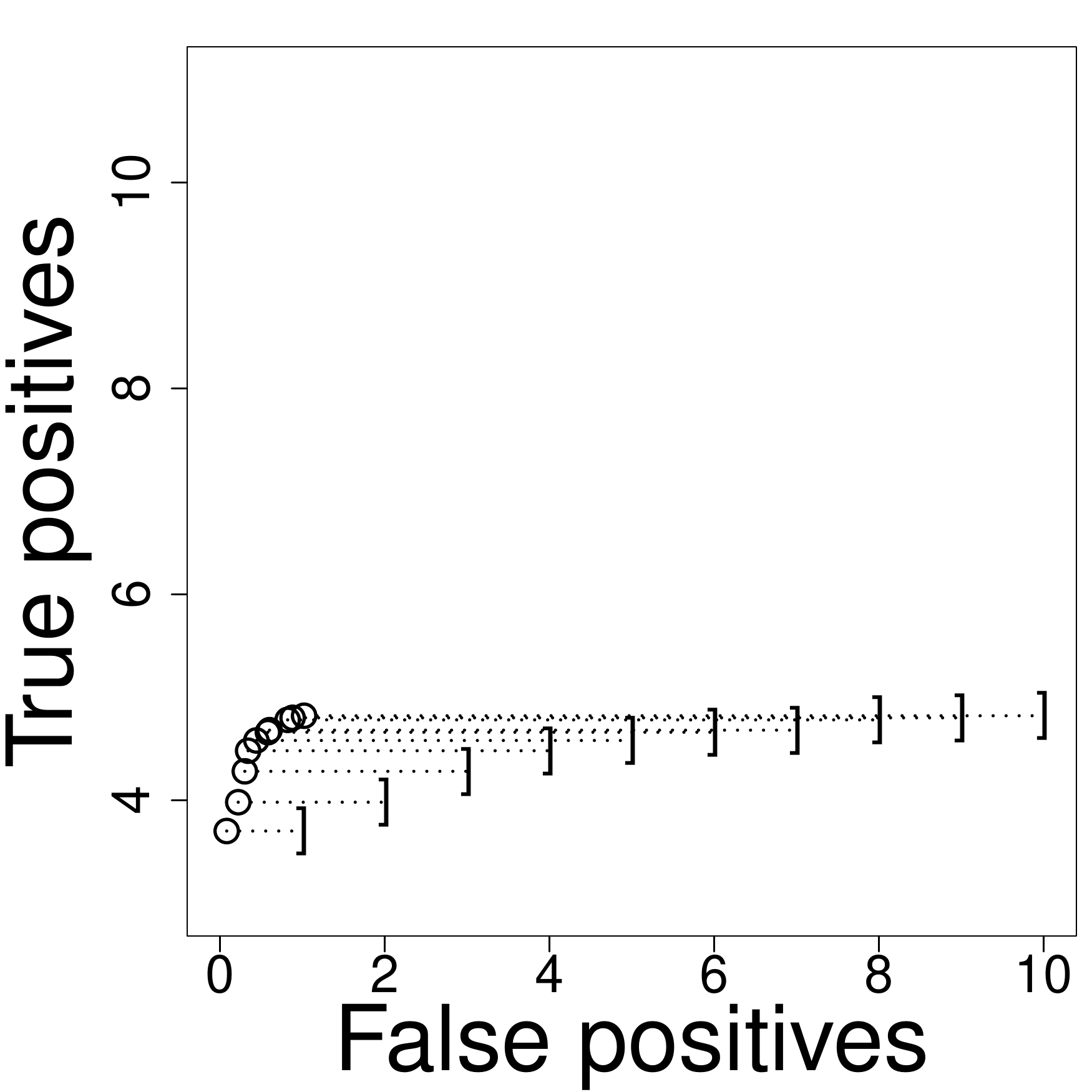}}
    \subfigure[Mixed:
    Rates]{\includegraphics[width=0.3\textwidth]{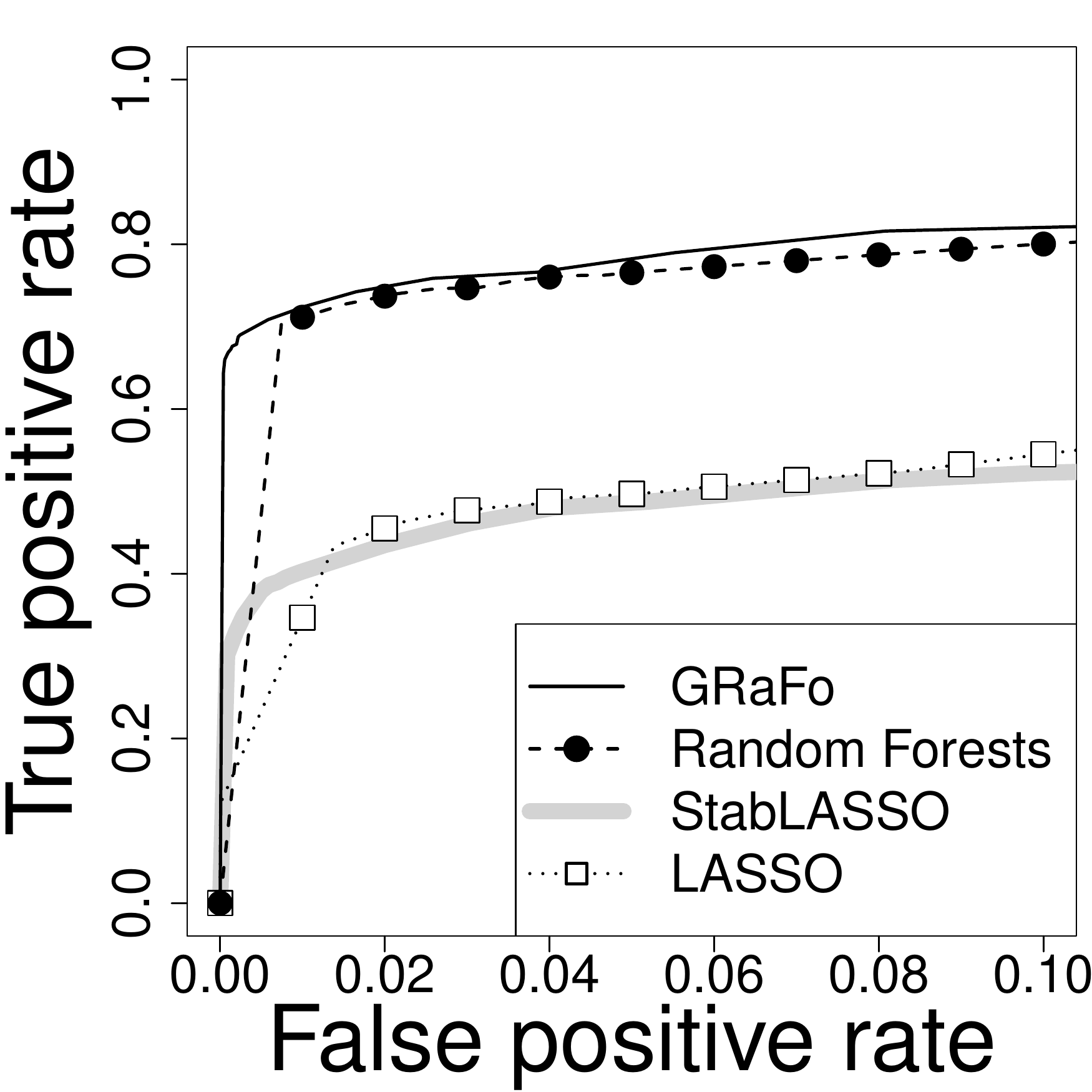}}
    \caption{The rows correspond to the multinomial and mixed-type
      model with $p=50$. Their true CIGs both have $16$ edges. The
      first two columns report the observed number of true and false
      positives (``o'') relative to the bound in
      (\ref{StabSelErrorControl}) for the expected number
      $\mathbb{E}[V]$ of false positives (``$]$'') for GRaFo and
      StabLASSO, respectively, averaged over 50 simulations.  The
      third column reports the averaged true and false positive rates
      of GRaFo and StabLASSO relative to the performance of their
      ``raw'' counterparts without Stability Selection.}
    \label{FigMMi50}
  \end{figure}

  \begin{figure}
    \centering
\textbf{Multinomial and mixed-type models, $p=100$}\\\vspace{0.5cm}
    \subfigure[Multinomial:
    GRaFo]{\includegraphics[width=0.3\textwidth]{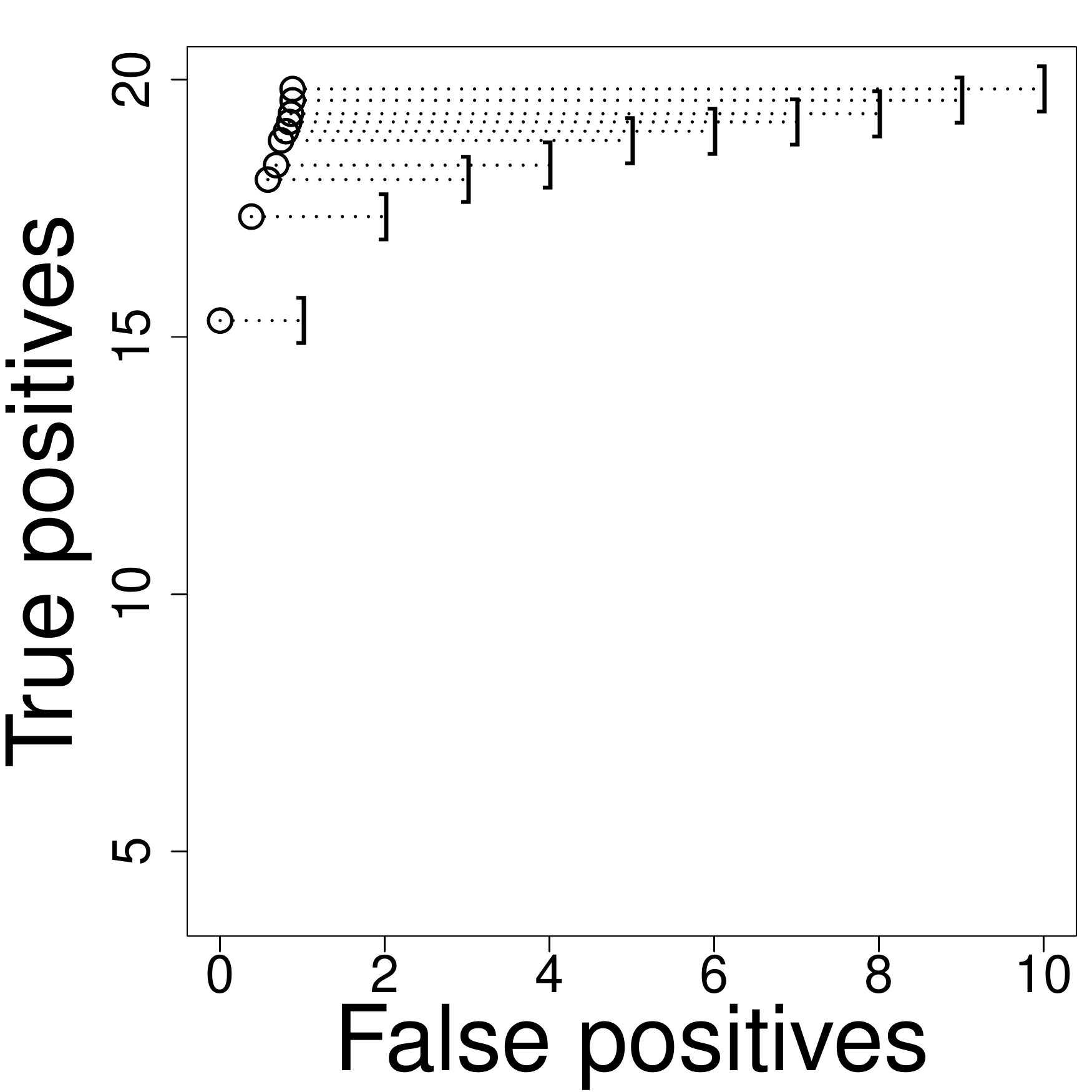}}
    \subfigure[Multinomial:\newline \textcolor{White}{(b) }StabLASSO (-1/1)]{\includegraphics[width=0.3\textwidth]{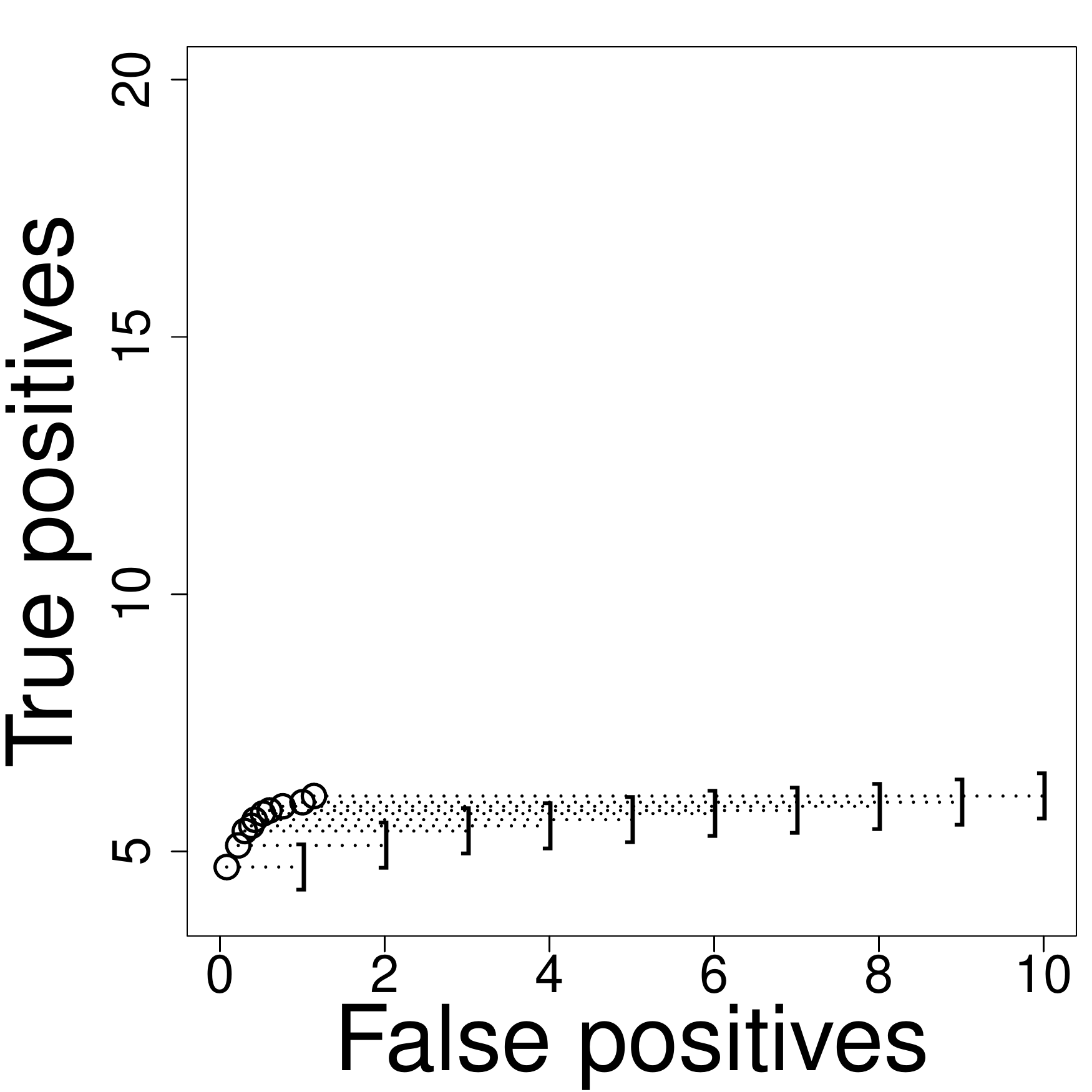}}
    \subfigure[Multinomial: Rates]{\includegraphics[width=0.3\textwidth]{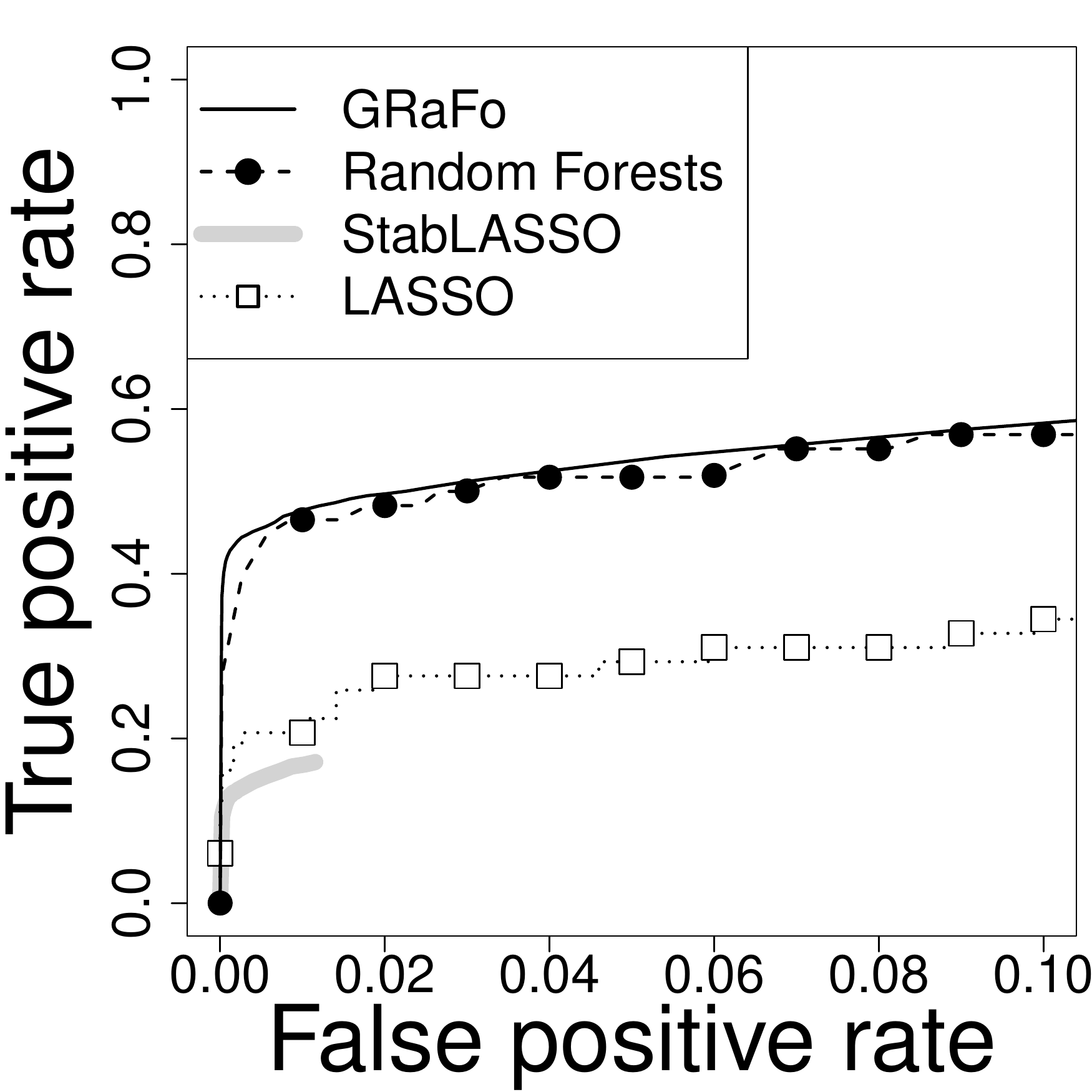}}\\
    \subfigure[Mixed:
    GRaFo]{\includegraphics[width=0.3\textwidth]{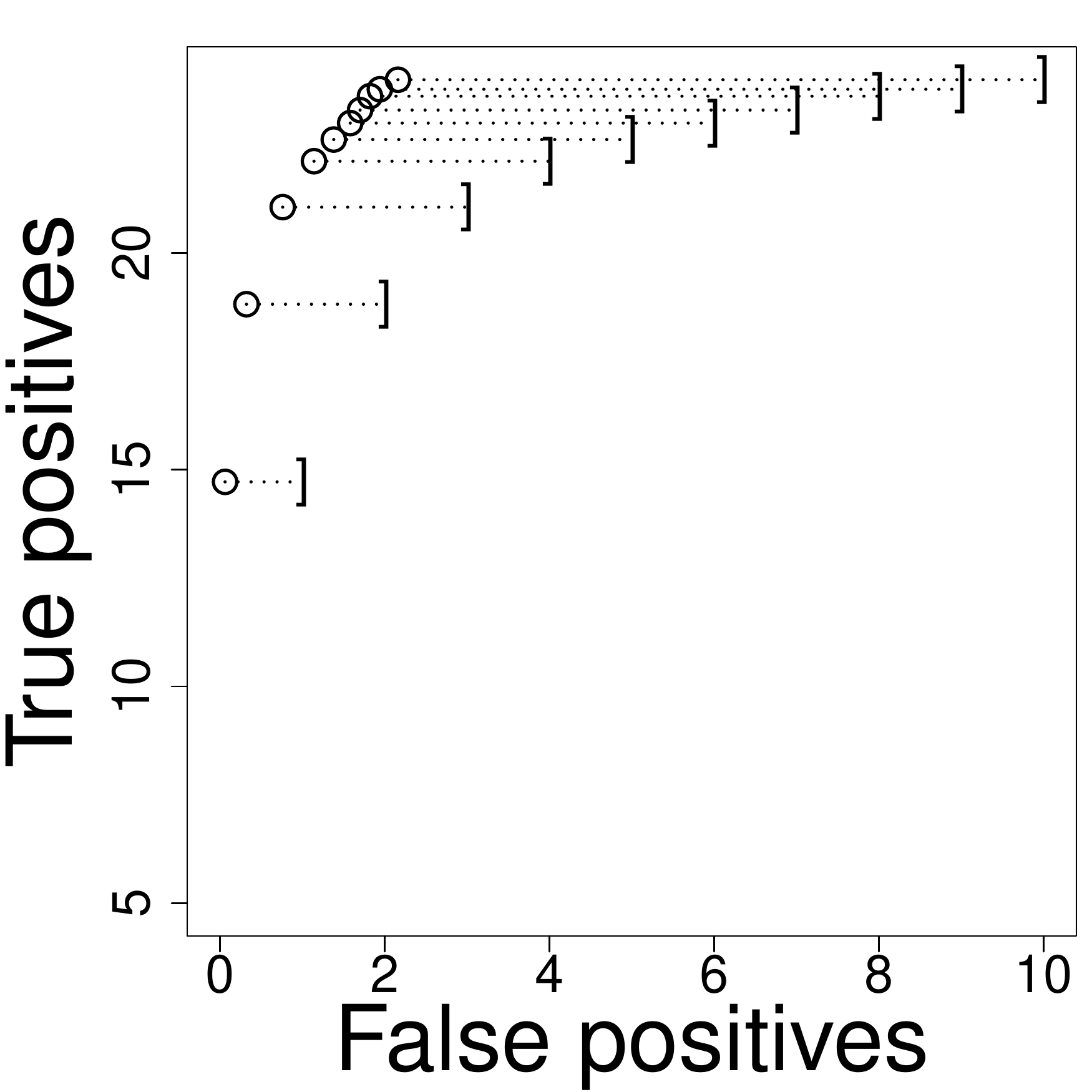}}
    \subfigure[Mixed:\newline \textcolor{White}{(b) }StabLASSO (-1/1)]{\includegraphics[width=0.3\textwidth]{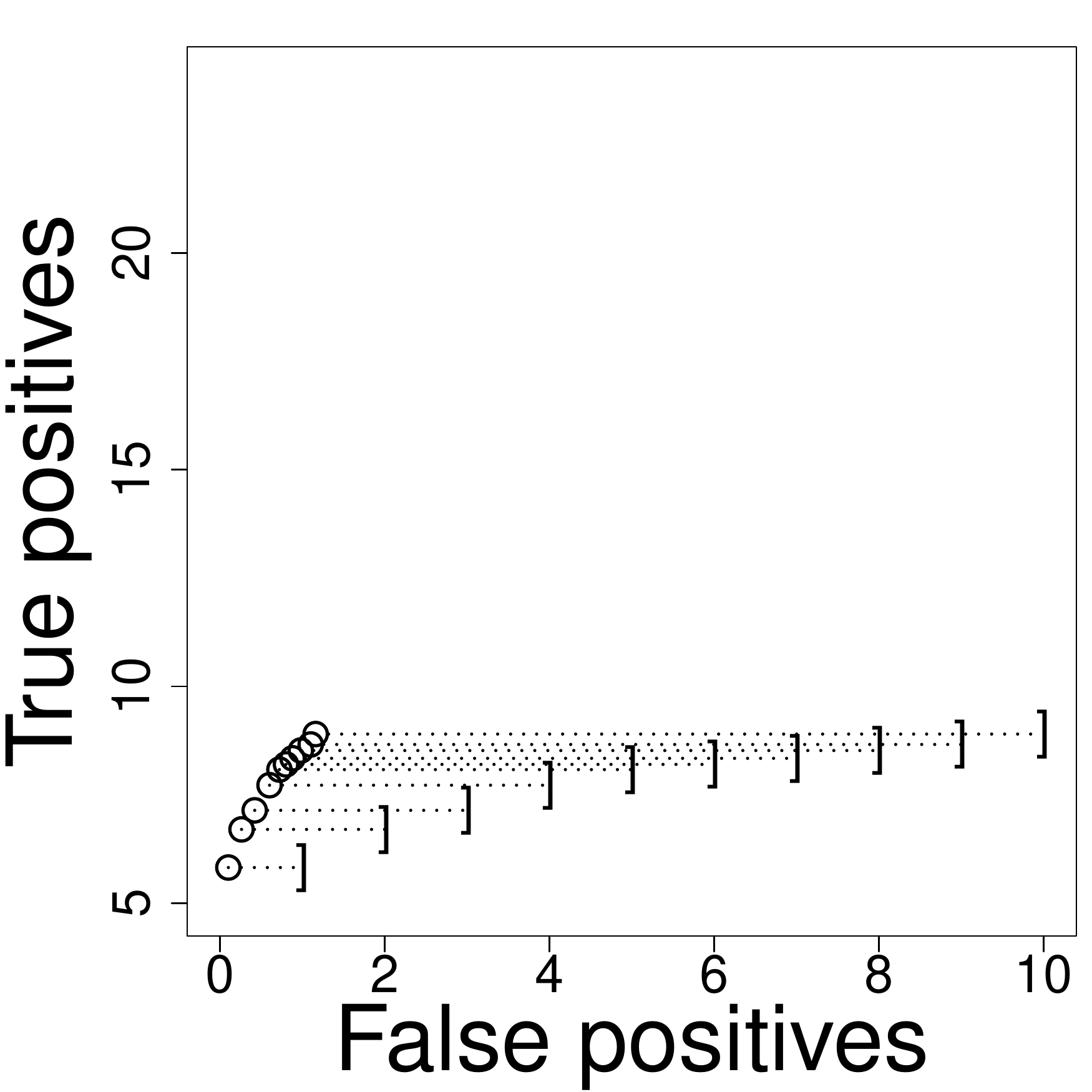}}
    \subfigure[Mixed: Rates]{\includegraphics[width=0.3\textwidth]{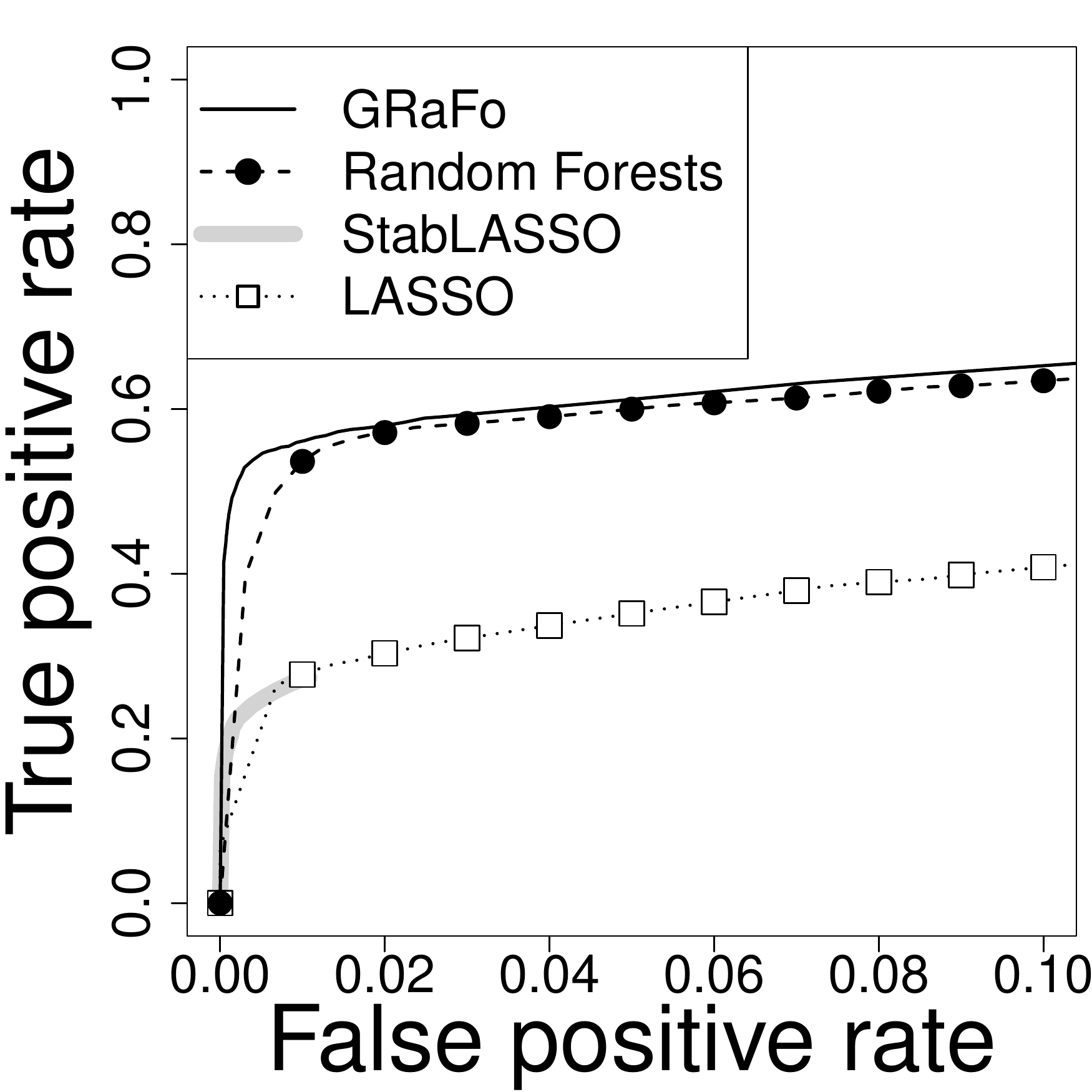}}
    \caption{The rows correspond to the multinomial and mixed-type
      model with $p=100$. Their true CIGs both have $58$ edges. The
      first two columns report the observed number of true and false
      positives (``o'') relative to the bound in
      (\ref{StabSelErrorControl}) for the expected number
      $\mathbb{E}[V]$ of false positives (``$]$'') for GRaFo and
      StabLASSO, respectively, averaged over 50 simulations.  The
      third column reports the averaged true and false positive rates
      of GRaFo and StabLASSO relative to the performance of their
      ``raw'' counterparts without Stability Selection.}
    \label{FigMMi100}
  \end{figure}

  \begin{figure}
    \centering
\textbf{Multinomial and mixed-type models, $p=200$}\\\vspace{0.5cm}
    \subfigure[Multinomial:
    GRaFo]{\includegraphics[width=0.3\textwidth]{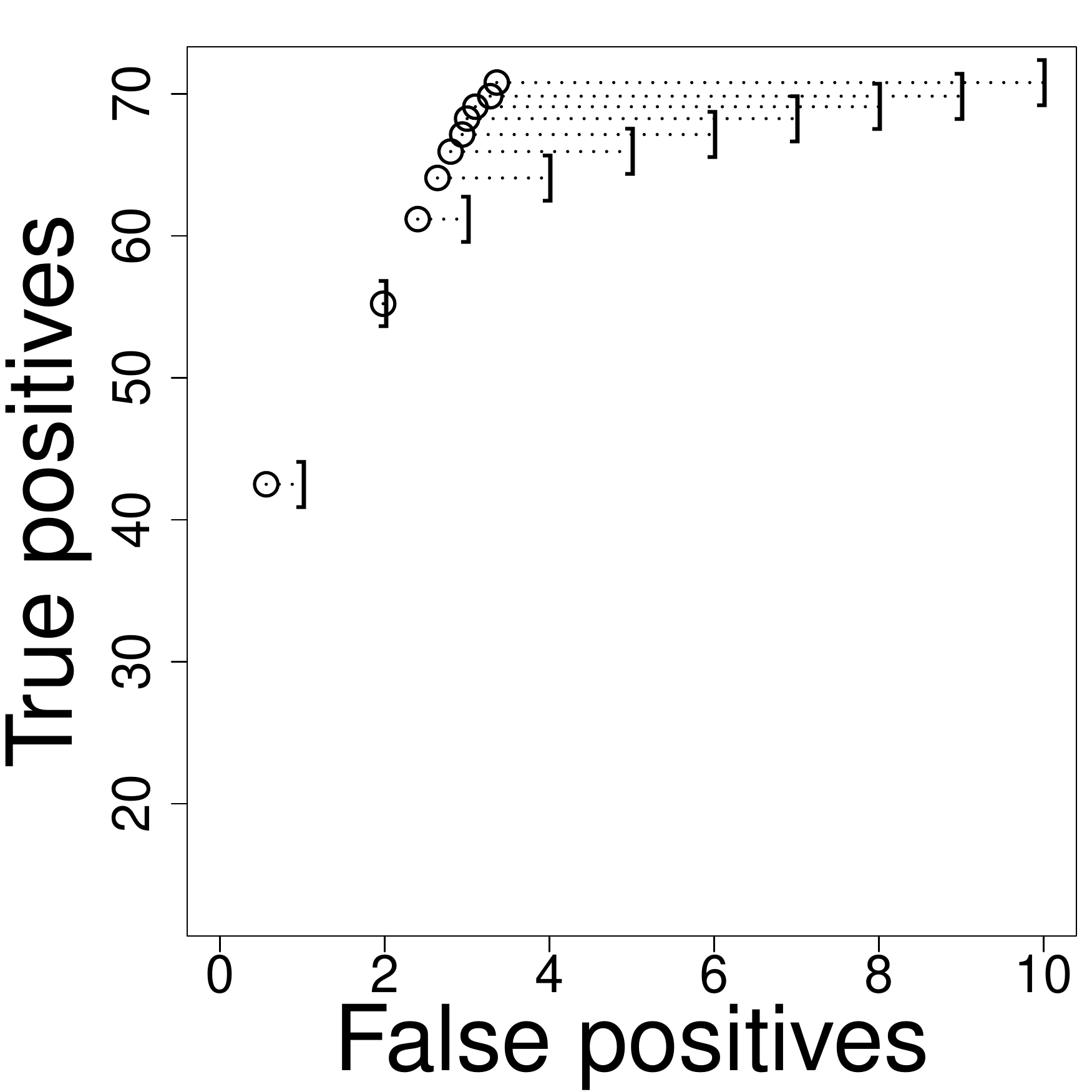}}
    \subfigure[Multinomial:\newline \textcolor{White}{(b) }StabLASSO (-1/1)]{\includegraphics[width=0.3\textwidth]{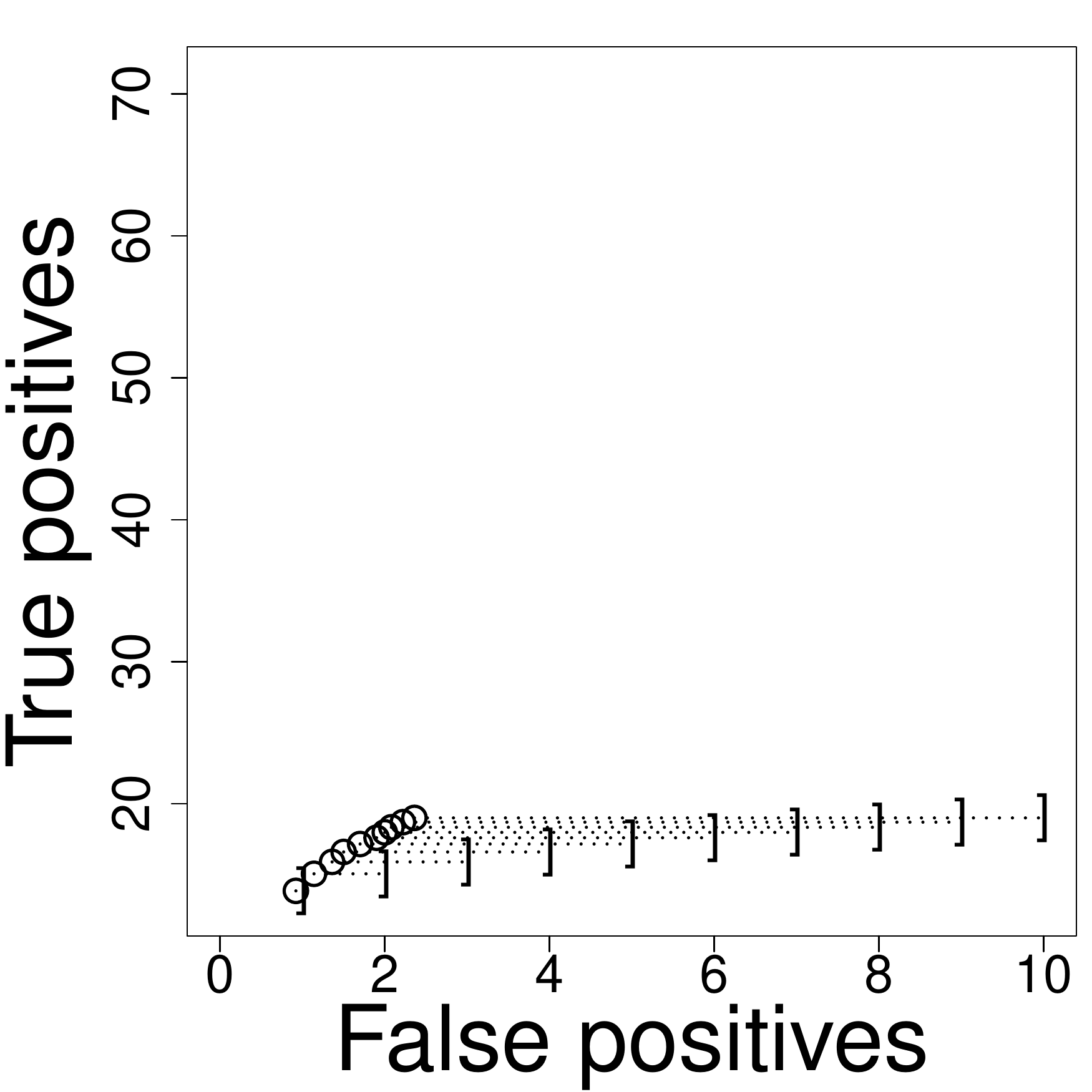}}
    \subfigure[Multinomial: Rates]{\includegraphics[width=0.3\textwidth]{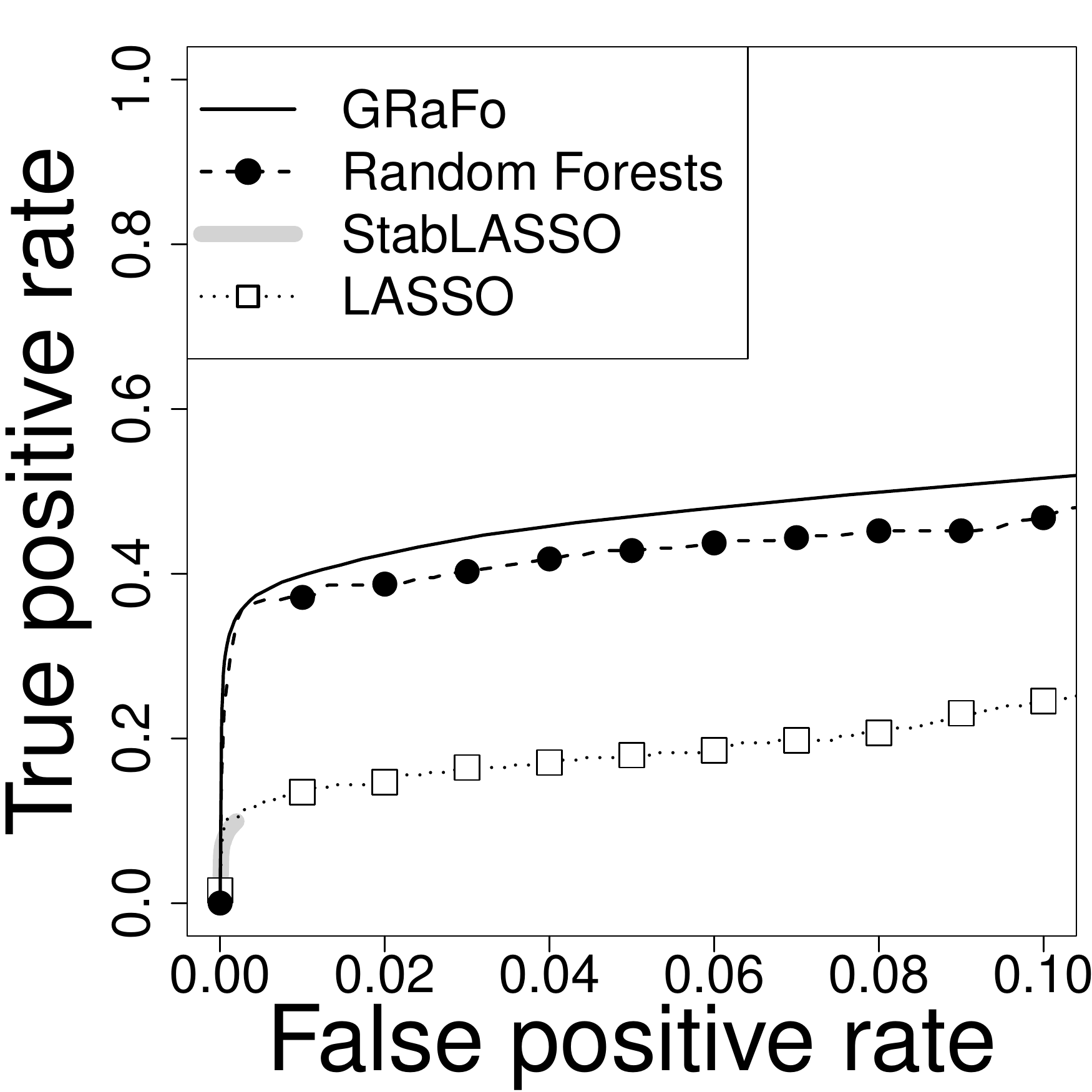}}\\
    \subfigure[Mixed:
    GRaFo]{\includegraphics[width=0.3\textwidth]{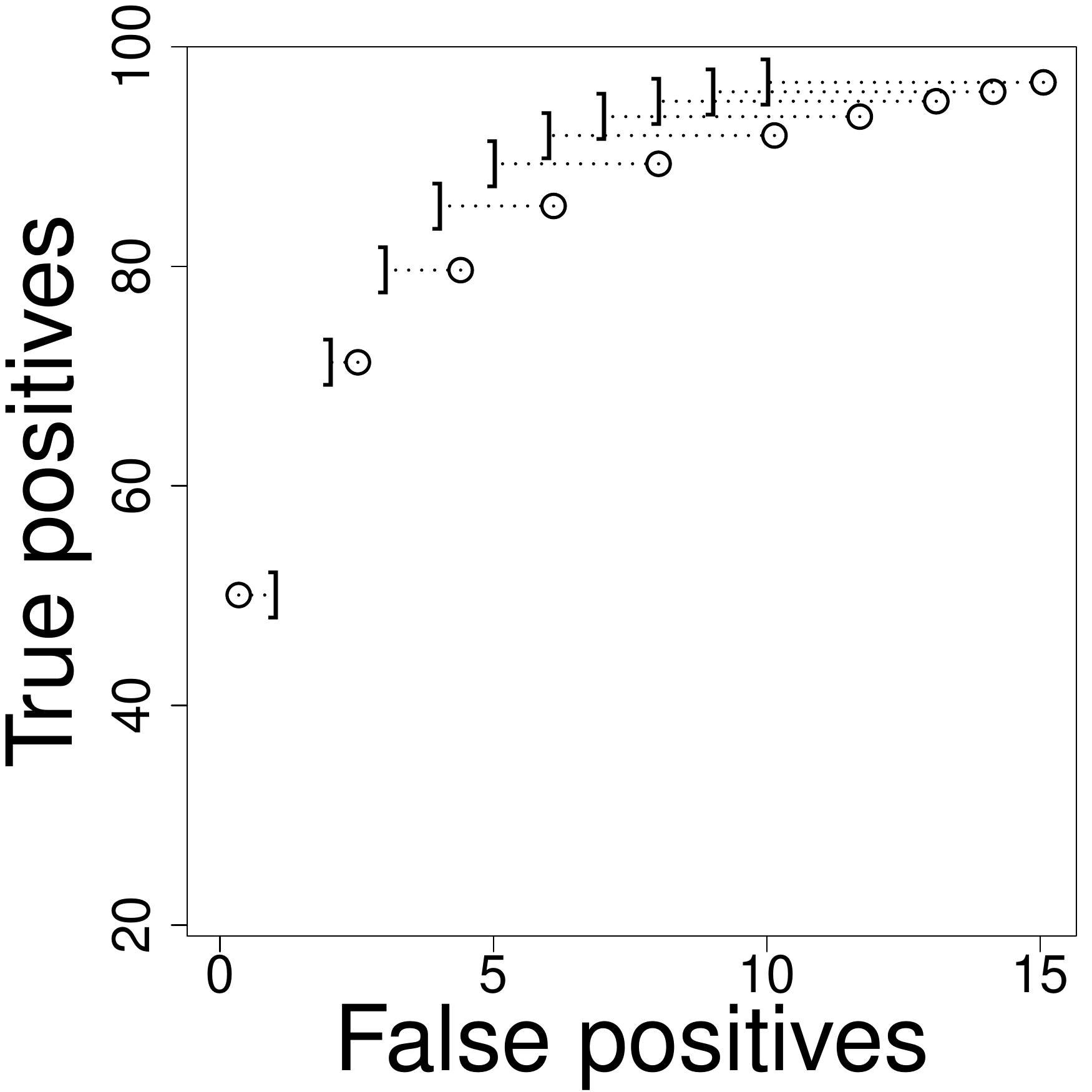}}
    \subfigure[Mixed:\newline \textcolor{White}{(b) }StabLASSO (-1/1)]{\includegraphics[width=0.3\textwidth]{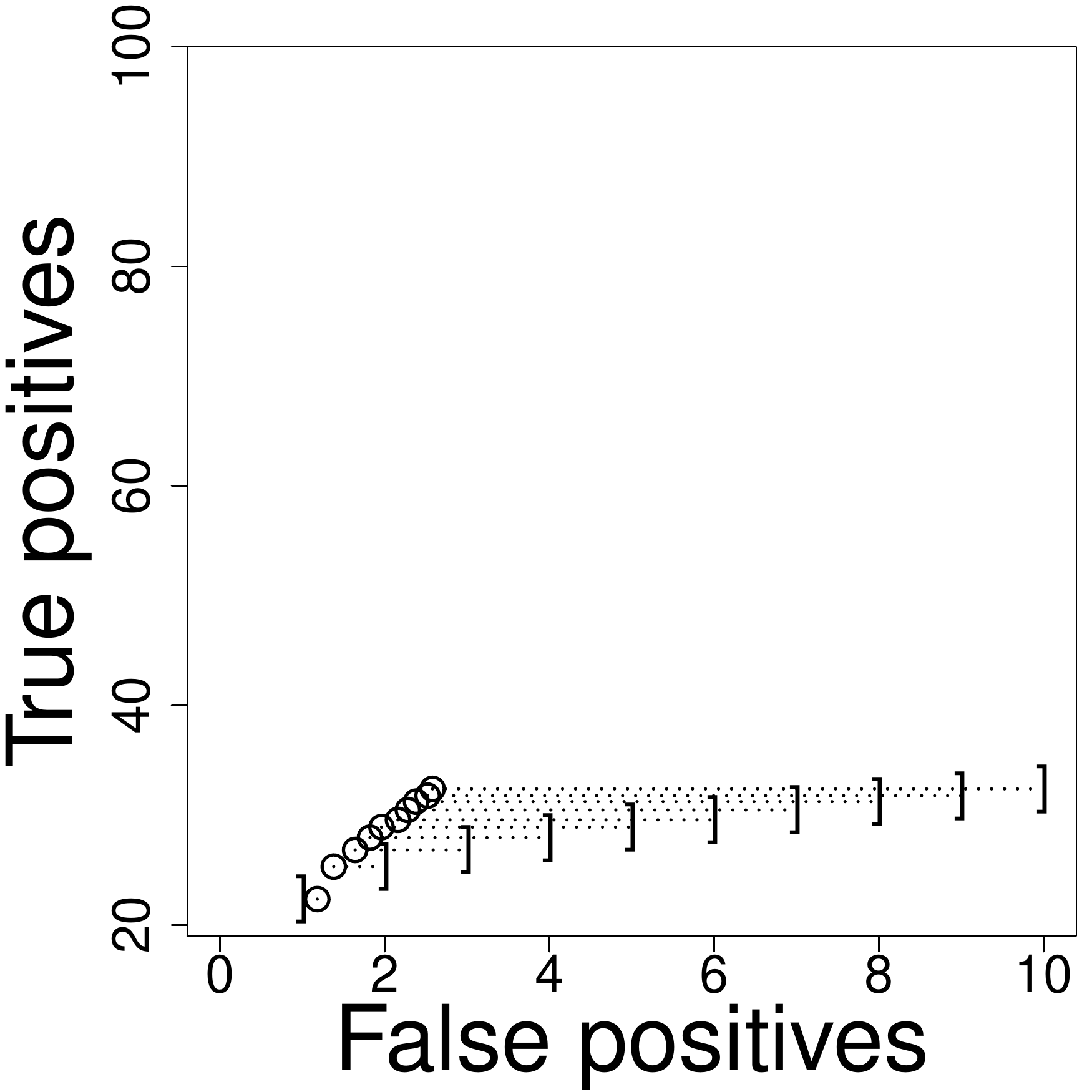}}
    \subfigure[Multinomial: Rates]{\includegraphics[width=0.3\textwidth]{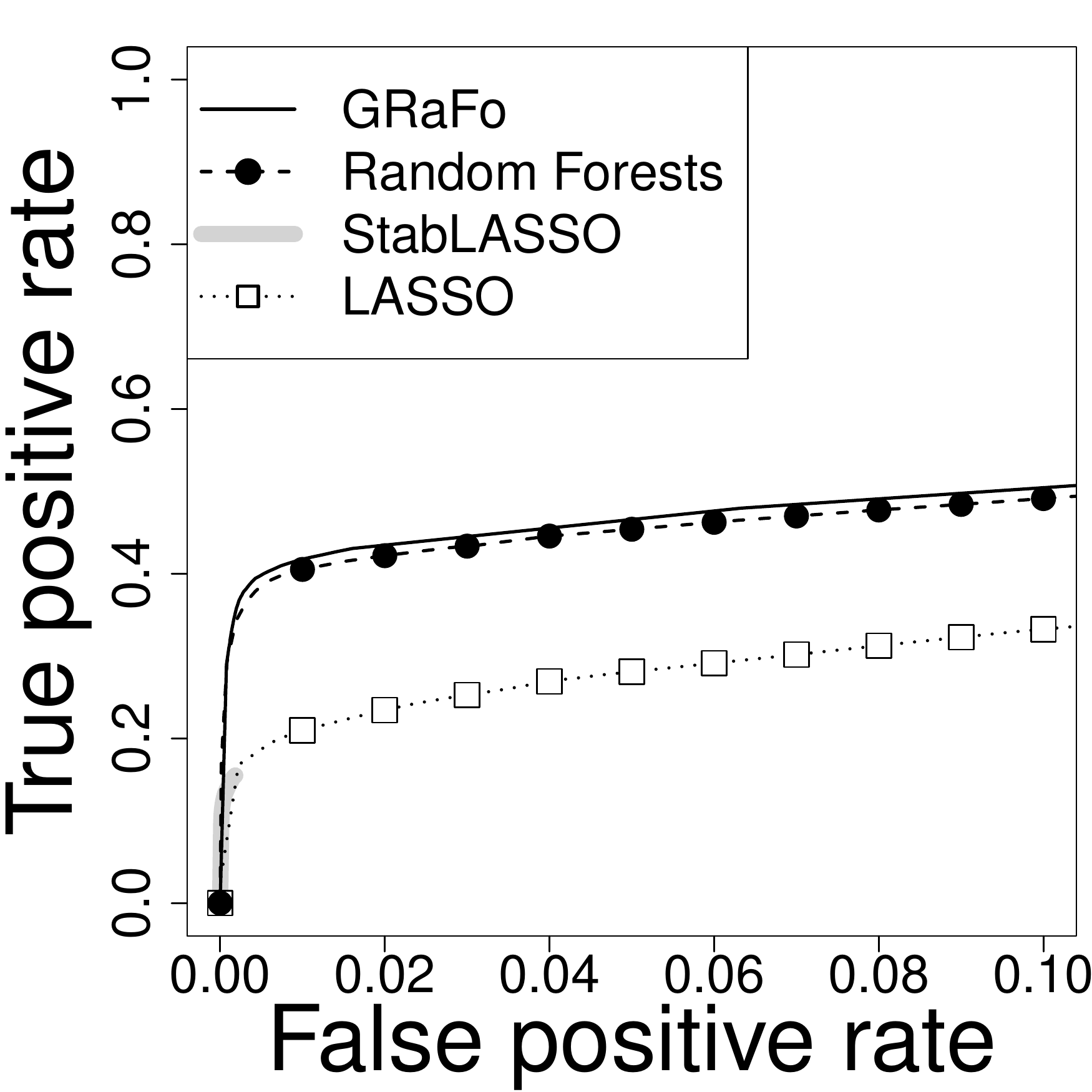}}
    \caption{The rows correspond to the multinomial and mixed-type
      model with $p=200$. Their true CIGs both have $334$ edges. The
      first two columns report the observed number of true and false
      positives (``o'') relative to the bound in
      (\ref{StabSelErrorControl}) for the expected number
      $\mathbb{E}[V]$ of false positives (``$]$'') for GRaFo and
      StabLASSO, respectively, averaged over 50 simulations.  The
      third column reports the averaged true and false positive rates
      of GRaFo and StabLASSO relative to the performance of their
      ``raw'' counterparts without Stability Selection.}
    \label{FigMMi200}
  \end{figure}
\fi

\subsection{Simulation Results: Gaussian with Interaction Effects}
\label{Sec:SimResInteraction}

\begin{figure}
    \centering
    \textbf{Gaussian with interaction effects $p=50$, $100$, and $200$}\\\vspace{0.5cm}
     \subfigure[Interaction, $p=50$:\newline \textcolor{White}{(a) }GRaFo]{\includegraphics[width=0.3\textwidth]{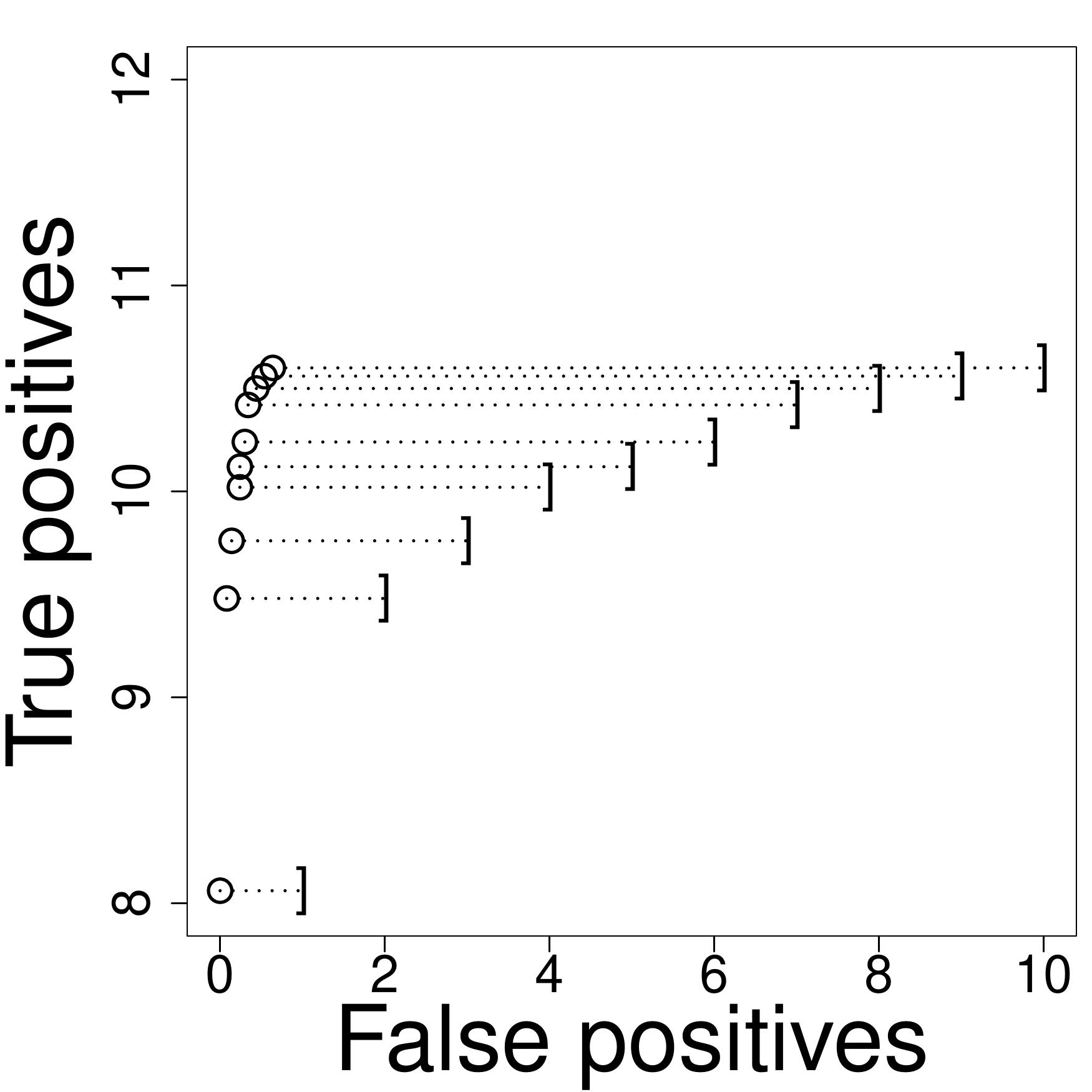}}
    \subfigure[Interaction, $p=50$:\newline \textcolor{White}{(b) }StabLASSO]{\includegraphics[width=0.3\textwidth]{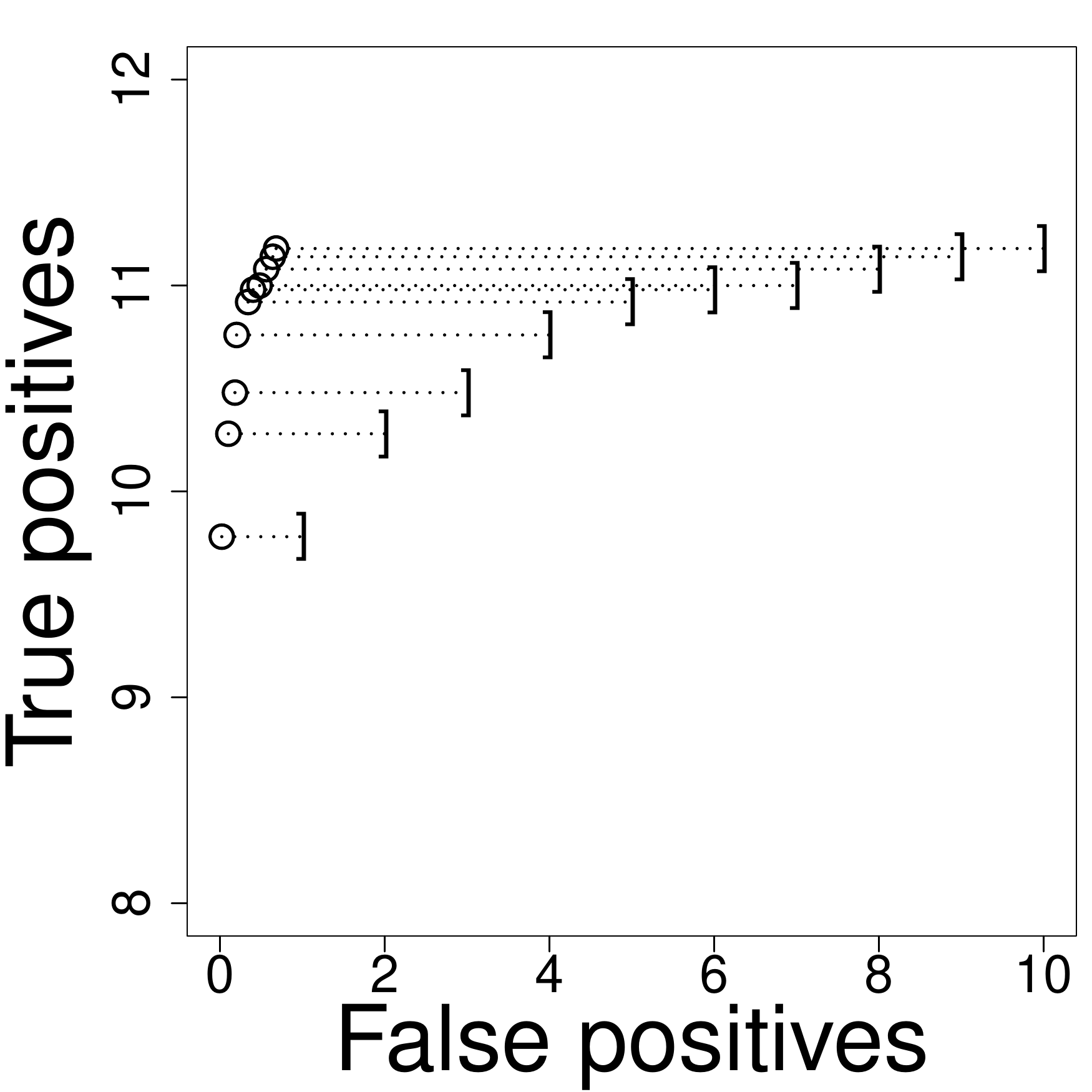}}
    \subfigure[Interaction, $p=50$:\newline \textcolor{White}{(c) }Rates]{\includegraphics[width=0.3\textwidth]{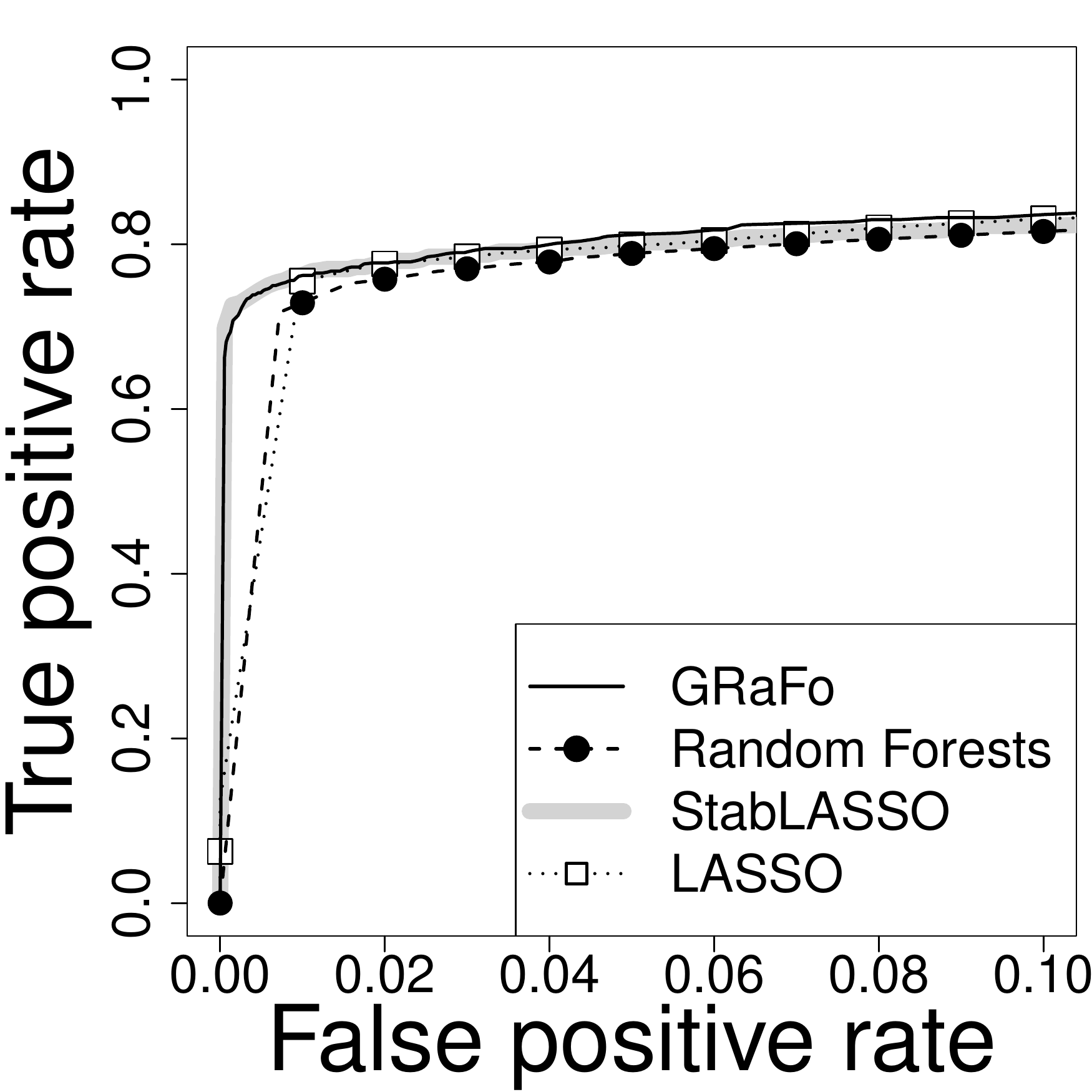}}\\
    \subfigure[Interaction, $p=100$:\newline \textcolor{White}{(d) }GRaFo]{\includegraphics[width=0.3\textwidth]{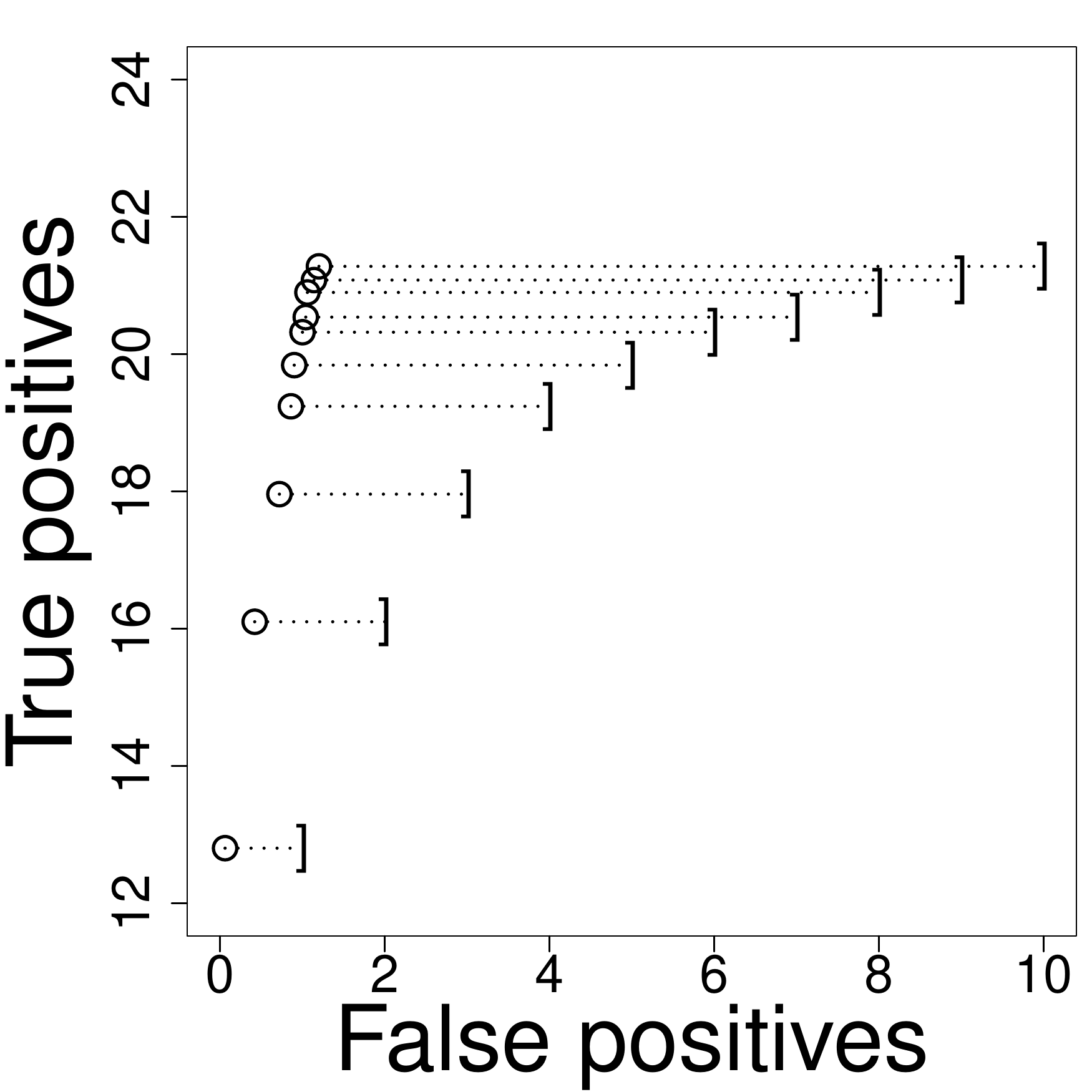}}
    \subfigure[Interaction, $p=100$:\newline \textcolor{White}{(e) }StabLASSO]{\includegraphics[width=0.3\textwidth]{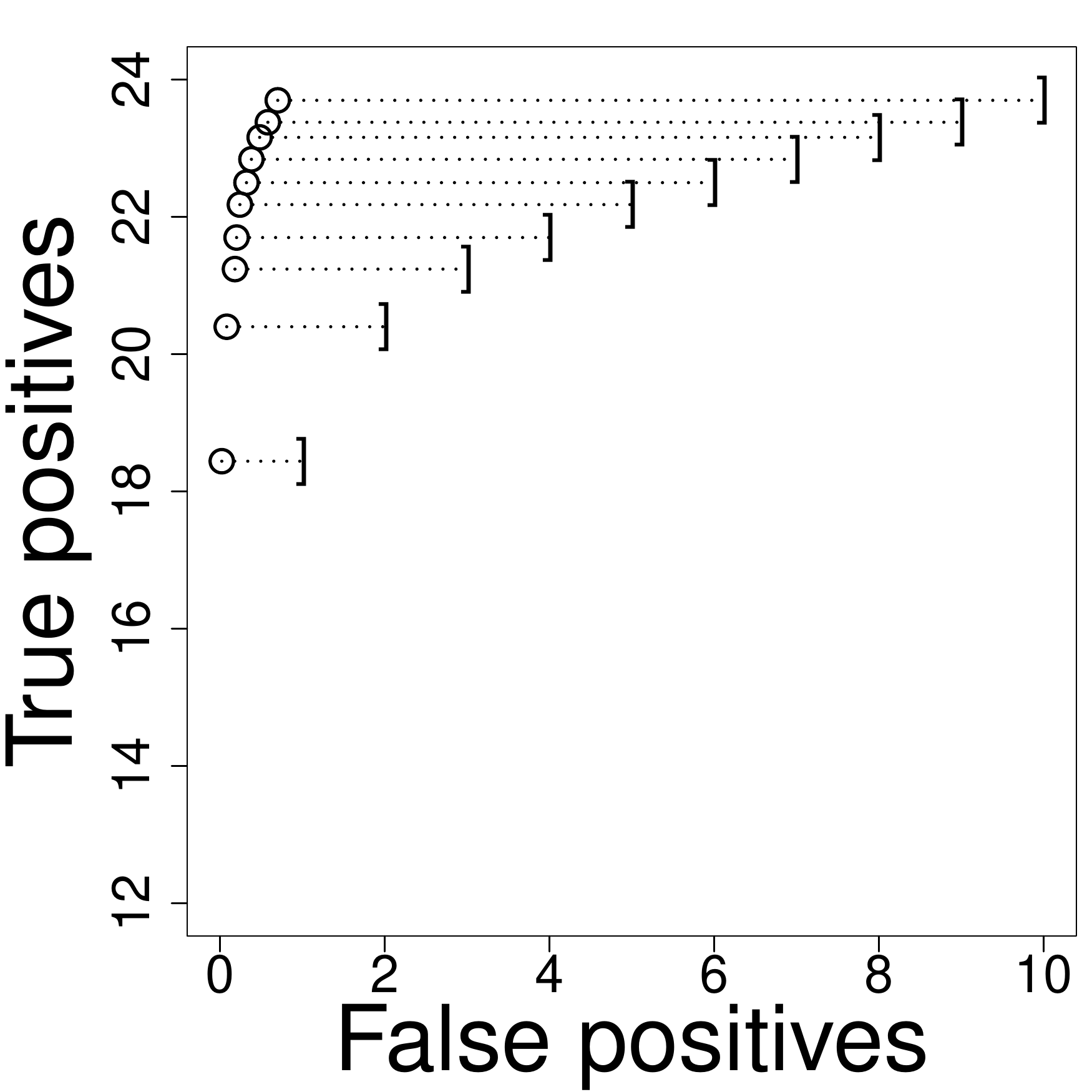}}
    \subfigure[Interaction, $p=100$:\newline \textcolor{White}{(e) }Rates]{\includegraphics[width=0.3\textwidth]{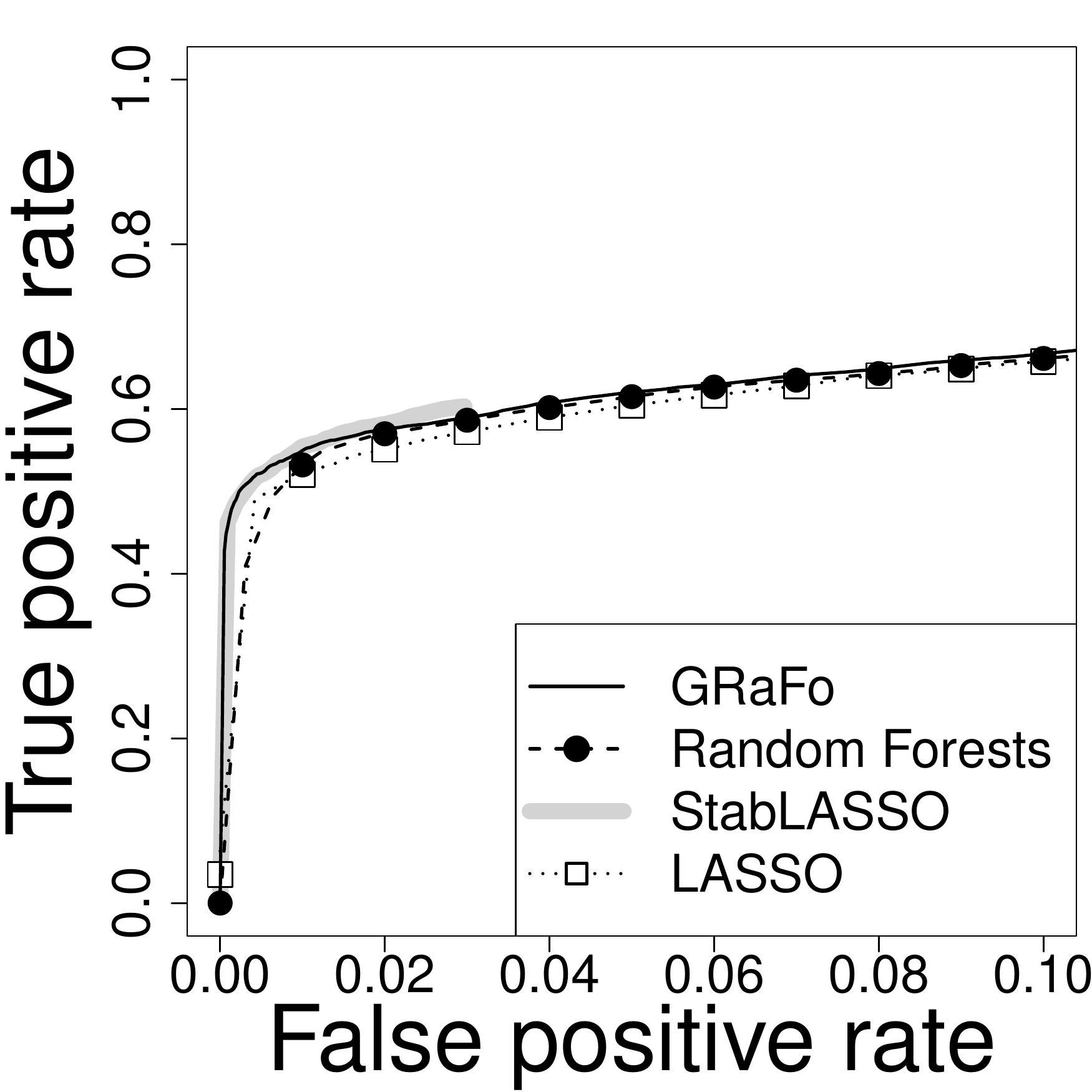}}\\
    \subfigure[Interaction, $p=200$:\newline \textcolor{White}{(f) }GRaFo]{\includegraphics[width=0.3\textwidth]{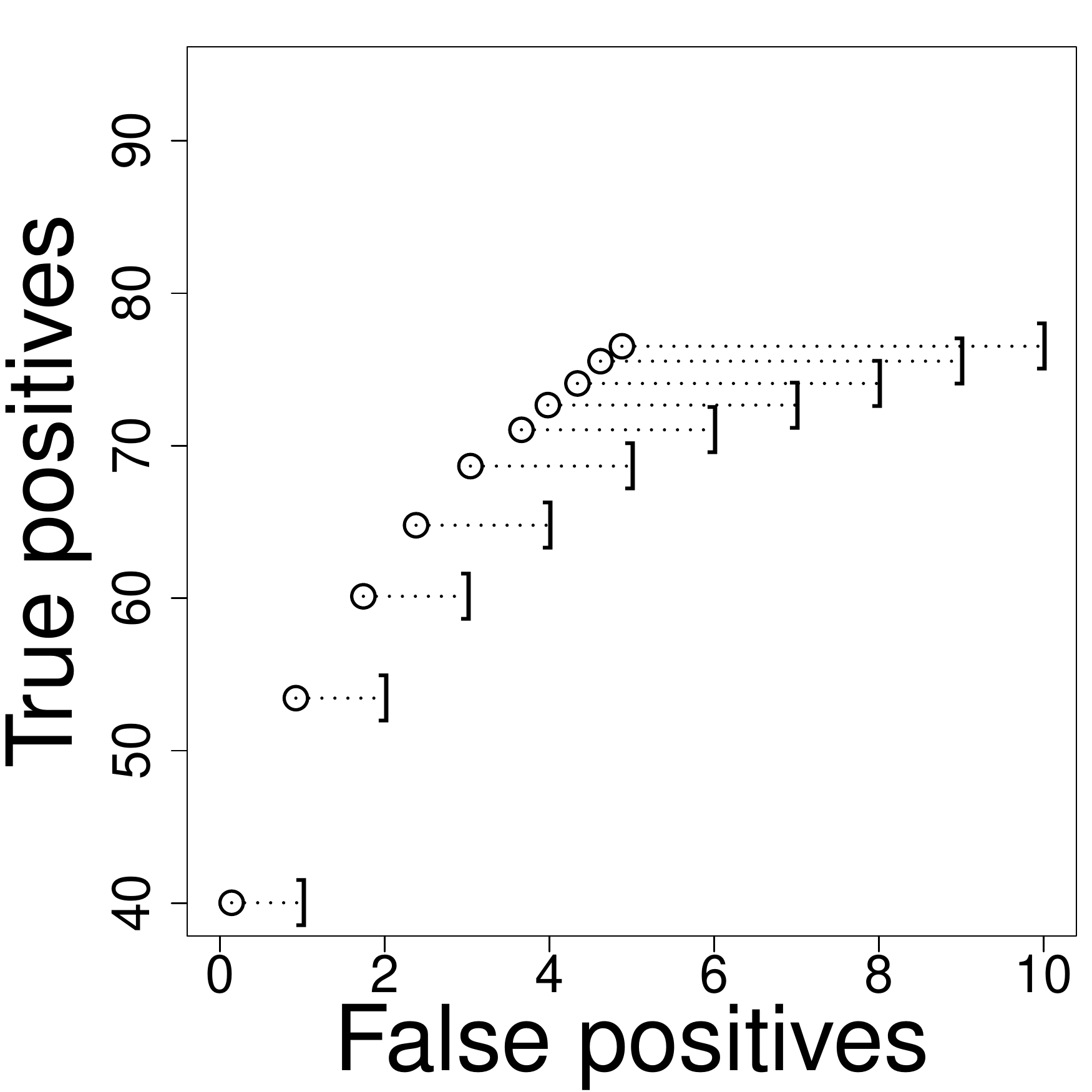}}
    \subfigure[Interaction, $p=200$:\newline \textcolor{White}{(h) }StabLASSO]{\includegraphics[width=0.3\textwidth]{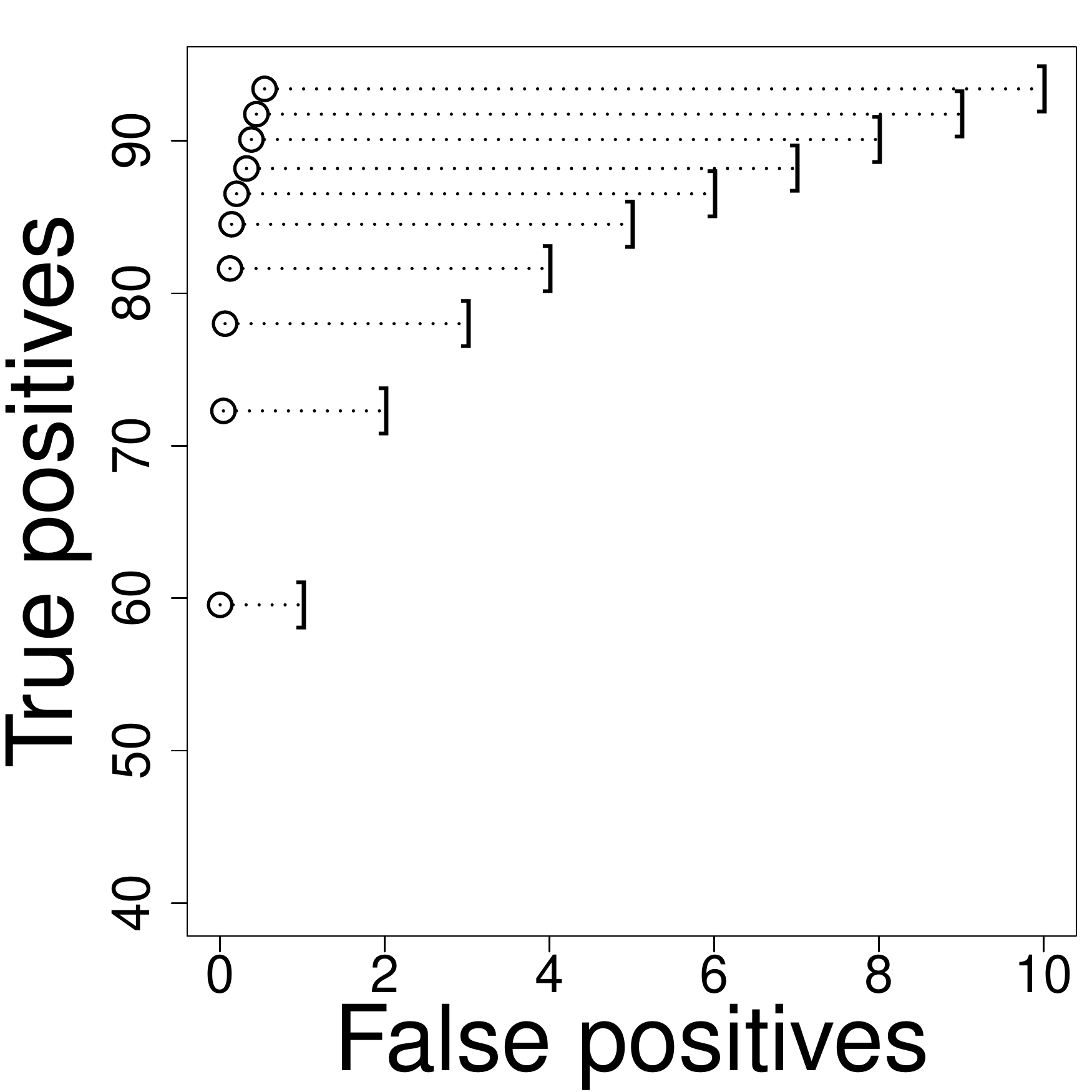}}
    \subfigure[Interaction, $p=200$:\newline \textcolor{White}{(i) }Rates]{\includegraphics[width=0.3\textwidth]{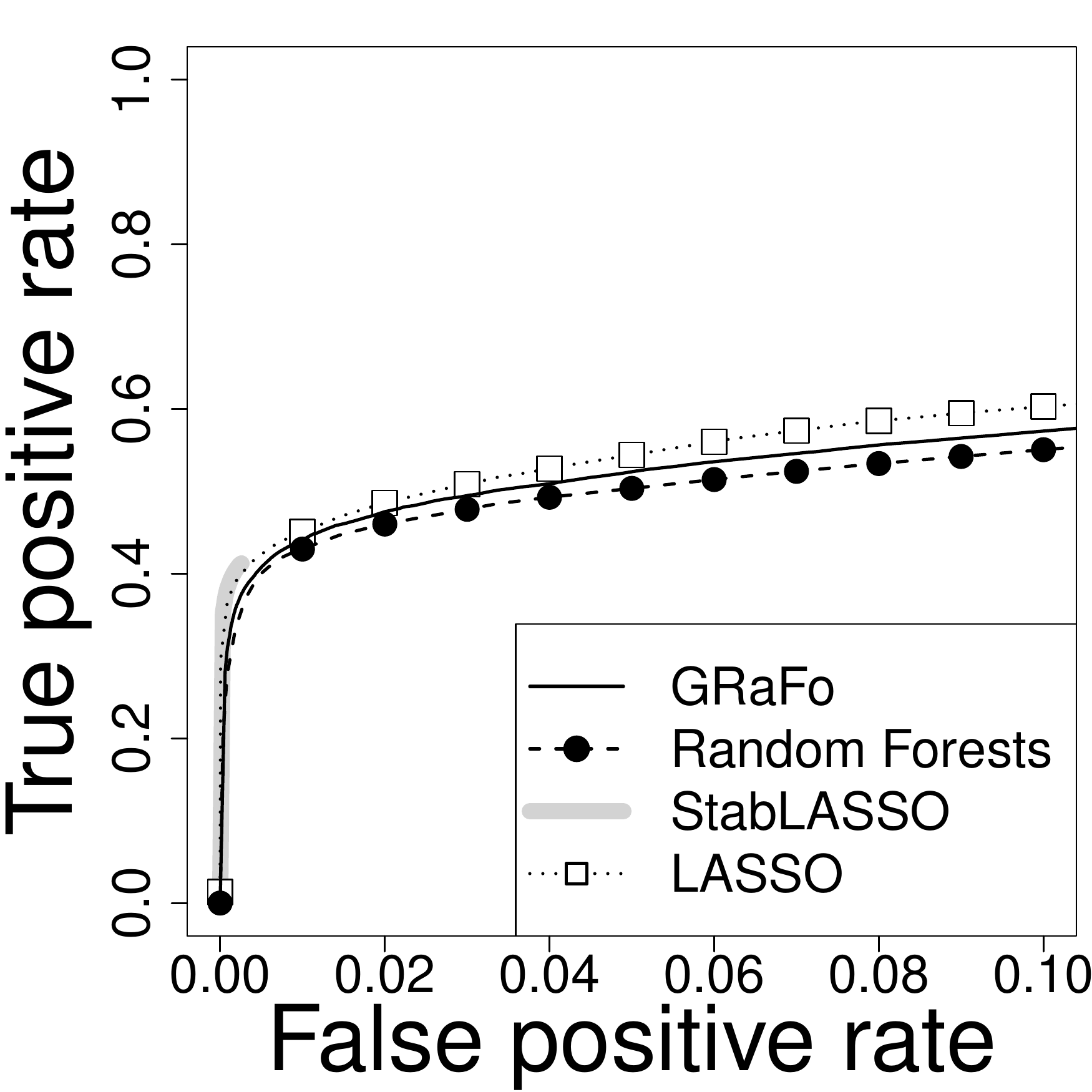}}
    \caption{Gaussian model with interactions with $p=50$, $100$, and
      $200$. Their true CIGs have $16$, $58$, and $334$ edges,
      respectively, with $1$, $6$, and $21$ first-order interaction
      terms. The first two columns report the observed number of true
      and false positives (``o'') relative to the bound in
      (\ref{StabSelErrorControl}) for the expected number
      $\mathbb{E}[V]$ of false positives (``$]$''), respectively,
    averaged over 50 simulations. The third column reports the
    averaged true and false positive rates of GRaFo and StabLASSO
    relative to the performance of their ``raw'' counterparts without
    Stability Selection.}
    \label{FigInt}
  \end{figure}

\textbf{For $p\in\{50,100,200\}$ variables and samples of size
  $n=100$, each graph in Figure~\ref{FigInt} was averaged over $50$
  repetitions. The results appear very similar to our findings for the
  Gaussian model without interactions and without nonlinear
  effects. However, here the number of true positives is somewhat
  lower for both GRaFo and StabLASSO with an (arguably) slightly
  smaller drop for the GRaFo procedure. This does not seem too
  surprising, given that Random Forests have the ability to incorporate
  interactions naturally, whereas they have to be specified explicitly
  for the LASSO (which has not been done here).}

\textbf{
However, overall the total number of interaction terms is relatively
small, ranging from roughly 5\% to 10\% of all model terms. For a
larger number of interaction terms, we would thus expect a further
gain of the GRaFo over the StabLASSO procedure.}

\subsection{Simulation Results: Gaussian with Nonlinear Effects}
\label{Sec:SimResNonlinear}
\begin{figure}
    \centering
    \textbf{Gaussian with nonlinear effects $p=50$, $100$, and $200$}\\\vspace{0.5cm}
     \subfigure[Nonlinear, $p=50$:\newline \textcolor{White}{(a) }GRaFo]{\includegraphics[width=0.3\textwidth]{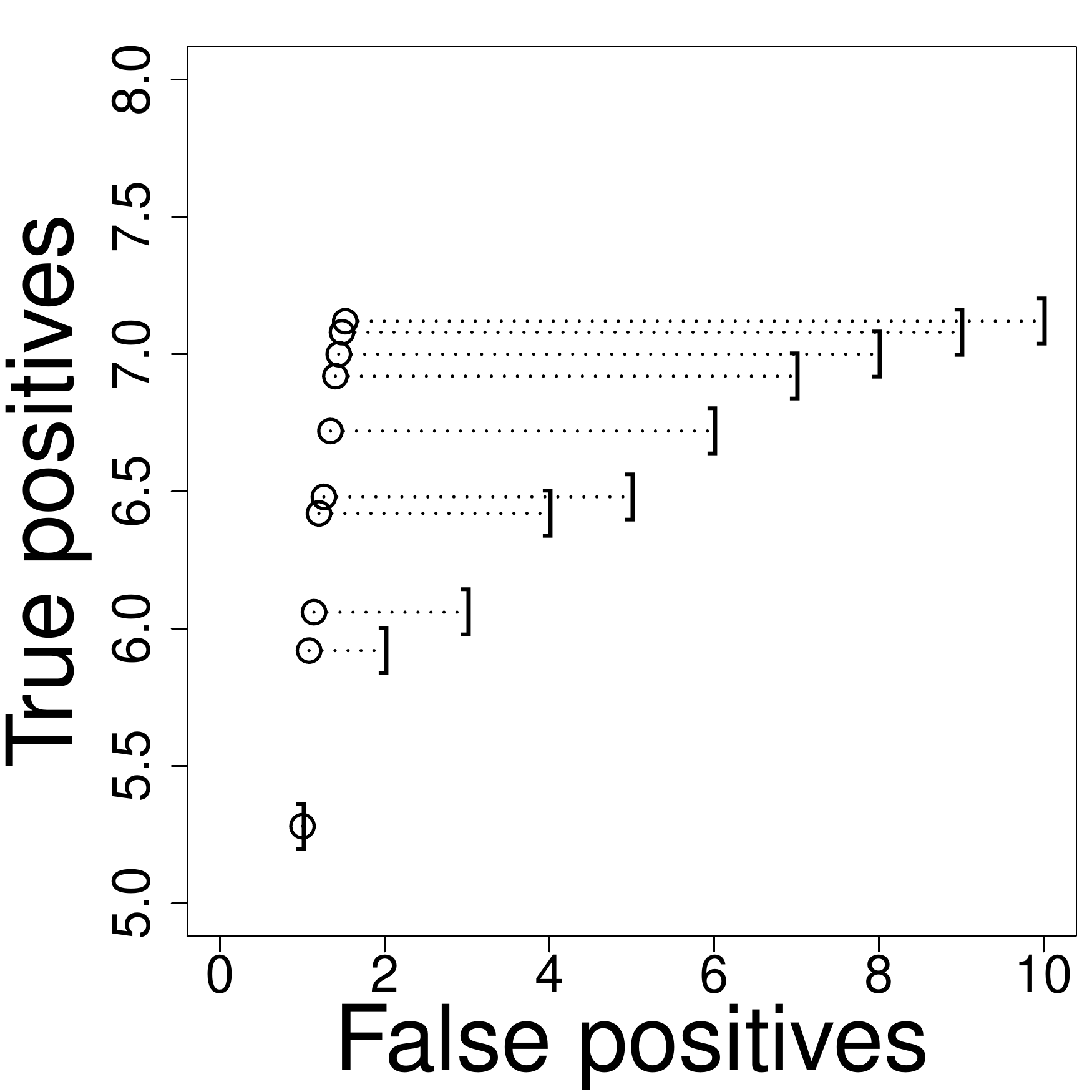}}
    \subfigure[Nonlinear, $p=50$:\newline \textcolor{White}{(b) }StabLASSO]{\includegraphics[width=0.3\textwidth]{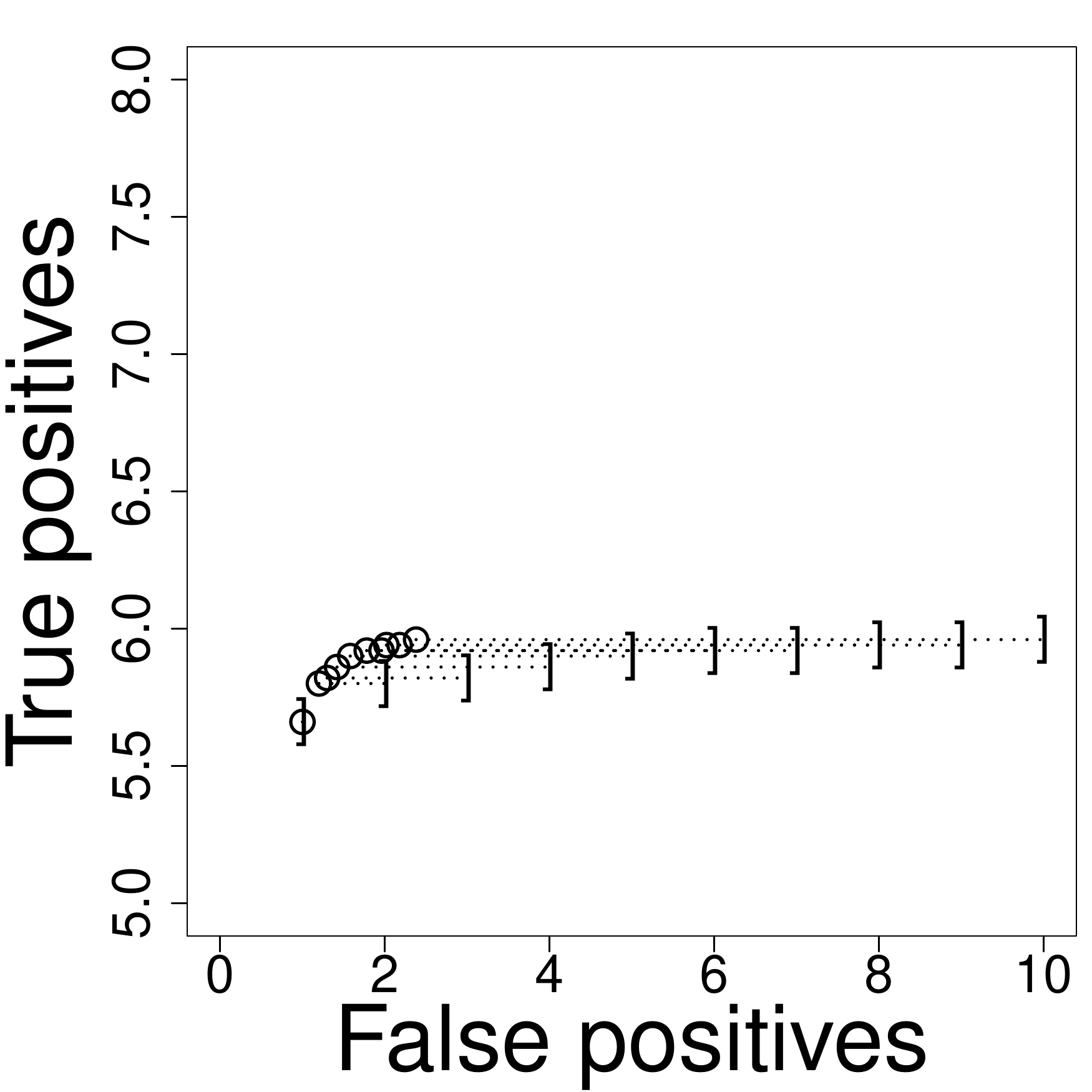}}
    \subfigure[Nonlinear, $p=50$:\newline \textcolor{White}{(c) }Rates]{\includegraphics[width=0.3\textwidth]{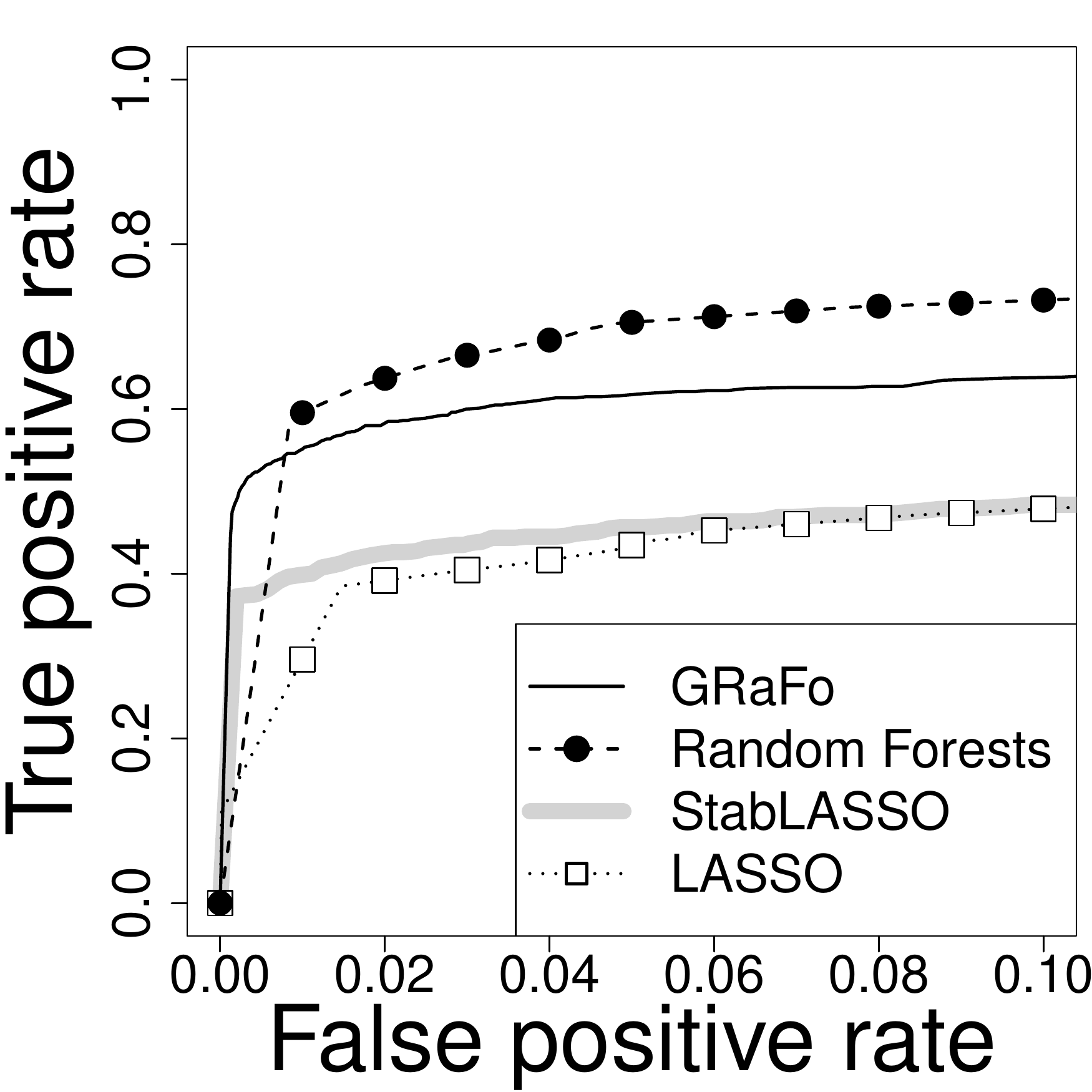}}\\
    \subfigure[Nonlinear, $p=100$:\newline \textcolor{White}{(d) }GRaFo]{\includegraphics[width=0.3\textwidth]{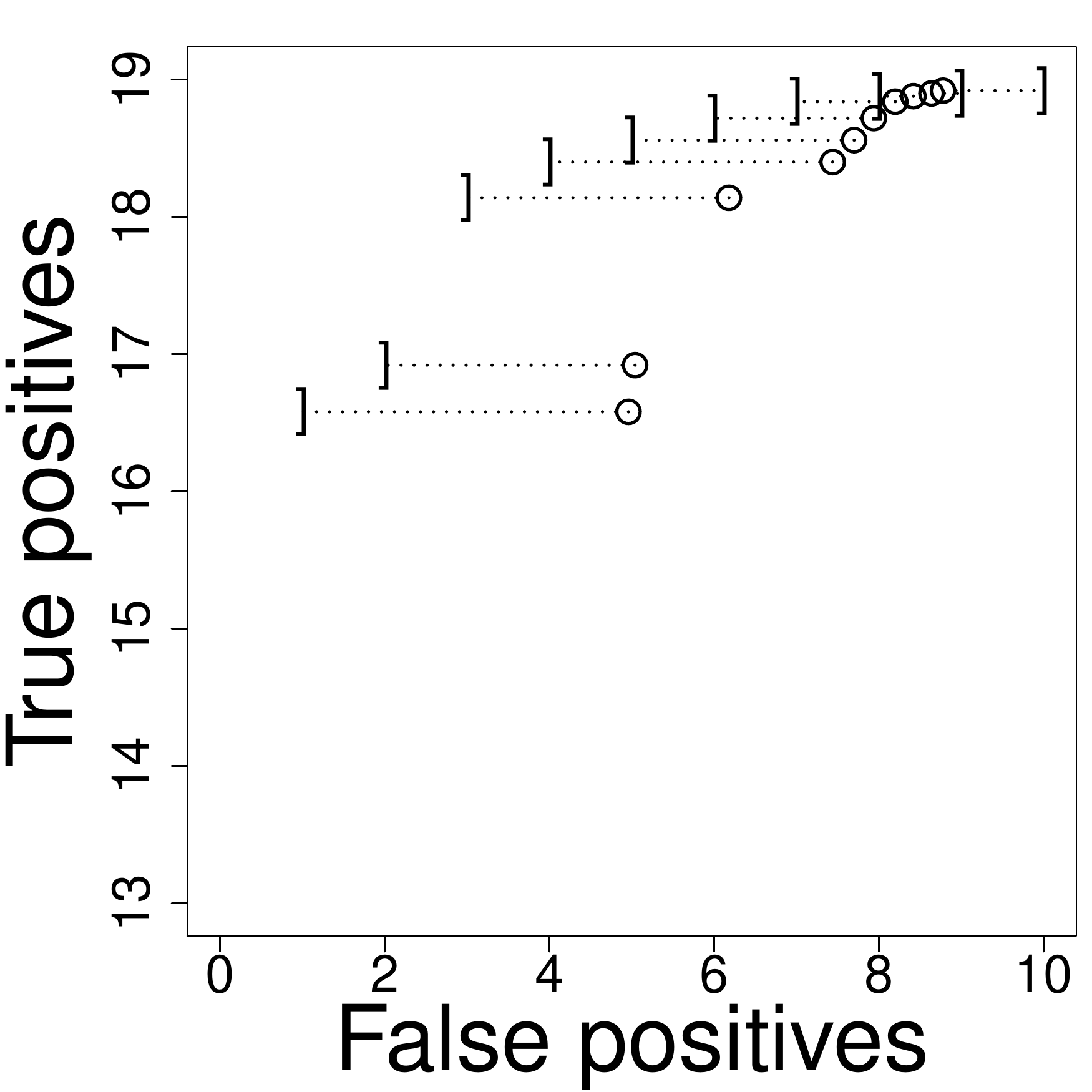}}
    \subfigure[Nonlinear, $p=100$:\newline \textcolor{White}{(e) }StabLASSO]{\includegraphics[width=0.3\textwidth]{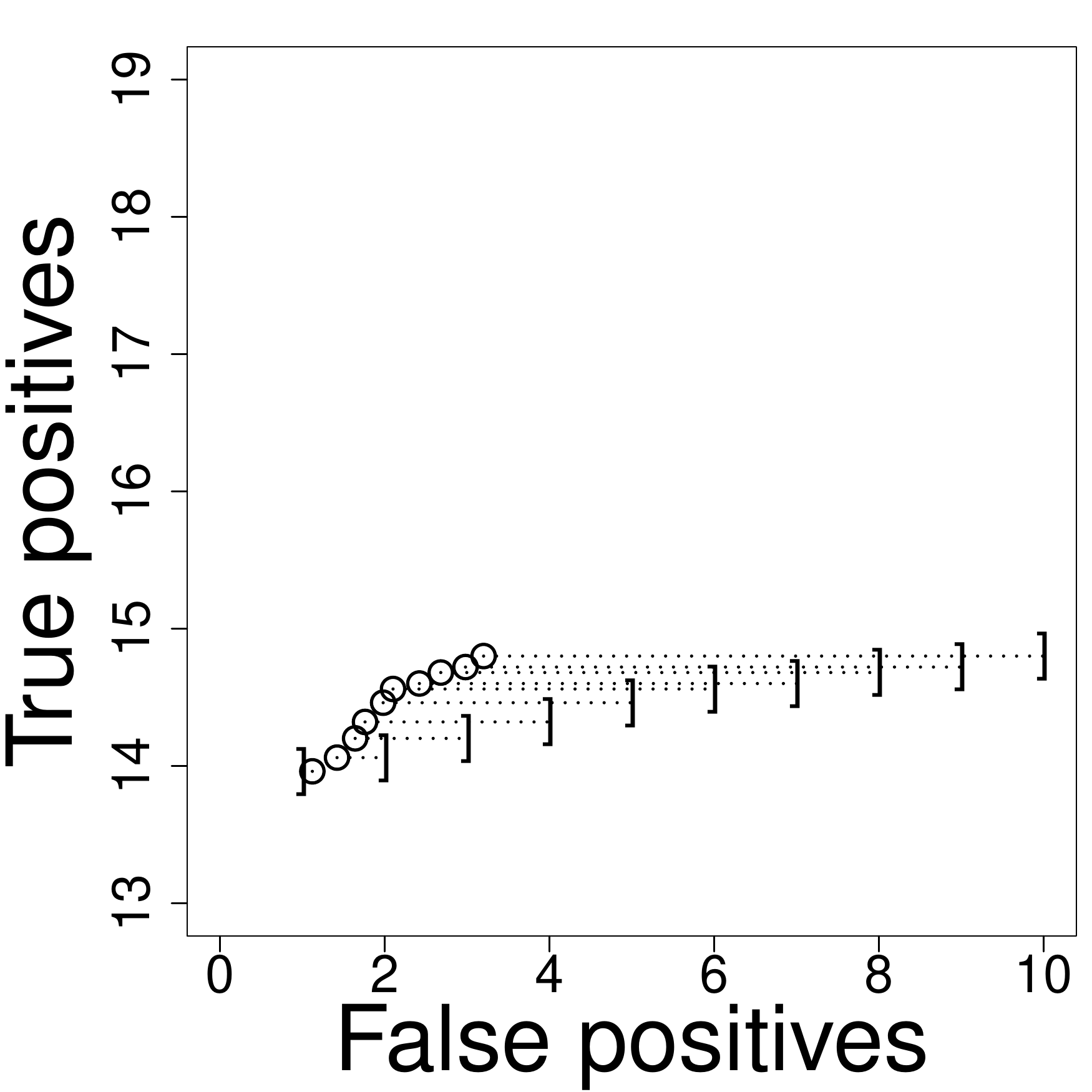}}
    \subfigure[Nonlinear, $p=100$:\newline \textcolor{White}{(f) }Rates]{\includegraphics[width=0.3\textwidth]{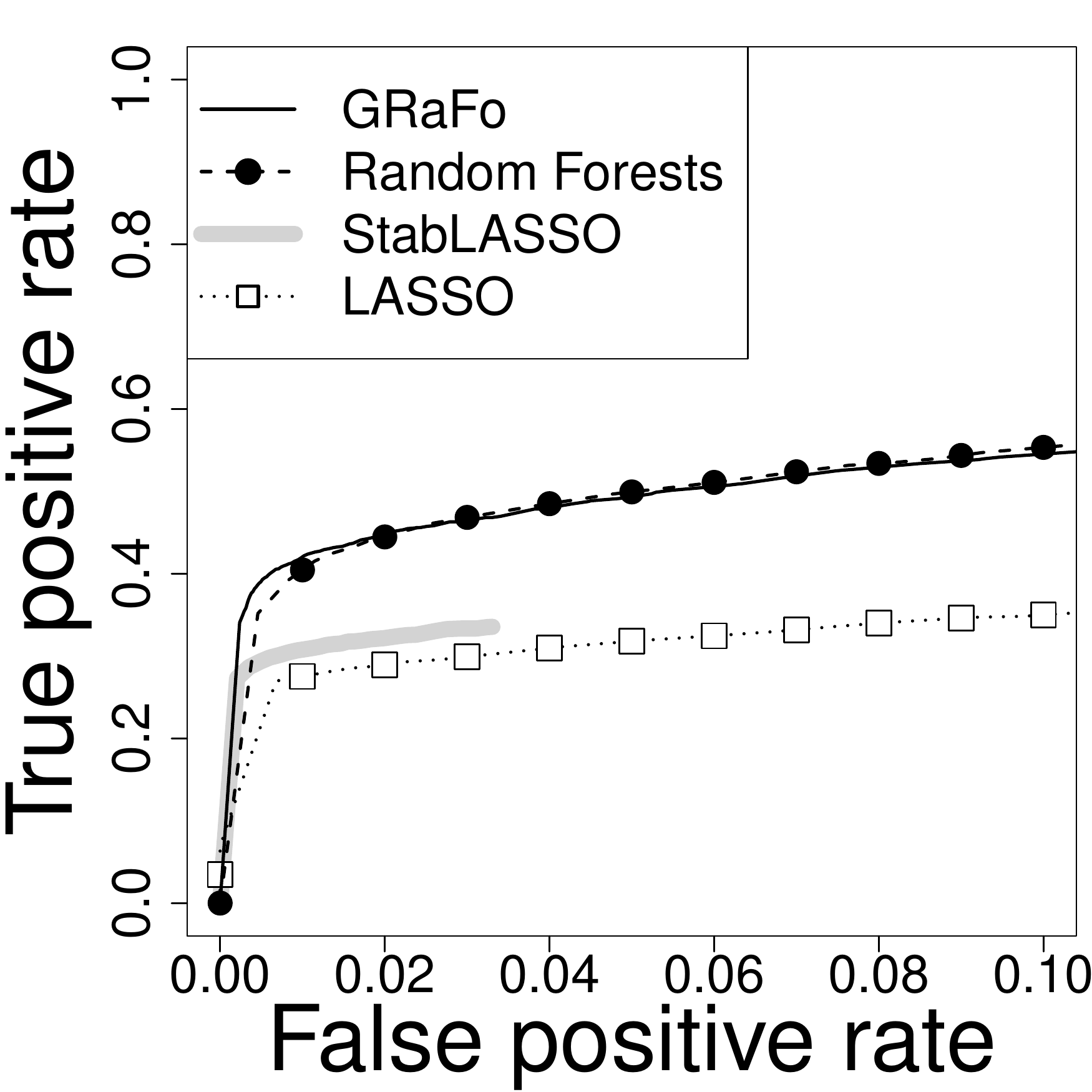}}\\
    \subfigure[Nonlinear, $p=200$:\newline \textcolor{White}{(g) }GRaFo]{\includegraphics[width=0.3\textwidth]{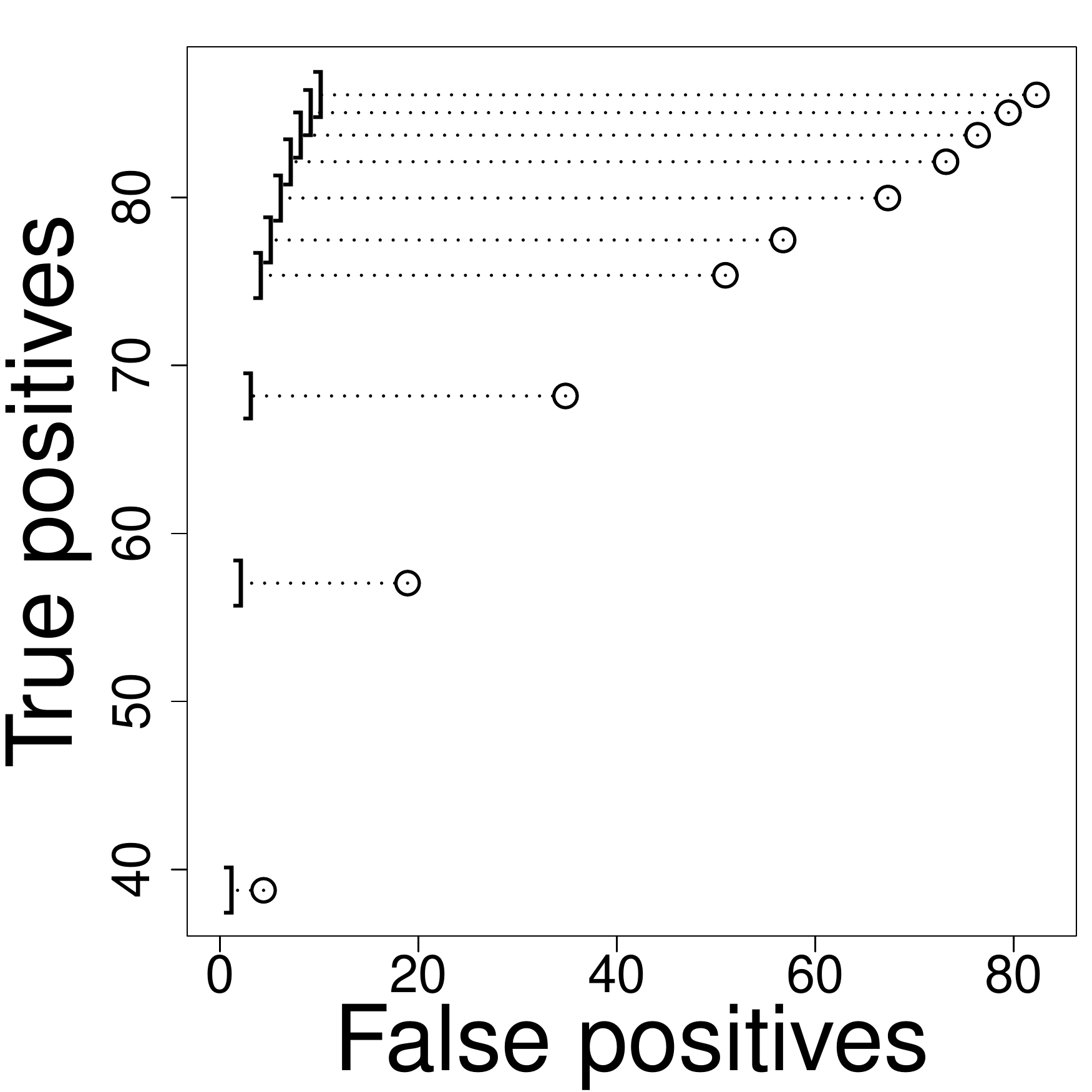}}
    \subfigure[Nonlinear, $p=200$:\newline \textcolor{White}{(h) }StabLASSO]{\includegraphics[width=0.3\textwidth]{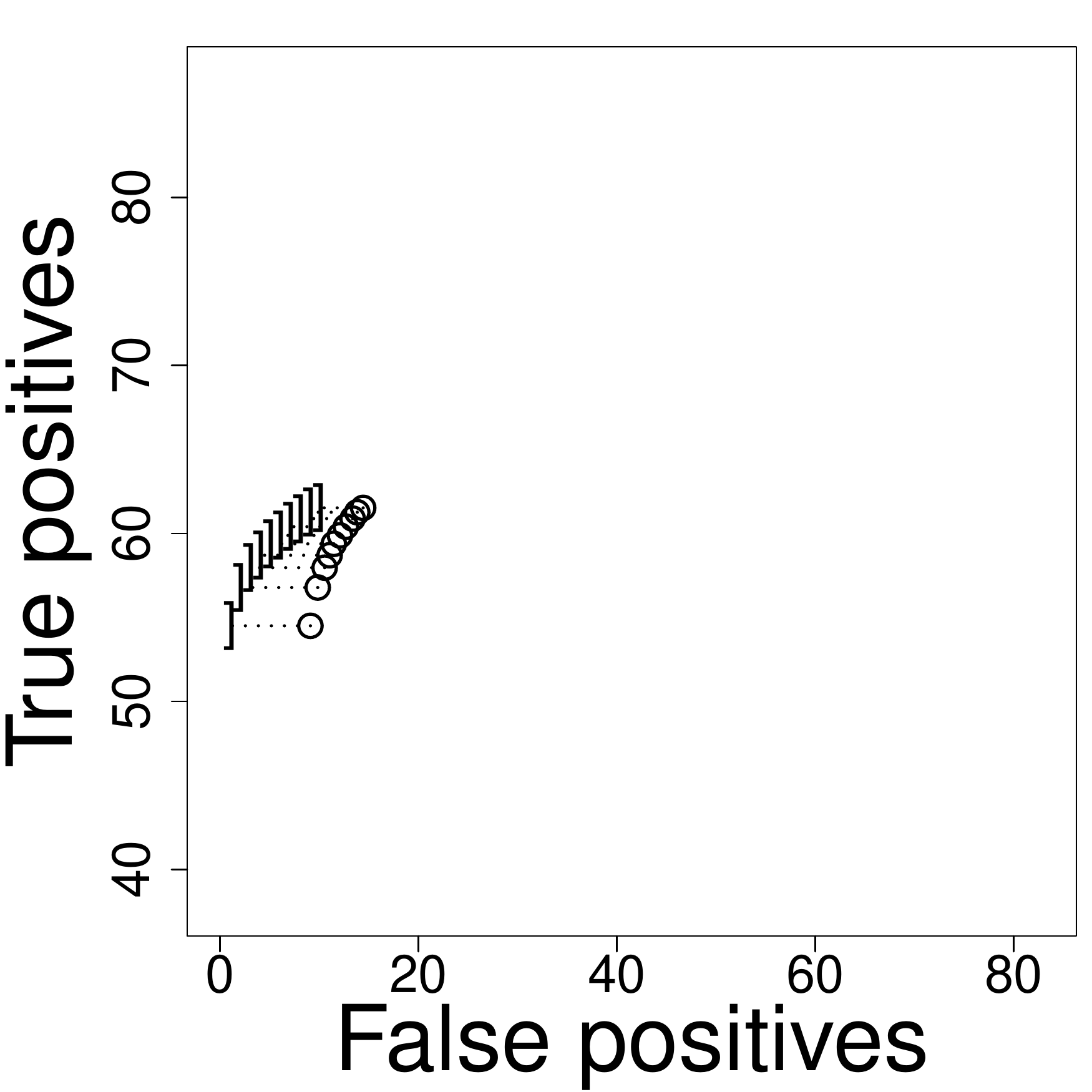}}
    \subfigure[Nonlinear, $p=200$:\newline \textcolor{White}{(i) }Rates]{\includegraphics[width=0.3\textwidth]{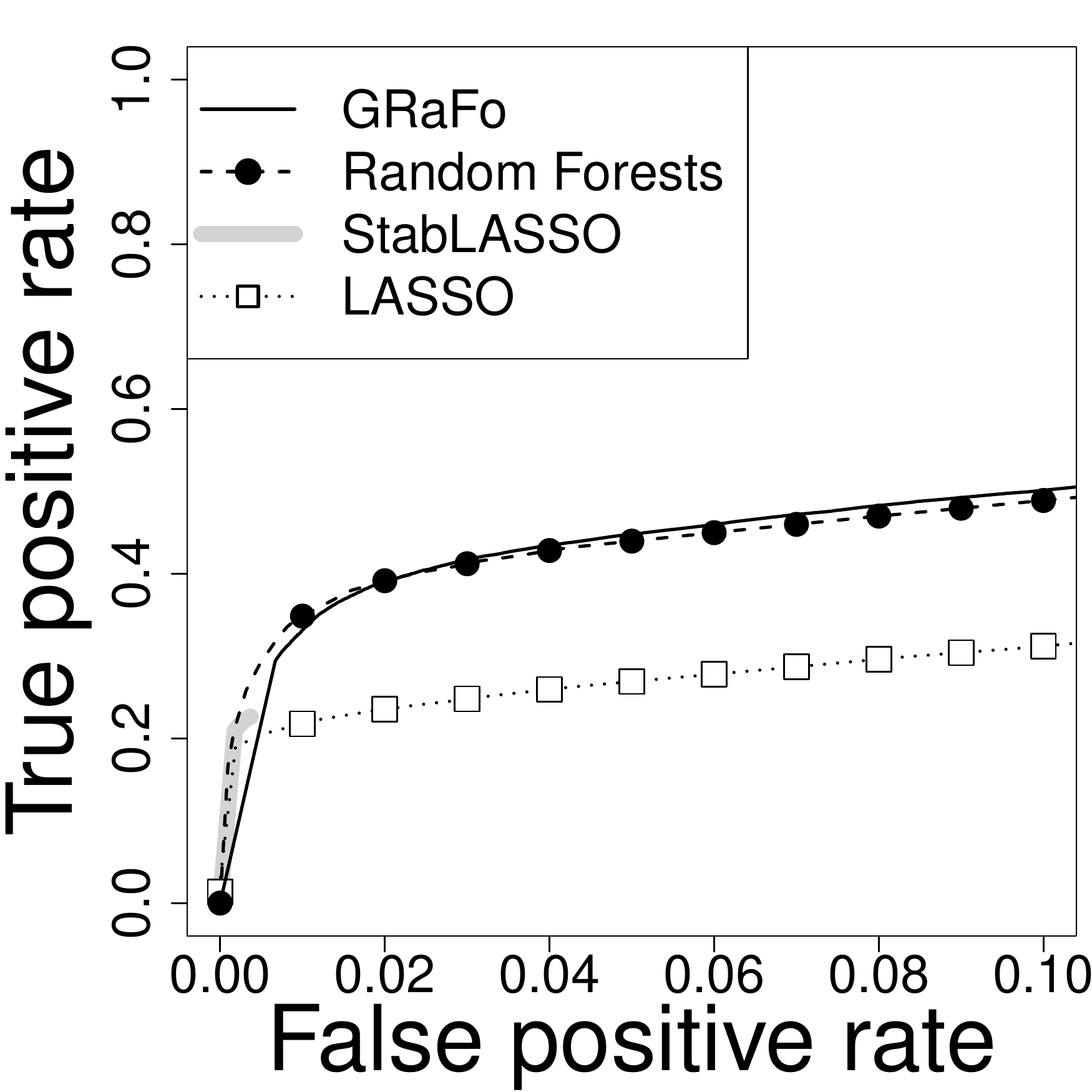}}
    \caption{The rows correspond to the Gaussian with nonlinear
      associations with $p=50$, $100$, and $200$. Their true CIGs have
      $16$, $58$, and $334$ edges, respectively. The first two columns
      report the observed number of true and false positives (``o'')
      relative to the bound in (\ref{StabSelErrorControl}) for the
      expected number $\mathbb{E}[V]$ of false positives (``$]$'') for
    GRaFo and StabLASSO, respectively, averaged over 50 simulations.
    The third column reports the averaged true and false positive
    rates of GRaFo and StabLASSO relative to the performance of their
    ``raw'' counterparts without Stability Selection.}
    \label{FigNonlinear}
  \end{figure}

\textbf{For $p\in\{50,100,200\}$ variables and samples of size
  $n=100$, each graph in Figure~\ref{FigNonlinear} was averaged over
  $50$ repetitions. Here, GRaFo clearly outperforms StabLASSO in terms
  of true positives for all considered $p$. However, for GRaFo the
  number of false positives is not controlled by a small bound on
  $\mathbb{E}[V]$ anymore for $p>50$, which is especially apparent in
  the case where $p=200$. For StabLASSO there seems to be a similar
  behavior, but only for $p=200$ the number of false positives clearly
  violates $\mathbb{E}[V]$. The ``raw'' Random Forests and LASSO
  estimates show very similar results to their Stability Selection
  counterparts. Note that the signal has been amplified by a factor of
  5 to achieve comparable performance of the estimation procedures to
  the linear Gaussian setting.}

\subsection{Simulation Results: Mixed-Setting with ML and StabCForests}
\label{Sec:SimResMixedGlmCforest}

\begin{figure}
    \centering
    \textbf{Mixed setting with ML and StabcForests with $p=50$, $100$, and $200$}\\\vspace{0.5cm}
    \subfigure[Mixed, $p=50$, $n=500$:\newline \textcolor{White}{(a) }GRaFo]{\includegraphics[width=0.3\textwidth]{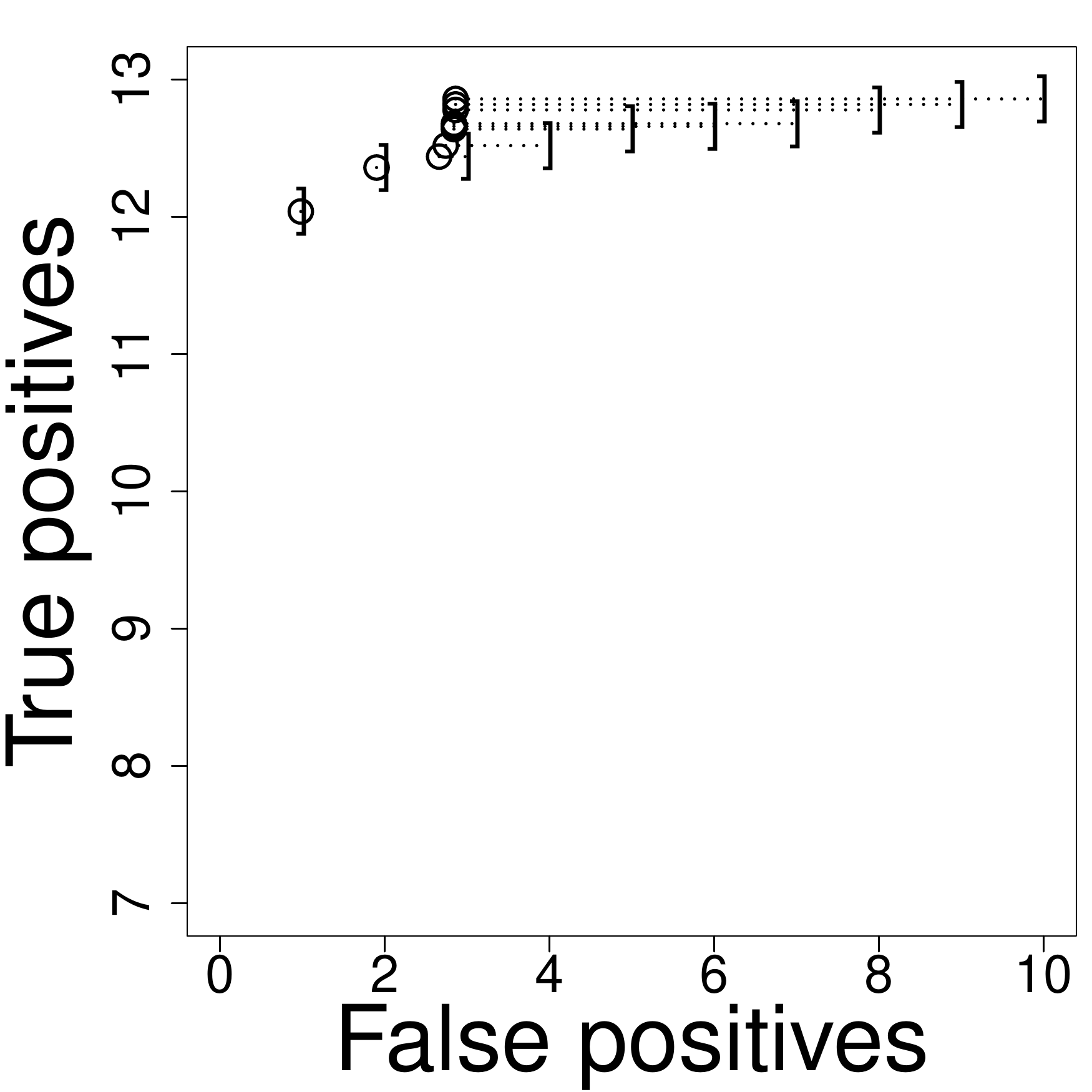}}
    \subfigure[Mixed, $p=50$, $n=500$:\newline \textcolor{White}{(b) }StabLASSO]{\includegraphics[width=0.3\textwidth]{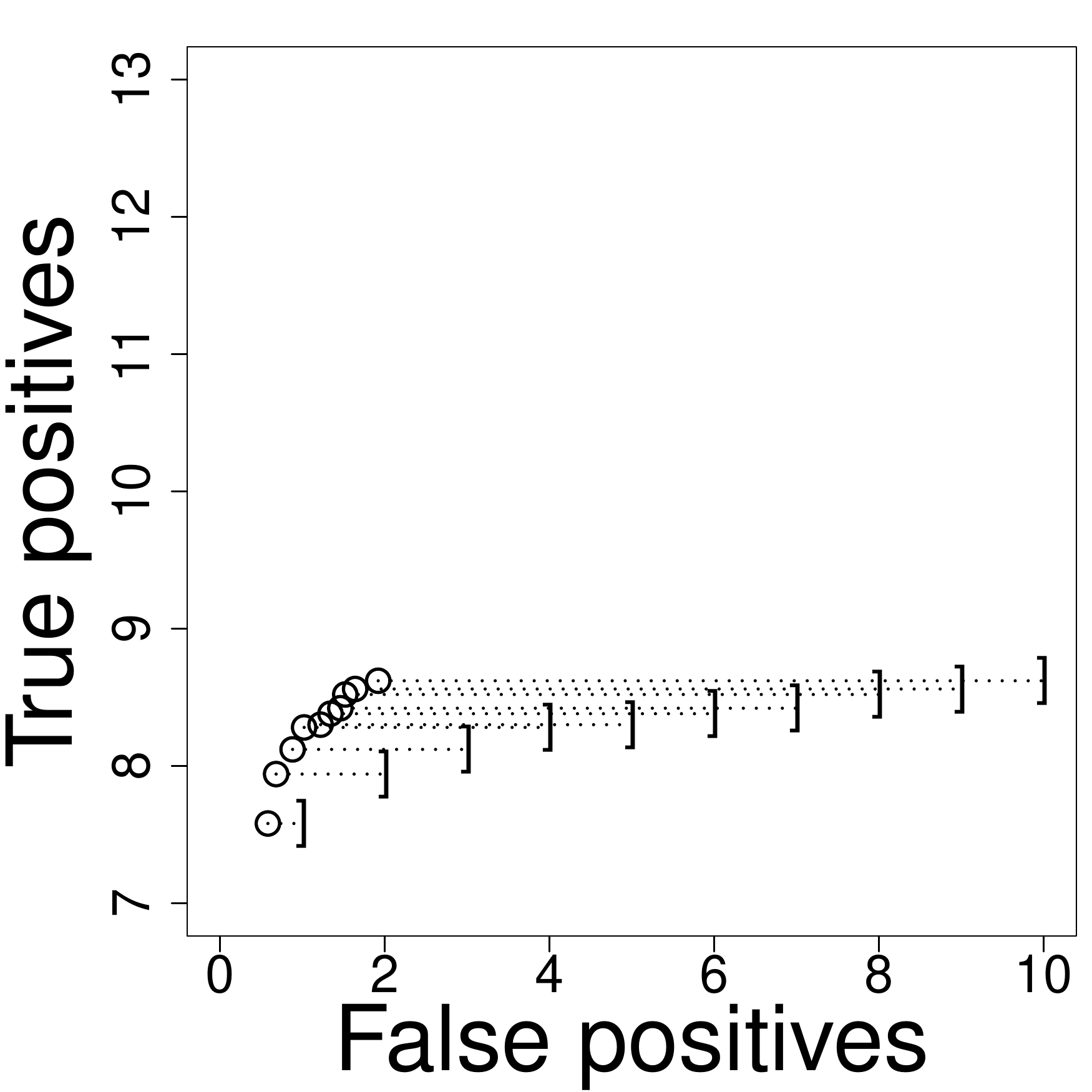}}
    \subfigure[Mixed, $p=50$, $n=500$:\newline \textcolor{White}{(c) }Rates]{\includegraphics[width=0.3\textwidth]{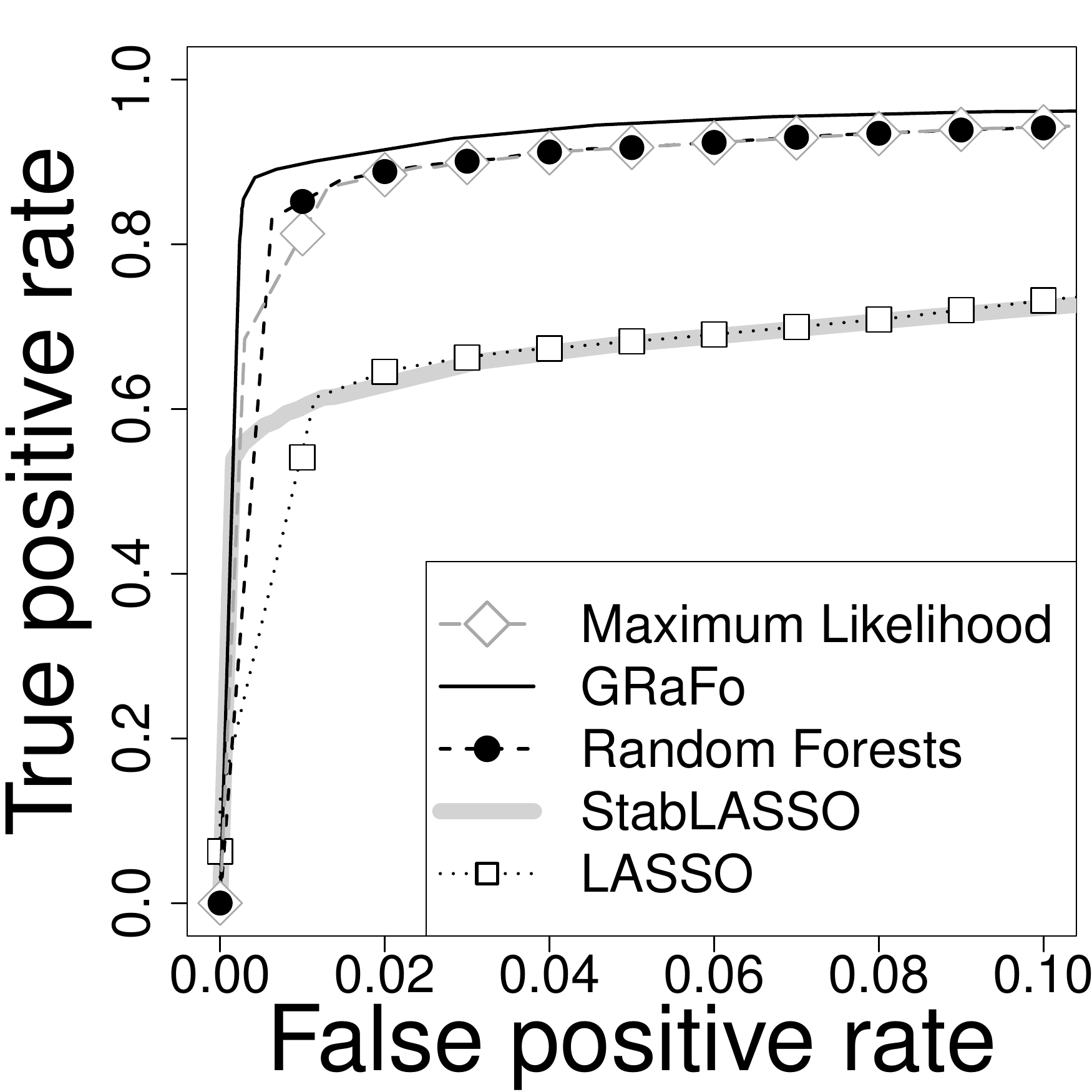}}\\
    \subfigure[Mixed, $p=50$, $n=100$:\newline \textcolor{White}{(d) }GRaFo]{\includegraphics[width=0.3\textwidth]{setting-Mi-gmwrf-p-50.pdf}}
    \subfigure[Mixed, $p=50$, $n=100$:\newline \textcolor{White}{(e) }StabcForests]{\includegraphics[width=0.3\textwidth]{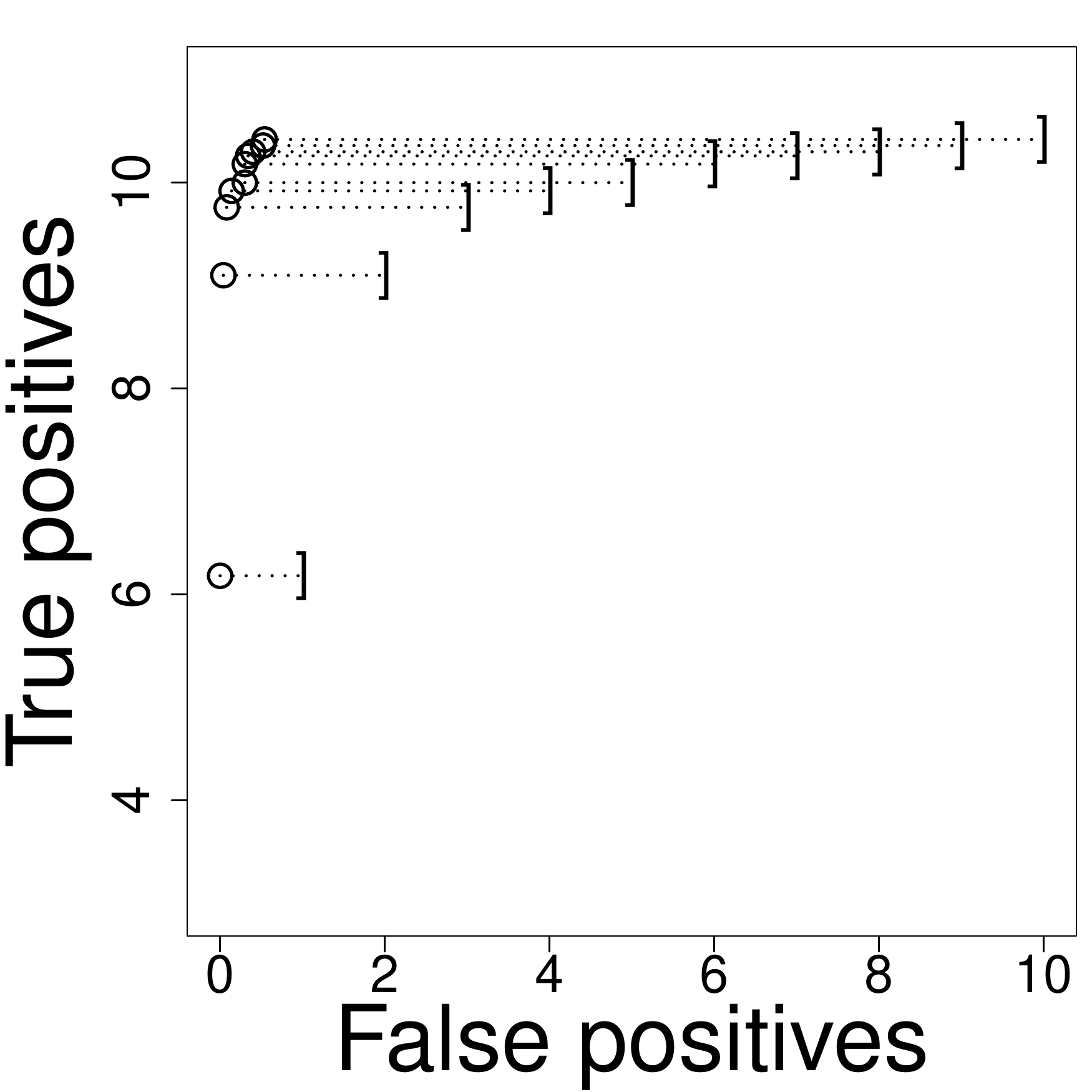}}
    \subfigure[Mixed, $p=50$, $n=100$:\newline \textcolor{White}{(f) }Rates]{\includegraphics[width=0.3\textwidth]{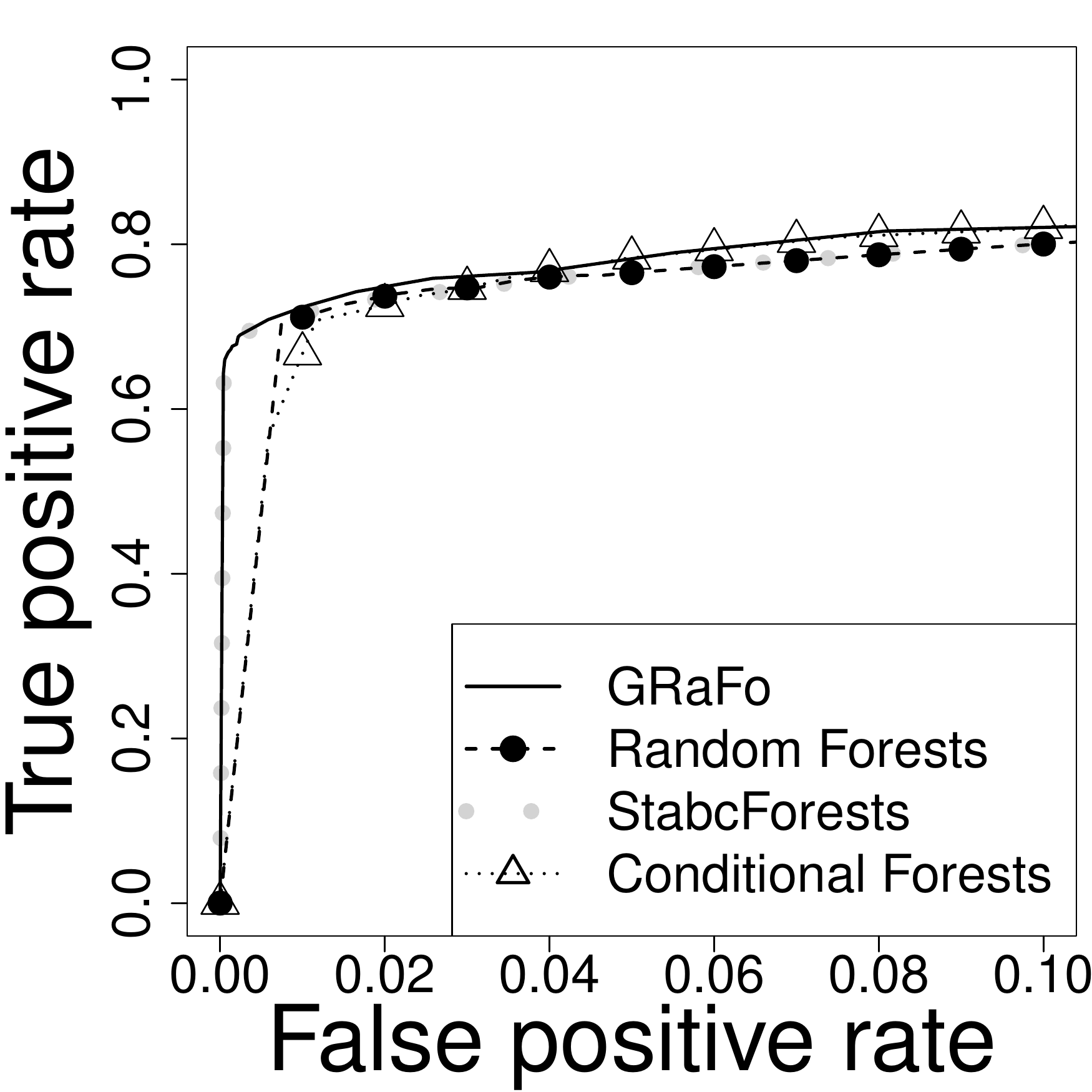}}\\
    \subfigure[Mix., $p=100$, $n=100$:\newline \textcolor{White}{(g) }GRaFo]{\includegraphics[width=0.3\textwidth]{setting-Mi-gmwrf-p-100.pdf}}
    \subfigure[Mix., $p=100$, $n=100$:\newline \textcolor{White}{(h) }StabcForests]{\includegraphics[width=0.3\textwidth]{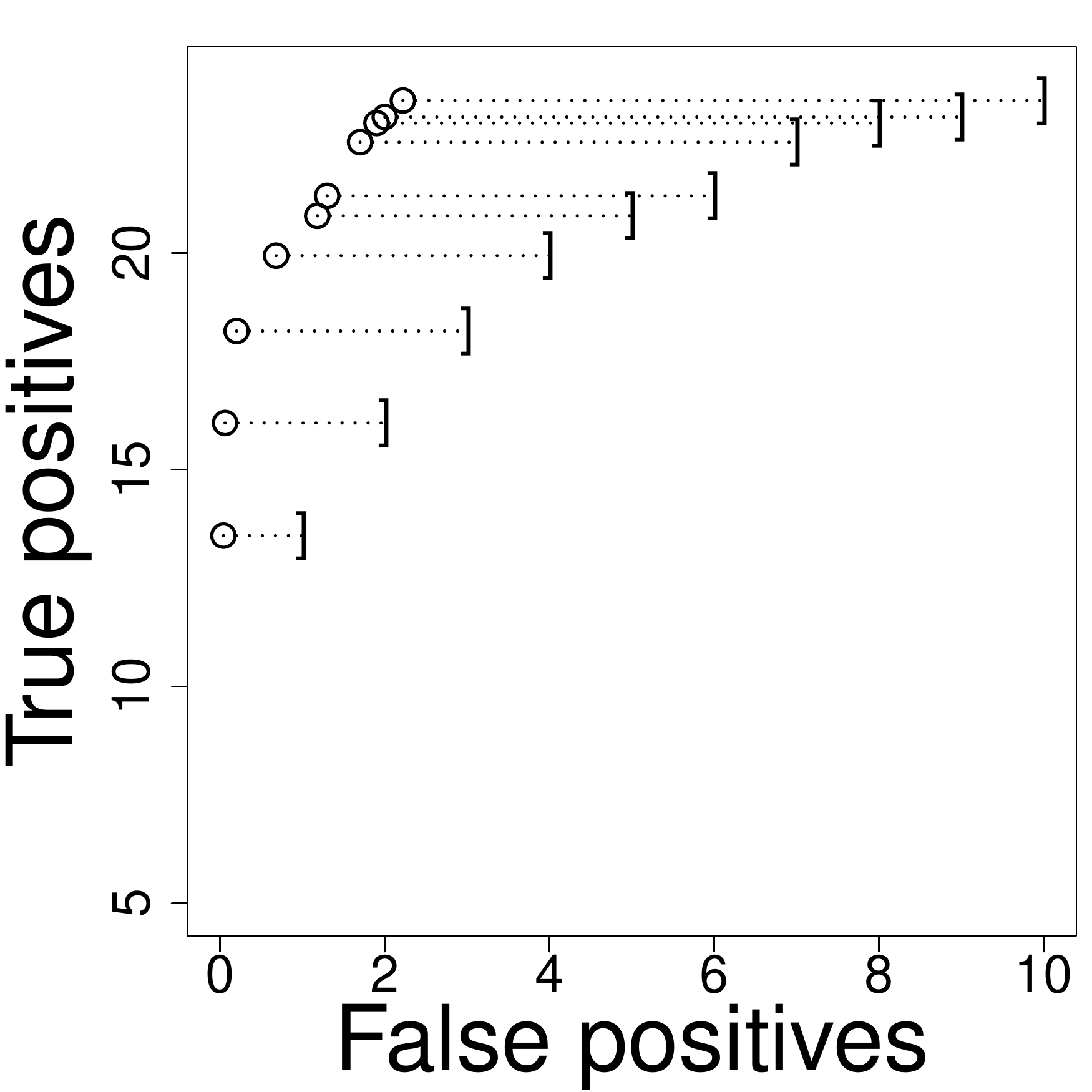}}
    \subfigure[Mix., $p=100$, $n=100$:\newline \textcolor{White}{(i) }Rates]{\includegraphics[width=0.3\textwidth]{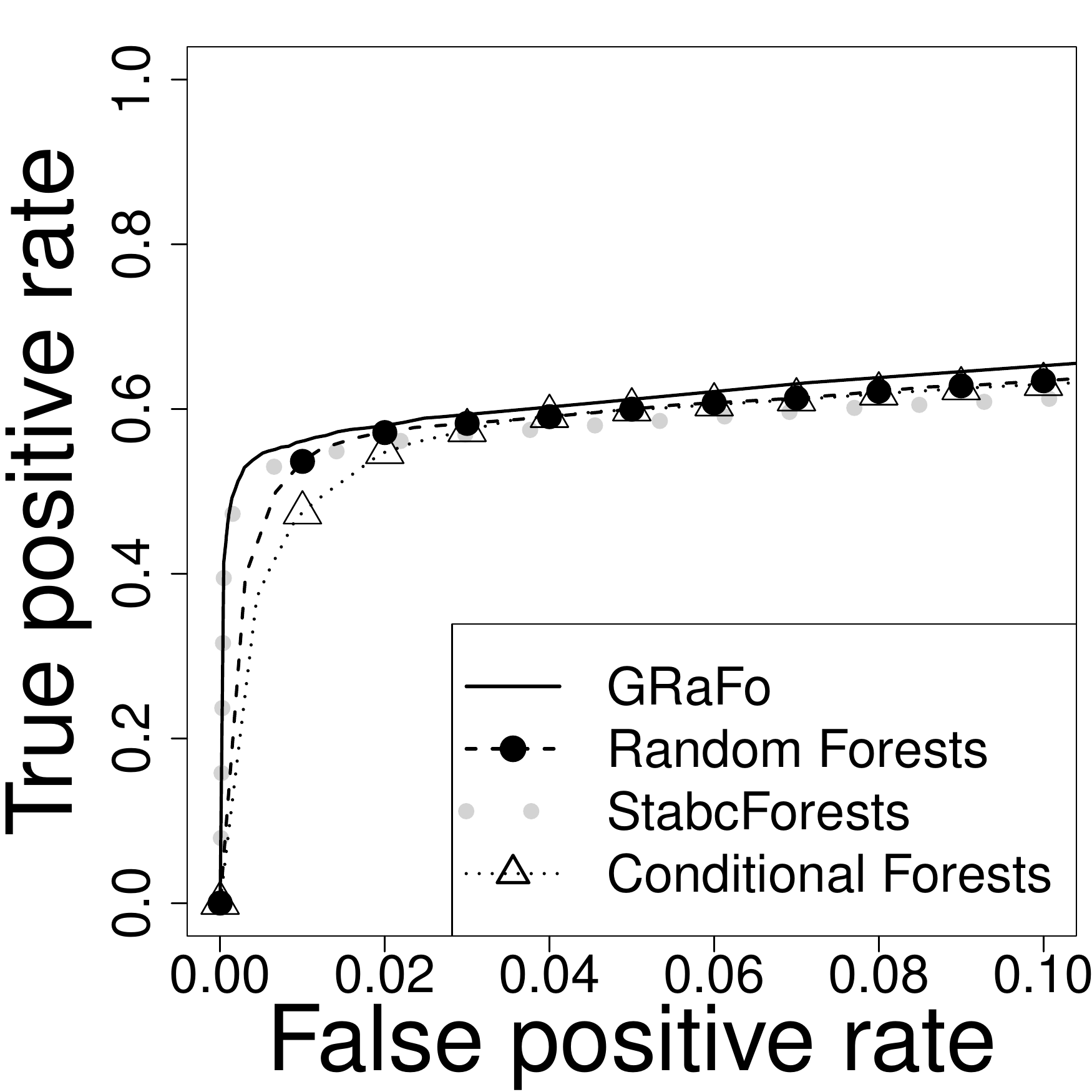}}
    \caption{The rows correspond to applications of ML
      and StabcForests to data from the mixed model with varying $p$
      and $n$. For $p=50$ ($p=100$), the true CIG has $16$ ($58$)
      edges. The first two columns report the observed number of true
      and false positives (``o'') relative to the bound in
      (\ref{StabSelErrorControl}) for the expected number
      $\mathbb{E}[V]$ of false positives (``$]$'') for GRaFo and
    StabLASSO or StabcForests, respectively, averaged over 50
    simulations.  The third column reports the averaged true and false
    positive rates.}
    \label{FigMixedLMCFor}
  \end{figure}

\textbf{ The first row of Figure~\ref{FigMixedLMCFor} reports for
  $p=50$ and $n=500$ the results of ML estimation, GRaFo, and
  StabLASSO, averaged over 50 runs. Not surprising, both GRaFo and
  StabLASSO perform better than in the setting where $n=100$, though
  StabLASSO remains at a clear disadvantage due to the unfavorable
  dichotomization. On the other hand, the performance of GRaFo (and
  also its ``raw'' Random Forests counterpart) is on par with the ML
  estimation. Stability Selection was not applied to ML estimation
  due to the immense computational burden and thus no bounds on
  $\mathbb{E}[V]$ could be specified. However, for
  both GRaFo and StabLASSO we find that the number of false positives
  are typically well below the specified bounds.}

\textbf{ The second and third row of Figure~\ref{FigMixedLMCFor}
  report the performance of StabcForests and GRaFo for $p=50$ and
  $p=100$ with $n=100$, averaged over 50 runs. The GRaFo results from
  above are reproduced for better readability. We find that both GRaFo
  and StabcForests show very similar results. In the first two columns
  we see that GRaFo seems to perform somewhat better for very small
  bounds on $\mathbb{E}[V]$. The performance of the two ``raw''
  methods is very similar to their stable counterparts.  }

\textbf{ The computational burden of StabcForests is much larger than
  for GRaFo and amounts to roughly 2 hours for $p=50$ and roughly 6
  hours for $p=100$. Also note that the reported results within the
  Conditional Forests framework use the marginal permutation
  importance due to the very heavy computational burden of the
  conditional variable importance.}

\section{Functional Health in the Swiss General Population}

\subsection{The Importance of Functional Health}
According to the World Health Organization's (WHO) new framework of the
International Classification of Functioning, Disability and Health
\citep[ICF; cf.,][]{ICF2001} the lived experience of health
\citep{Stucki2008} can be structured in experiences related to body
functions and structures as well as to activity and participation in
society.  All of these are, in turn, influenced by a variety of so-called
personal factors such as gender, income, or age and environmental factors
including individual social relations and supports as well as properties of
larger macro social systems such as the economy (see Figure
\ref{ICF-Model}). Also, the WHO and The World Bank recommend in their
recent World Report on Disability (2011) that functional health state
descriptors are analyzed in conjunction with other health outcomes and,
particularly, that more research is conducted on ``[...] the interactions
among environmental factors, health conditions, and disability [...]''
\citep[p.~267][]{WorldRepDisab2011}. Under these prerequisites it is of
interest which variables are conditionally dependent on each other. For
instance, ``Does the income distribution affect participation, conditional
on known impairments, environmental, and personal factors?''.

\ifpdf
  \begin{figure}
    \centering
    \includegraphics[width=0.85\textwidth]{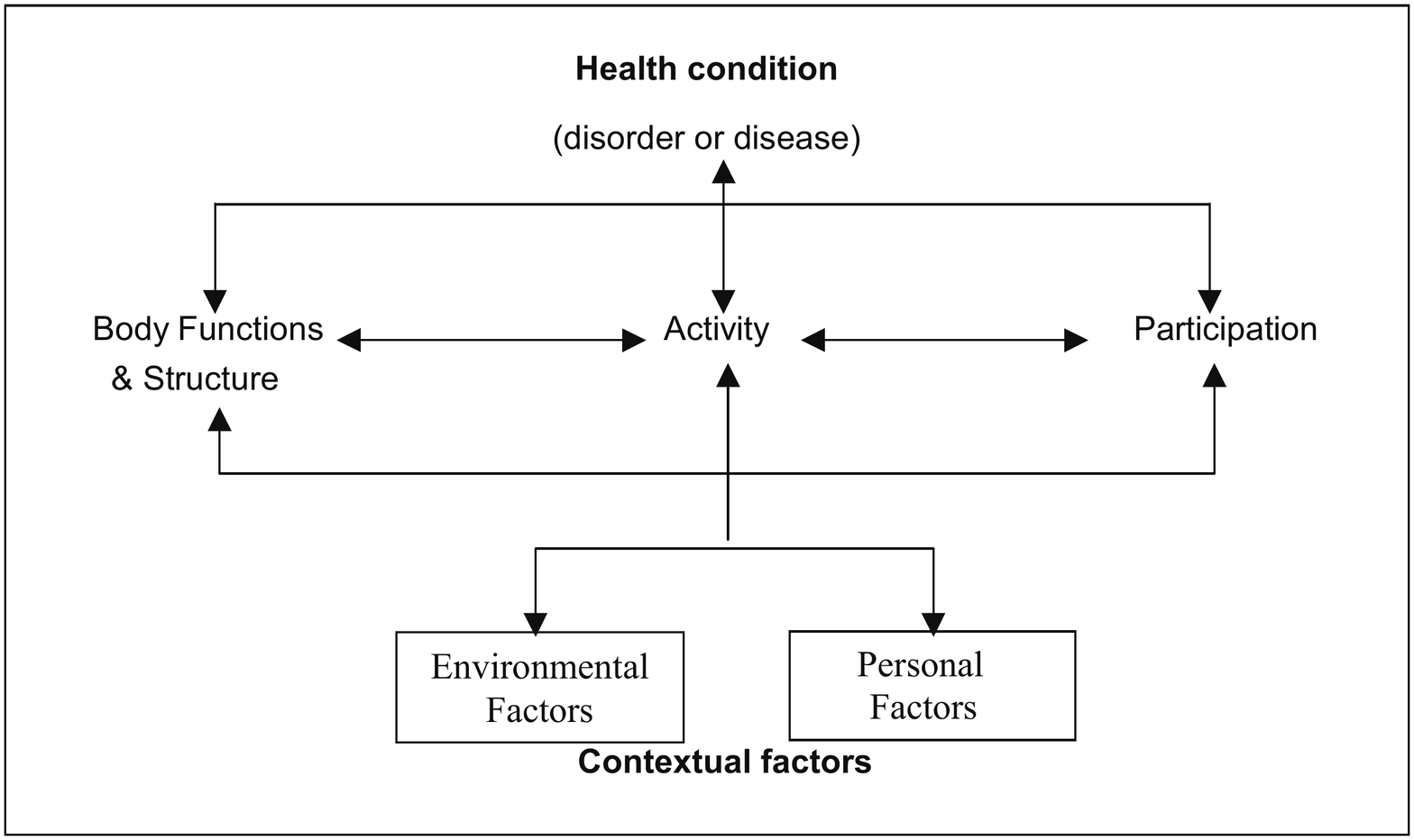}
    \caption{The International Classification of Functioning, Disability
      and Health (ICF) model relates aspects of human functioning and
      provides a common language for practitioners.}
    \label{ICF-Model}
  \end{figure}
\fi

\subsection{Study Population}
We use GRaFo for a secondary analysis of cross-sectional observational data
on functional health from the Swiss Health Survey (SHS) in 2007. Data were
obtained from the Federal Statistics Office of Switzerland.  The original
study was based on a stratified random sample of all private Swiss
households with fixed line telephones. Within each household one household
member aged 15 or older was randomly selected. The survey was completed by
a total of 18760 persons, corresponding to a participation rate of 66
percent \citep{Graf2010}. The mean age of study participants was 49.6 years
$(\pm 18.5)$. The data were mostly collected with computer assisted
telephone interviews. Further information is available elsewhere
\citep{SHS-Web}.

\subsection{Variables}
The SHS included various information on symptoms (in particular pain),
impairments, and activity limitations. Since the respective items were
sometimes nominal, sometimes ordinal, and sometimes (e.g.~body mass index)
metric, we dichotomized each item so that 1 was indicative of having any
kind of problem. As overall summary scores on functioning and disability
were not recommendable \citep{Reinhardt2010b}, we followed the framework of
the WHO's biopsychosocial model of health, outlined in the ICF \citep[][see
Figure~\ref{ICF-Model}]{ICF2001}, and other theoretical considerations
\citep{WorldRepDisab2011,Reinhardt2010b} in constructing sum indices (see
Table~\ref{tab:SumIndices}). The plausibility of all indices was checked
using the Stata 11 confirmatory factor analysis module confa
\citep{Kolenikov2009}. In each case the index construction was tested and
the null hypothesis of a diagonal structure of the covariance matrix
rejected.

We created a dummy variable for labor market participation
restrictions such that 1 identified persons who gave up work, reduced
the number of working hours, or changed jobs because of health
reasons. We also created a dummy variable for participation in leisure
physical activity (LPA) differentiating between people participating
in leisure activities leading to sweating at least once a week and
those who do not.  General health perception was measured with the
following question and answer options: ``How would you rate your
health in general? Very good, good, fair, poor, or very poor?''. We
further included indicators of socio-economic status (SES) in our
analysis: equivalence household income, years of formal education,
employment status, and migration background (foreign origin of at
least one parent).  On the macro- or cantonal-level we obtained
information on the Swiss counties' (cantons) gross domestic products
(GDP), Gini coefficients, and crime rates for 2006. Moreover, we
considered information on gender, age, marital status (being married),
alcohol consumption (in grams per day), and current smoking (yes/no).

Of these, in total, 20 mixed-type variables (see Table \ref{tab:AllVars}),
income had the highest number of missing values with roughly 6
percent. Overall, less than 0.85 percent of replies were missing
corresponding to 2687 cases with one or more missing values. To assess
their effect, we estimated the CIG once with casewise deletion and once
with imputation of missing values with the missForest procedure
\citep{Stekhoven2011} available in R. An alternative would be to use
  surrogate splits, which may be particularly feasible if the speed of the
  imputation method is of importance \citep{Hapfelmeier2012}.

\begin{table}
\begin{tiny}
\begin{center}
\begin{tabular}[H]{ll}
  \toprule
  Construct & Variable specification\\
  \midrule
Impairment       & Problems with$\ldots$ \\
                   & $\ldots$vision\\
		   & $\ldots$hearing\\
		   & $\ldots$speaking\\
		   & $\ldots$body mass index (i.e.~over 30 or under 16)\\
		   & $\ldots$urinary incontinence\\
		   & $\ldots$defecation\\
		   & $\ldots$feeling weak, tired, or a lack of energy\\
		   & $\ldots$sleeping\\
		   & $\ldots$tachycardia\\
                   & Range of sum index: 0-9\\

Pain     	 & Pain in$\ldots$\\
                 & $\ldots$head\\
		 & $\ldots$chest\\
		 & $\ldots$stomach\\
		 & $\ldots$back\\
		 & $\ldots$hands\\
		 & $\ldots$joints\\
                 & Range of sum index: 0-6\\

Activity \&       & Problems with independently$\ldots$\\
participation			     & $\ldots$walking\\
limitation				     & $\ldots$eating\\
				     & $\ldots$getting up from bed or chair\\
					     & $\ldots$dressing\\
					     & $\ldots$using the toilet\\
					     & $\ldots$taking a shower or bath\\
					     & $\ldots$preparing meals\\
					     & $\ldots$using a telephone\\
					     & $\ldots$doing the laundry\\
					     & $\ldots$caring for finances/accounting\\
					     & $\ldots$using public transport\\
					     & $\ldots$doing major household tasks\\
					     & $\ldots$doing shopping\\
			                     & Range of sum index: 0-13\\

Social support	& Having$\ldots$\\
                & $\ldots$no feelings of loneliness\\
		& $\ldots$no desire to turn to someone\\
		& $\ldots$at least one supportive family member\\
		& $\ldots$someone to turn to\\
                & Range of sum index: 0-4\\
Social network utilization & At least weekly$\ldots$\\
                           & $\ldots$visits from family\\
			   & $\ldots$phone calls with family\\
			   & $\ldots$visits from friends\\
			   & $\ldots$phone calls with friends\\
			   & $\ldots$participation in clubs/associations/parties\\
                           & Range of sum index: 0-5\\
  \bottomrule
\end{tabular}
\caption{Construction rules of sum indices for functioning (pain,
  impairment, activity and participation limitation) and social integration
  (social support and social network utilization) from 37 dichotomous
  (yes=1/no=0) variables.}
\label{tab:SumIndices}
\end{center}
\end{tiny}
\end{table}

\begin{table}
\begin{center}
\begin{tabular}[H]{lll}
  \toprule
  Type & Variable & \% Missing\\
  \midrule
$>2$ categories & Impairment index & 5.92\\
& Pain index & 0.37\\
& Activity limitation index & 0.69\\
& Social support index & 5.84\\
& Social network utilization index & 2.32\\
& General health perception & 0.05\\

Dichotomous & Male & 0.00\\
& Married & 0.09\\
& Paid work & 0.03\\
& Migration background & 4.73\\
& Smoker & 0.07\\
& Work restriction & 0.00\\
& Leisure physical activity & 0.00\\

Continuous & Age & 0.00\\
& Years of formal education & 0.07\\
& Income & 5.94\\
& Alcohol consumption (in grams per day) & 2.59\\
& Gross domestic product & 0.00\\
& Gini coefficient & 0.00\\
& Crime rate  & 0.00\\
  \bottomrule
\end{tabular}
\caption{List of all 20 variables used in the CIG estimation, their type, and their percentage of missing values.}
\label{tab:AllVars}
\end{center}
\end{table}

\subsection{Research Hypothesis}

From the WHO's ICF model \citep[][see Figure~\ref{ICF-Model}]{ICF2001}, we
hypothesized that all variables on functional and general health
perception, and all variables on social status, networks, and supports were
connected via paths within the same component of the CIG.

\subsection{Findings}
\ifpdf
  \begin{figure}
    \centering
    \includegraphics[width=\textwidth]{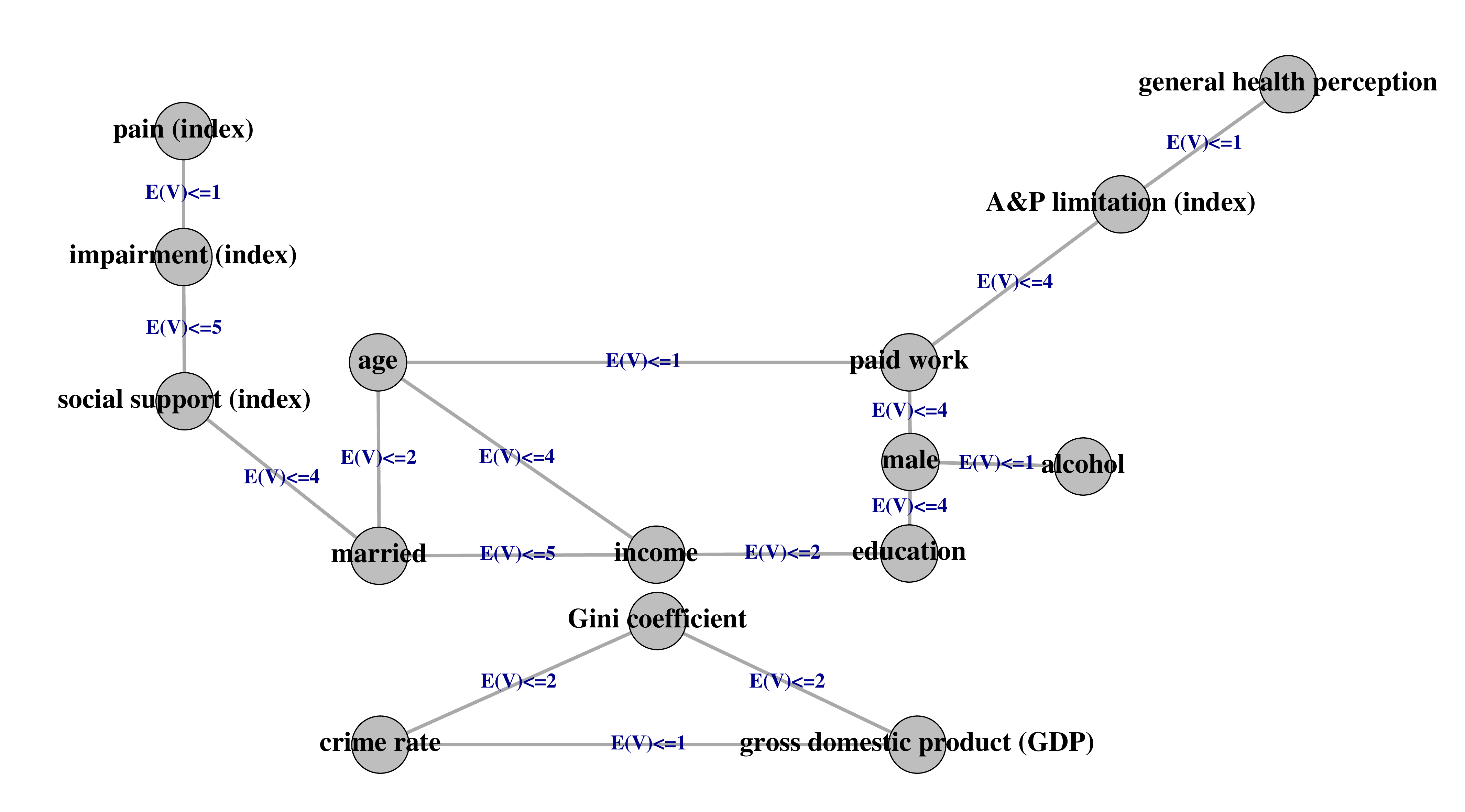}
    \caption{Conditional independence graph of the $p=20$ variables
      (nodes) remaining after construction of indices based on the
      2007 Swiss Health Survey estimated with GRaFo. Edges were
      selected with respect to an upper bound of 5 on the expected
      number of false positives, see formula
      (\ref{StabSelErrorControl}). Five nodes (social network
      utilization, migration background, smoker, work restriction, and
      LPA) were isolated (no edges) and thus
      neglected.}
    \label{SHS-NON-imputed}
  \end{figure}
\fi

Figure \ref{SHS-NON-imputed} shows the resulting graph from our application
of GRaFo to the (non-imputed) data on functional health from the SHS with
casewise deletion of missing values regularized for a bound (as in formula
(\ref{StabSelErrorControl})) for an expected number of false positives
$\mathbb{E}[V]\leq 5$. The selected edge sets for the imputed and casewise
deleted data were quite similar for various bounds on $\mathbb{E}[V]$ and
even identical for $\mathbb{E}[V]\leq 5$ (not shown). In the following, we
thus focus on the CIG derived from the complete observations remaining
after casewise deletion of missing values. As the data contains
  mixed-type variables we did not perform a similar analysis with the
  LASSO (clearly non-favorable dichotomization was used in the simulations
  in Section~\ref{Sec:SimRes}).

The resulting edges for $\mathbb{E}[V] \leq 1$ depict relatively obvious
associations known from everyday observations. Interestingly, general
health perception is conditionally dependent on activity limitation but
conditionally independent of impairment and pain. In the larger graph for
$\mathbb{E}[V]\leq 5$, one sees that general health perception,
impairments, and pain are connected through a path of several environmental
and personal factors such as social support, being married, age, etc. That
implies, for instance, that we do not need information on impairment to
predict general health perception if we have information on activity
limitation and the remaining predictors, whereas activity limitation is an
essential predictor of general health perception even if information on all
the remaining predictors is provided. For instance, a person with a spinal
cord injury who has no activity limitation because of social and
technological supports, could thus still report good health. This finding
is supported by other sources reporting that many people with disabilities
do not consider themselves to be unhealthy
\citep{WorldRepDisab2011,Watson2002}. In the 2007-2008 Australian National
Health Survey, 40 percent of people with a severe or profound impairment
rated their health as good, very good, or excellent \citep{AHS2009}.

As regards our hypothesis derived from the ICF model
\citep{ICF2001}, we can confirm that the bulk of individual level
variables form one component and support the biopsychosocial model of
health: Functional and general health influence each other and are
connected with a variety of environmental and personal
factors. However, not all candidate personal and environmental factors
were related in our study. This may be due to our conservative upper
bound on the error that is likely to favor false negatives,
i.e.~missing edges. There may also be an issue with our selection of
variables that was restricted by the choices of the original survey
team. In particular, macro-level variables pertaining information
about the counties, in which the individuals are nested, form a second
component. It may be that their effect is already contained in the
individual-level variables, for example paid work. Five variables do
not appear in the graph entirely: social network utilization,
migration background, smoker, work restriction, and LPA. If we remove
the three macro-level variables GDP, Gini, and crime rate from the
model, the connectivity of the individual-level component does not
change. Instead, the two variables migration background and social
network utilization are now present as a separate component (not shown).

Unfortunately, lack of information on the directions of relationships is a
weakness of CIGs. Also, condition (A) of Theorem 1 and the exchangeability
condition have likely been violated. One disadvantage of the
  randomForest implementation is the inability to model continuous
  variables with $<6$ unique values, which may oftentimes be an issue for
  the sum indices in combination with subsampling. Consequently, we chose
  to model them as categorical variables. Regardless, given the high face
validity of the findings and the achievement of error control in the mixed
setting for small $p$ in Section~\ref{Sec:SimRes}, the results seem
satisfactory.

The runtime of GRaFo depends also on $n$, even if $p$ is small. Hence,
estimation of the SHS graph was executed in parallel on 10 cores of
the BRUTUS cluster with a runtime of roughly 8 hours.

\section{Modeling Neurodevelopment in Children Experiencing Open-Heart Surgery}

\textbf{Here we demonstrate an application of GRaFo to a research
  question, where $p$ is much larger than $n$. It is thus of
  particular interest, whether GRaFo can suggest meaningful
  associations or tends to produce seemingly spurious associations.
}

\subsection{Neurodevelopment after Open-Heart Surgery}

In children with complex congenital heart disease (CHD) neurological and
developmental alterations are common
\citep{Bellinger2003,Snookes2010,Ballweg2007}. The observed cognitive,
behavioral, and motor deficits can significantly impact daily routine and
educational perspectives and lead to a high rate of special schooling and
supportive therapies in this population
\citep{vonRhein2012,HH2006,HH2008}. In severe congenital heart disease
requiring open-heart surgery, factors can be further subdivided into pre-,
peri-, and post-operative factors. One of the major limitations of studies
on patient specific risk-factors \citep{Ballweg2007,HH2006,HH2008},
treatment and bypass protocols \citep{Bellinger2003,Snookes2010}, and
post-operative complications \citep{Bellinger2003,Snookes2010} is the
inability to provide a full picture of the interplay of all potentially
relevant risk-factors available in the data. Thus, understanding their
common association structure is of large interest.

\subsection{Study Population}

A group of 221 infants with a congenital heart disease that underwent open-heart
surgery with full-flow cardiopulmonary bypass prior to their first birthday
from a study of the Children Hospital Zurich from 2004 to 2008
\citep{vonRhein2012}. We restricted our sample to a more homogeneous
sub-population of 34 infants suffering from trisomy 21 of whom 14 were male
and 31 caucasian.

\subsection{Variables}

In total, 133 variables were available for modeling. They can further be
subdivided into 40 variables describing basic characteristics (e.g.~birth
parameters, family information), 10 variables characterizing a child's
neurodevelopment prior to surgery, 69 peri-operative factors
(i.e.~data on pre-operative, intra-operative, and post-operative course), 13
variables characterizing a child's neurodevelopment 1 year post surgery,
and 1 variable summarizing quality of life based on the TAPQOL
questionnaire \citep{TNO2004}.

To ease interpretation, we focus in Table~\ref{tab:AllVarsKiSpiGraph} on
the 29 variables which had at least one adjacent node in the resulting
graph which we discuss below. These variables are of mixed-type, with 23
continuous variables and 6 factors with more than 2 levels.

Outcome-variables of primary interest are the Bayley score for motor development and
the Bayley score for cognitive development \citep{Bayley1993}. Both scores
were assessed at one year of age.

\begin{table}
\begin{center}
\begin{tabular}[H]{llll}
  \toprule
  Scale & Group & Variable & Missing\\
  \midrule
  cont & birth/family & Apgar score 5 mins & 11\\
  cont & birth/family & Apgar score 1 mins & 11\\
  cont & birth/family & Apgar score 10 mins & 10\\
  cont & birth/family & birth weight & 1\\
  cont & birth/family & gestational age & 0\\
  cont & birth/family & birth length & 1\\
  cont & birth/family & father age & 1\\
  cont & birth/family & mother age & 0\\
  $>2$ & birth/family & father school education & 2\\
  $>2$ & birth/family & father professional education & 2\\
  cont & birth/family & socio economic status & 1\\
  $>2$ & birth/family & mother school education & 1\\
  $>2$ & birth/family & mother number pregnancies & 1\\
  $>2$ & birth/family & mother number births gestational age $> 24$ weeks & 1\\
  cont & peri-operative & time aorta occlusion & 0\\
  $>2$ & peri-operative & operation risk & 0\\
  cont & peri-operative & lactate max during surgery & 1\\
  cont & peri-operative & lactate max 24h post surgery& 0\\
  cont & peri-operative & age at surgery & 0\\
  cont & peri-operative & lowest SO2 during surgery & 0\\
  cont & peri-operative & lowest SO2 24h post surgery & 0\\
  cont & peri-operative & length at surgery & 0\\
  cont & peri-operative & weight at surgery & 0\\
  cont & peri-operative & head circumference at surgery & 0\\
  cont & 1 year post surgery & weight at 1 year & 5\\
  cont & 1 year post surgery & length at 1 year & 5\\
  cont & 1 year post surgery & head circumference at 1 year & 5\\
  cont & 1 year post surgery & Bayley motor score & 5\\
  cont & 1 year post surgery & Bayley cognitive score & 6\\
  \bottomrule
\end{tabular}
\caption{List of all 29 variables which appear in the graph, their
  scale type \textbf{($>2$ for categorical; cont.~for continuous)} , variable
  group, and their percentage of missing values.}
\label{tab:AllVarsKiSpiGraph}
\end{center}
\end{table}

In total, 3.4 percent of the data were missing, ranging from 87 completely
observed variables to 3 variables with 11 missing observations (two Apgar
score variables \citep[see also][]{Apgar1953} and the child's head
circumference at birth (not in graph)). Case-wise exclusion of children
with missing values seems infeasible as this would result in the loss of 26
children. Data were thus imputed using the missForest procedure
\citep{Stekhoven2011}.

\subsection{Objective}

To identify risk-factors associated with the cognitive and motor development
of infants that have undergone open-heart surgery in the first 12 months
after birth due to a congenital heart disease using GRaFo.

\textbf{Due to the large number of variables, many methods of
  analysis (such as bivariate correlations) may be prone to yield
  various spurious associations. It is here thus also of interest to
  demonstrate that, whenever GRaFo suggests an association, it tends
  to have a high face validity (which is judged by the collaborating health
  professionals).
}

\subsection{Findings}

\begin{figure}
    \centering
      \includegraphics[width=\textwidth]{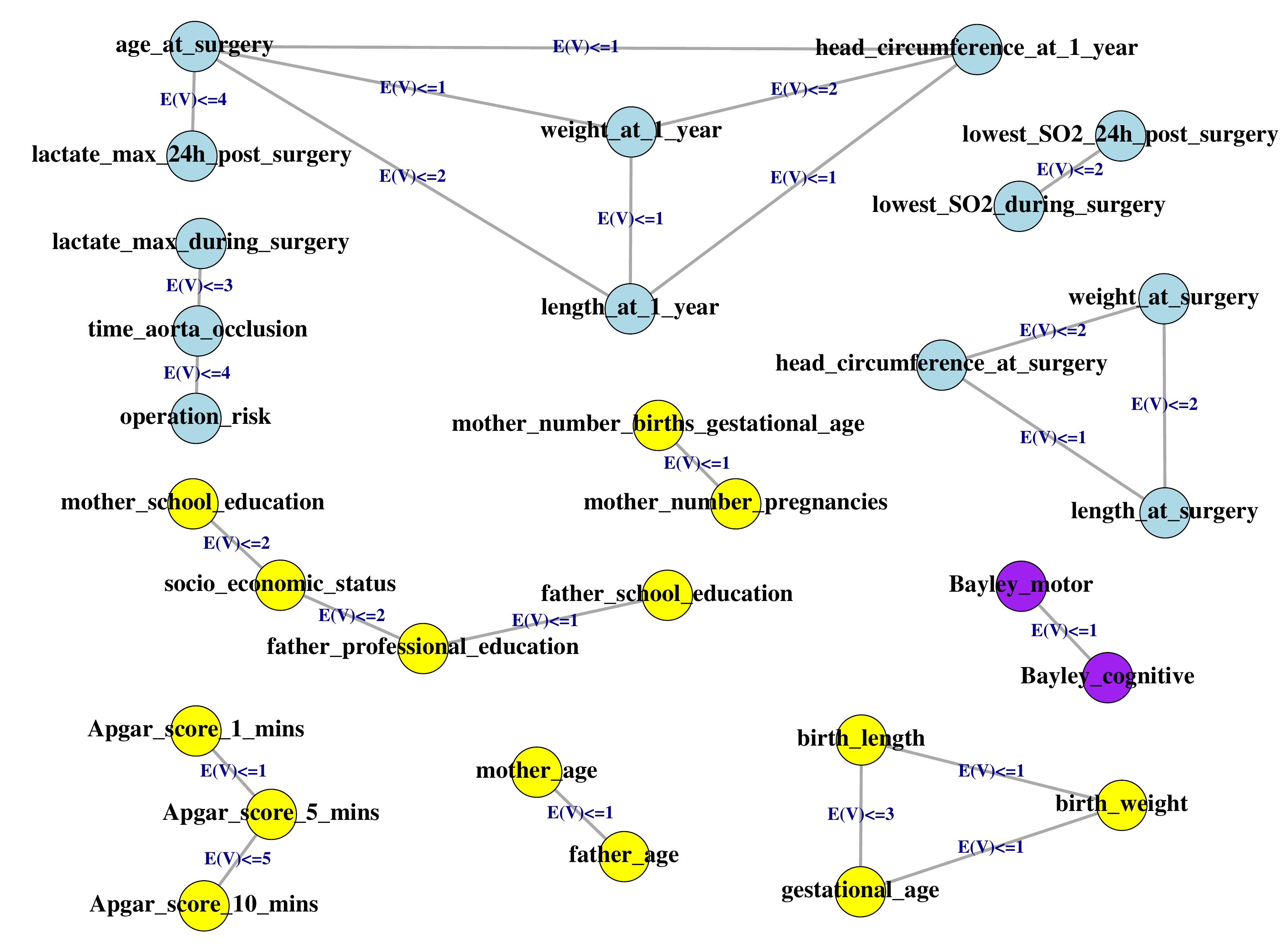}
    \caption{The figure shows the conditional independence graph of
      children with trisomy 21 experiencing open-heart surgery. The reported
      $p=29$ variables (nodes) have at least one adjacent node for an upper
      bound of 5 on the expected number of false positives $\mathbb{E}[V]$.}
    \label{Fig:KiSpi-Graph}
  \end{figure}

For an upper bound of 5 on the expected number of false positives
$\mathbb{E}[V]$ we find that the Balyey scores for motor and cognitive
development are only associated with each other, but not with any other
node in the graph (conditional the remainder) in
Figure~\ref{Fig:KiSpi-Graph}. We do, however, find 10 small clusters of
high face-validity. For example, the age of each child's father and mother
form a common cluster. Likewise, the children's Apgar score after 1 minute
is connected with the Apgar score after 5 minutes. The latter furthermore
connects with the Apgar score after 10 minutes. It thus seems that GRaFo
manages to identify many edges which appear intuitively correct, but it
fails to provide new insights into the association structure of the Bayley
scores. \textbf{On the other hand, no apparent ``odd'' associatons
  were suggested}.

 This result mirrors current knowledge about the neurodevelopment of infants
 after open heart surgery: genetic defects
 \citep{Bellinger2003,Snookes2010,Ballweg2007} and ethnicity
 \citep{Ballweg2007} have been described as relevant risk-factors for
 adverse neurodevelopment. As we mostly worked with caucasian
 children, all of whom have trisomy 21, these factors have already
 been controlled for by the design. Even if we increase the upper
 bound on $\mathbb{E}[V]$ to 50 we still cannot find any additional
 variables connected to the Bayley scores. The plausibility of the
 other observed clusters would thus suggest, that no stable
 associations with the Bayley scores can be identified using GRaFo.

  However, potential bias induced by the imputation method which also
  utilizes Random Forests cannot be excluded. For example, all Apgar scores
  showed a large number of missing values. The identified cluster may thus
  also be an artifact of the missing value imputation. Furthermore, our
  choice of variables was determined by the original study design. Also, we
  cannot guarantee that the exchangeability assumption
  \citep{Meinshausen2010} and assumption (A) from Theorem 1 hold.

The small number of children ($n\ll p$) allowed to run this analysis on an
AMD Athlon 64 X2 5600+ PC with 6 GB of memory in just under 14 minutes.

\section{Conclusion}

We propose GRaFo (Graphical Random Forests) performed satisfactory,
mostly on par or superior to StabLASSO, \textbf{StabcForests, LASSO,
  Conditional Forests, Random Forests, and ML estimation}. Error
control of false positive edges could be achieved in all but the
mixed-type simulation with $p=200$ \textbf{and the nonlinear Gaussian
  setting with $p\geq 100$}. Violation of assumption~(A) in Theorem~1 and of the
exchangeability condition might be responsible for this behavior. In
contrast, in most of the other settings GRaFo was very conservative
and observed false positive edges were well below their expected upper
bound. The Ising model, the sole model not based on DAGs, was
particularly hard for both GRaFo and StabLASSO resulting in few true
positives if error bounds were chosen very small.

\textbf{Results in the Gaussian setting with interactions were very
  similar to the main effects Gaussian setting, which is likely due to
  the small number of interactions in our simulation model. On the
  contrary, GRaFo shows a clear gain over StabLASSO in the nonlinear
  setting, where half of the associations were nonlinear in nature.  }

Poor results for the LASSO in the multinomial and mixed case, where
  we need dichotomization, may be improved by feasible modifications of the
  LASSO, such as an extension of the group LASSO \citep{Meier2008} to
  multinomial responses \citep{Dahinden2010}. However, penalization if both discrete and
  continuous variables are included is not a straightforward task
  (including the issue of scaling).

\textbf{The ML results indicate that both GRaFo and StabcForests
  perform very well in the mixed setting, though the computational
  cost of StabcForests notably exceeds the cost of GRaFo. Both
  Forests-based algorithms used marginal permutation importance as the
  conditional permutation importance turned out impractical due to its
  high computational cost.
}

The Swiss Health Survey graph consists of an individual- and a macro-level
variable cluster which were highly stable with respect to the way of
handling missing values. Exclusion of the macro-level cluster did not
affect the individual-level cluster. For a small error bound, our
hypothesis that all factors should connect could not be fully confirmed,
though a strong tendency toward the ICF's biopsychosocial model of health
was evident in the individual-level cluster.

The children hospital graph consists of many clusters of high
face-validity. \textbf{We believe this emphasizes GRaFo's potential to
  isolate true and stable associations}. However, we failed to
identify any new potential risk factors that may help to explain
adverse neurodevelopment (since no edges connect to the corresponding
outcome measures). The known risk factors ethnicity and genetic
defects were controlled for by the design. \textbf{This may be a
  consequence of the available pool of variables. Also, it is
  imaginable, that some associations are only of importance for a
  sub-group of the study. In this case, they would appear to be
  instable to GRaFo and consequently not be reported.}

\section{Proof of Theorem \ref{TheoremCigMeanEst}}
\label{Sec:Proofs}

Proof: We know that $X_j\independent X_i|{\bf
  X}\setminus\{X_j,X_i\}$ is equivalent to
\begin{align}
\label{eq:equalProb} \mathbb{P}[X_j\leq x_j|\{x_h; h\neq j\}] = \mathbb{P}[X_j\leq x_j|\{x_h; h\neq j,i\}]
\end{align}
for all realizations $x_j$ of $X_j$ and $\{x_h; h\neq j\}$ of ${\bf X}\setminus
\{X_j\}$.  Due to assumption (A) we can rewrite (\ref{eq:equalProb}):
\begin{align}
\label{eq:equalCumDist}
F_j(x_j|m_j(\{x_h; h\neq j\}))=F_j(x_j|m_j(\{x_h; h\neq j,i\}))
\end{align}
for all $x_j$ and all $\{x_h; h\neq j\}$. But (\ref{eq:equalCumDist})
is equivalent to
\begin{align}
m_j(\{x_h; h\neq j\})=m_j(\{x_h; h\neq j,i\})
\end{align}
for all $\{x_h; h\neq j\}$. This completes the proof.\\

\section{Acknowledgment}
The authors would like to thank three anonymous reviewers, Gerold
Stucki, Markus Kalisch, Marloes Maathuis, Philipp R\"utimann, and
Holger H\"ofling for valuable feedback and discussion.

\section{References}

\bibliographystyle{model2-names}
\bibliography{references}







\end{document}